\documentclass[twocolumn]{aa}  

\usepackage{graphicx}
\usepackage[varg]{txfonts}

\usepackage{ulem}

\usepackage[utf8]{inputenc}

\usepackage{supertabular}
\usepackage{longtable}
\usepackage{lscape}

\usepackage{multicol}
\usepackage{natbib,twoopt}
 \bibpunct{(}{)}{;}{a}{}{,}    

\begin{document}

	\title{The applicability of FIR fine-structure lines as Star Formation Rate tracers over wide ranges of metallicities and galaxy types} 
         \titlerunning{The applicability of FIR fine-structure lines as Star Formation Rate tracers}

\author{Ilse De Looze\inst{1}
  \and Diane Cormier\inst{2}
 \and Vianney Lebouteiller\inst{3}
  \and Suzanne Madden\inst{3} 
 \and Maarten Baes\inst{1}
 \and George J. Bendo\inst{4}
  \and M{\'e}d{\'e}ric Boquien\inst{5}
  \and Alessandro Boselli\inst{6}
  \and David L. Clements\inst{7}  
  \and Luca Cortese\inst{8,9}
  \and Asantha Cooray\inst{10,11}
  \and Maud Galametz\inst{8}
  \and Fr{\'e}d{\'e}ric Galliano\inst{3}
  \and Javier Graci{\'a}-Carpio\inst{12}
  \and Kate Isaak\inst{13}
  \and Oskar {\L}. Karczewski\inst{14}
  \and Tara J. Parkin\inst{15}
  \and Eric W. Pellegrini\inst{16}
  \and Aur{\'e}lie R{\'e}my-Ruyer\inst{3}
  \and Luigi Spinoglio\inst{17}
  \and Matthew W. L. Smith\inst{18}
  \and Eckhard Sturm\inst{12}
  } 

\institute{Sterrenkundig Observatorium, Universiteit Gent, Krijgslaan 281 S9, B$-$9000 Gent, Belgium
\and Zentrum f{\"u}r Astronomie der Universit{\"a}t Heidelberg, Institut f{\"u}r Theoretische Astrophysik, Albert-Ueberle Str. 2, D-69120 Heidelberg, Germany
\and Laboratoire AIM, CEA, Universit{\'e} Paris VII, IRFU/Service d$'$Astrophysique, Bat. 709, 91191 Gif-sur-Yvette, France
\and UK ALMA Regional Centre Node, Jodrell Bank Centre for Astrophysics, School of Physics and Astronomy, University of Manchester, Oxford Road, Manchester M13 9PL, United Kingdom
\and Institute of Astronomy, University of Cambridge, Madingley Road, Cambridge, CB3 0HA, UK
\and Laboratoire d$'$Astrophysique de Marseille $-$ LAM, Universit{\'e} Aix-Marseille \& CNRS, UMR7326, 38 rue F. Joliot-Curie, 13388 Marseille CEDEX 13, France
\and Astrophysics Group, Imperial College, Blackett Laboratory, Prince Consort Road, London SW7 2AZ, United Kingdom
\and European Southern Observatory, Karl Schwarzschild Str. 2, 85748 Garching, Germany
\and Centre for Astrophysics and Supercomputing, Swinburne University of Technology, PO Box 218, Hawthorn, VIC 3122, Australia
\and Department of Physics \& Astronomy, University of California, Irvine, CA 92697, USA
\and Division of Physics, Astronomy and Mathematics, California Institute of Technology, Pasadena, CA, 91125, USA
\and Max-Planck-Institute for Extraterrestrial Physics (MPE), Giessenbachstra$\ss$e 1, 85748 Garching, Germany
\and ESA Research and Scientific Support Department, ESTEC/SRE-SA, Keplerlaan 1, 2201 AZ Noordwijk, The Netherlands
\and Department of Physics \& Astronomy, University of Sussex, Brighton, BN1 9QH, UK
\and Department of Physics \& Astronomy, McMaster University, Hamilton, Ontario, L8S 4M1, Canada
\and Department of Physics \& Astronomy, University of Toledo, Toledo, OH 43606, USA
\and Istituto di Astrofisica e Planetologia Spaziali, INAF-IAPS, Via Fosso del Cavaliere 100, I-00133 Roma, Italy
\and School of Physics and Astronomy, Cardiff University, Queens Buildings, The Parade, Cardiff, CF24 3AA, UK}

\date{Received 2013 August 14 / Accepted 2014 May 7}

\abstract{}
{We analyze the applicability of far-infrared fine-structure lines [C{\sc{ii}}] 158\,$\mu$m, [O{\sc{i}}] 63\,$\mu$m and [O{\sc{iii}}] 88\,$\mu$m to reliably trace the star formation rate (SFR) in a sample of low-metallicity dwarf galaxies from the \textit{Herschel} Dwarf Galaxy Survey and compare with a broad sample of galaxies of various types and metallicities in the literature.}
{We study the trends and scatter in the relation between the SFR (as traced by GALEX $FUV$ and MIPS\,24\,$\mu$m) and far-infrared line emission, on spatially resolved and global galaxy scales, in dwarf galaxies. We assemble far-infrared line measurements from the literature and infer whether the far-infrared lines can probe the SFR (as traced by the total-infrared luminosity) in a variety of galaxy populations.}
{In metal-poor dwarfs, the [O{\sc{i}}]$_{63}$ and [O{\sc{iii}}]$_{88}$ lines show the strongest correlation with the SFR with an uncertainty on the SFR estimates better than a factor of 2, while the link between [C{\sc{ii}}] emission and the SFR is more dispersed (uncertainty factor of 2.6). The increased scatter in the SFR-$L_{\text{[CII]}}$ relation towards low metal abundances, warm dust temperatures and large filling factors of diffuse, highly ionized gas suggests that other cooling lines start to dominate depending on the density and ionization state of the gas. \\
For the literature sample, we evaluate the correlations for a number of different galaxy populations. The [C{\sc{ii}}] and [O{\sc{i}}]$_{63}$ lines are considered to be reliable SFR tracers in starburst galaxies, recovering the star formation activity within an uncertainty of factor 2. For composite and AGN sources, all three FIR lines can recover the SFR with an uncertainty factor of 2.3. The SFR calibrations for ULIRGs are similar to starbursts/AGNs in terms of scatter but offset from the starburst/AGN SFR relations due to line deficits relative to their total-infrared luminosity. While the number of detections of the FIR fine-structure lines is still very limited at high-redshift for [O{\sc{i}}]$_{63}$ and [O{\sc{iii}}]$_{88}$, we provide a SFR calibration for [C{\sc{ii}}]. }
{}
\keywords{Galaxies: abundances --
Galaxies: dwarf --
Galaxies: ISM --
Galaxies: star formation
}
\maketitle
%

\section{Introduction}
Star formation encompasses the birth of new stars through the fragmentation and contraction of cold, dense molecular gas, hereby recycling the ISM content of galaxies \citep{2007ARA&A..45..565M}. 
Knowing the instantaneous level of star formation (i.e. the star formation rate) not only sheds light on the conditions in the ISM, but also on the evolution of galaxies and their formation processes.
If we want to understand the physical processes that control galaxy evolution, being able to probe the star formation activity is of great importance. Tracing back star formation in some of the first objects in the early Universe until the present-day could even allow us to probe the star formation activity throughout cosmic times.

Since the main coolants in metal-rich galaxies such as the Milky Way are mostly metal-based ([C{\sc{ii}}], [O{\sc{i}}], [O{\sc{iii}}], CO, dust), the metal abundance is considered a fundamental parameter in the regulation of star formation through its influence on the initial cooling of diffuse gas (e.g. \citealt{2014MNRAS.437....9G}) and the survival of clouds through shielding. Star formation in the early Universe is considered to differ significantly from the present day's gas consumption in galaxies. Due to the extremely low metal abundances in the early Universe, the gas coolants and initiation processes of star formation were likely to be different from star formation conditions in the Local Universe (e.g. Ly $\alpha$ cooling becomes more important). Nearby low-metallicity galaxies might be important laboratories to investigate the connection between the chemical enrichment and star formation processes. Although the present-day metal-poor dwarfs will have experienced some evolution throughout cosmic time (e.g. \citealt{1991ApJ...379..621H,2007IAUS..235...65T}), their slow chemical evolution makes them important testbeds to understand metallicity effects potentially applicable to galaxies in the early Universe.

Star formation rates (SFRs) on global galaxy scales are typically estimated from scaling relations between diagnostic tracers of the star formation activity, calibrated against the most up to date stellar population synthesis models and characterization of the initial mass function (IMF).
Up to the present day, the most widely used SFR diagnostics have been based on continuum bands and optical/near-IR emission lines (see \citealt{1998ARA&A..36..189K} and \citealt{2012ARA&A..50..531K} for a detailed overview). The brightest cooling lines in the atomic and molecular medium are emitted from mid-infrared to radio wavelengths, which have been probed extensively with the \textit{Herschel} Space Observatory \citep{2010A&A...518L...1P}. Follow-up is guaranteed with SOFIA in the local Universe or, at high spatial resolution, with ground-based interferometers such as ALMA and the future NOEMA in the high-redshift Universe. These facilities open up a whole new spectral window, which favors the use of far-infrared (FIR) and submillimeter (submm) continuum bands (e.g. PACS\,70\,$\mu$m, \citealt{2013ApJ...768..180L}) and emission lines as SFR diagnostics across a large variety of galaxy populations.

Here, we investigate the utility of the three brightest fine-structure cooling lines, [C{\sc{ii}}], [O{\sc{i}}]$_{63}$ and [O{\sc{iii}}]$_{88}$ (e.g. \citealt{2001ApJ...553..121H,2008ApJS..178..280B}; Cormier et al. in prep.), as tracers of the star formation activity in a sample of low-metallicity dwarf galaxies from the \textit{Herschel} Dwarf Galaxy Survey \citep{2013PASP..125..600M}. We, furthermore, asses the influence of metallicity, which is an important parameter controlling star formation in galaxies, constraining the reservoir of dust grains for the formation of molecules and regulating the attenuation of the FUV photons necessary for shielding molecules.

The [C{\sc{ii}}] 157.74 $\mu$m line has been put forward as a potential powerful tracer of the star formation activity in the nearby as well as the more distant Universe \citep{1991ApJ...373..423S,2002A&A...385..454B,2010ApJ...724..957S,2011MNRAS.416.2712D,2012ApJ...755..171S} and we, now, aim to expand this analysis to nearby, low-metallicity dwarf galaxies. [C{\sc{ii}}] is considered to be the dominant coolant for neutral atomic gas in the interstellar medium \citep{1985ApJ...291..722T,1985ApJ...291..747T,1995ApJ...443..152W} and, therefore, among the brightest emission lines originating from star-forming galaxies (e.g. \citealt{1991ApJ...373..423S,1997ApJ...491L..27M,2008ApJS..178..280B}). In particular, low-metallicity galaxies show exceptionally strong [C{\sc{ii}}] line emission (e.g. \citealt{1995ApJ...454..293P}, \citealt{1997ApJ...483..200M}, \citealt{2000NewAR..44..249M}, \citealt{2001ApJ...553..121H}, \citealt{2010A&A...518L..57C}, \citealt{2011A&A...531A..19I}).
Carbon has an ionization potential of 11.3 eV (compared to 13.6 eV for hydrogen), implying that line emission can originate from neutral and ionized gas components (see Table \ref{critical}).
A changing contribution of different gas phases on global scales can prevent a correlation between the [C{\sc{ii}}] line emission and level of star formation. The excitation of C$^{+}$ atoms might, furthermore, saturate at high temperatures, where the line becomes insensitive to the intensity of the radiation field at temperatures well above the excitation potential \citep{1999ApJ...527..795K}. The [C{\sc{ii}}] emission can also saturate in neutral gas media with hydrogen densities $n_{\text{H}}$ $\gtrsim$ 10$^{3}$ cm$^{-3}$, where the recombination of C$^{+}$ into neutral carbon and, eventually, CO molecules is favored \citep{1999ApJ...527..795K}. Self-absorption can also affect the [C{\sc{ii}}] line excitation in large column densities of gas ($N_{\text{H}}$ $\sim$ 4 $\times$ 10$^{22}$ cm$^{-2}$, \citealt{1997ApJ...491L..27M}). Although the [C{\sc{ii}}] line is usually not affected by extinction, optical depth effects might become important in extreme starbursts \citep{1998ApJ...504L..11L,2000isat.conf..337H} and edge-on galaxies \citep{1994ApJ...436..720H}. On top of this, deficits in the [C{\sc{ii}}]/FIR ratio towards warm dust temperatures \citep{1985ApJ...291..755C,1991ApJ...373..423S,1997ApJ...491L..27M,2001ApJ...561..766M,2003ApJ...594..758L,2005SSRv..119..355V,2008ApJS..178..280B,2011ApJ...728L...7G,2012ApJ...747...81C,2013ApJ...774...68D,2013ApJ...776...38F} suggest that the [C{\sc{ii}}] conditions might be different in galaxies which are offset from the main sequence of star-forming galaxies. Since carbon can also be significantly depleted on carbon-rich dust grains, the use of O-based gas tracers might instead be preferred.

The [O{\sc{i}}]$_{63}$ line has a critical density $n_{\text{crit,H}}$ $\sim$ 5 $\times$ 10$^{5}$ cm$^{-3}$ and upper state energy $E_{\text{u}}/k$ $\sim$ 228 K (see Table \ref{critical}), which makes it an efficient coolant in dense and/or warm photo-dissociation regions (PDRs).
Although generally observed to be the second brightest line (after [C{\sc{ii}}]), the [O{\sc{i}}]$_{63}$ line is observed to be brighter in galaxies with warm FIR colors and/or high gas densities \citep{2001ApJ...561..766M,2008ApJS..178..280B,2012A&A...548A..91L}.
The applicability of [O{\sc{i}}]$_{63}$ as a SFR calibrator might, however, be hampered by self-absorption \citep{1996AAS...189.2102K,1996ApJ...462L..43P}, optical depth effects (more so than [C{\sc{ii}}]) and the possible excitation of [O{\sc{i}}]$_{63}$ through shocks \citep{1989ApJ...342..306H}. In the situation that the gas heating is not longer dominated by the photo-electric effect but has an important contribution from other heating mechanisms (e.g. mechanical heating, soft X-ray heating), the origin of the line emission might differ from the paradigm of warm and/or dense PDRs. A possible origin of [O{\sc{i}}]$_{63}$ emission different from PDRs can be, in particular, expected in the most metal-poor galaxies characterized by overall low metal content and diminished PAH abundances.

With an energy of the upper state $E_{\text{u}}/k$  $\sim$ 163 K, critical density $n_{\text{crit,e}}$ $\sim$ 510 cm$^{-3}$ and a high ionization potential of 35.1 eV for O$^{+}$, the [O{\sc{iii}}]$_{88}$ line originates from diffuse, highly-ionized regions near young O stars. 
Ionized gas tracers such as [O{\sc{iii}}]$_{88}$ might gain in importance in low-metallicity environments where PDRs occupy only a limited volume of the ISM judging from their weak PAH emission \citep{2004A&A...428..409B,2005ApJ...628L..29E,2006ApJ...646..192J,2006A&A...446..877M,2007ApJ...663..866D,2008ApJ...678..804E,2008ApJ...672..214G} and low CO abundance \citep{1995ApJ...454..293P,1996ApJ...465..738I,1997ApJ...483..200M,2011A&A...531A..19I}. Based on our \textit{Herschel} observations, we focus on [O{\sc{iii}}]$_{88}$ as ionized gas tracer. Due to the low critical densities for [O{\sc{iii}}]$_{88}$ excitation with electrons, other lines (e.g. [O{\sc{iii}}]$_{52}$ but also optical lines such as [O{\sc{iii}}] 5007$\AA$ and H$\alpha$) will likely dominate the cooling of ionized gas media for intermediate and high gas densities. The brightness of the [O{\sc{iii}}]$_{88}$ emission line in metal-poor galaxies (Cormier et al. in prep.), however, hints at a SFR tracer with great potential for the high-redshift Universe.

On top of the SFR calibrations for single FIR lines, we try to combine the emission of [C{\sc{ii}}], [O{\sc{i}}]$_{63}$ and [O{\sc{iii}}]$_{88}$ lines to trace the SFR.
The total gas cooling in galaxies scales with the SFR assuming that the ISM is in thermal equilibrium and the total cooling budget balances the gas heating. Therefore, any FIR fine-structure line can be considered a reliable tracer of the SFR if it plays an important role in the cooling of gas that was heated by young stellar photons. Other heating mechanisms unrelated to the UV radiation (e.g. mechanical heating, cosmic ray heating and X-ray heating) and, thus, not directly linked to SFR, might disperse this link.  Ideally, we combine the emission of several cooling lines in the ultraviolet/optical (e.g. Ly $\alpha$, H$\alpha$, [O{\sc{iii}}] 5007$\AA$) and infrared (e.g. [Ne{\sc{iii}}] 16\,$\mu$m, [S{\sc{iii}}] 19, 33\,$\mu$m, [C{\sc{ii}}] 158\,$\mu$m,  [O{\sc{i}}] 63, 145\,$\mu$m, [N{\sc{ii}}] 122, 205\,$\mu$m, [O{\sc{iii}}] 52, 88\,$\mu$m, [N{\sc{iii}}] 57\,$\mu$m and [Si{\sc{ii}}] (35\,$\mu$m)) wavelength domains to cover the total gas cooling budget in galaxies. Given our focus on the far-infrared coolants, we attempt to obtain a more complete picture of the overall cooling budget by combining the [C{\sc{ii}}], [O{\sc{i}}]$_{63}$ and [O{\sc{iii}}]$_{88}$ lines.

The \textit{Herschel} Dwarf Galaxy Survey (DGS) and observations are presented in Section \ref{data}, together with the acquisition and processing of ancillary data, which will be used as reference SFR calibrators. In Section \ref{Res.sec}, we take advantage of the high spatial resolution attained for the most nearby galaxies to study the trends and scatter in the spatially resolved relation between the SFR and FIR line emission.
The SFR calibrations and trends with the scatter in the SFR-$L_{\text{line}}$ relations based on global galaxy measurements for the entire DGS sample are presented in Section \ref{Int.sec}. 
SFR calibrations for each of the fine-structure lines [C{\sc{ii}}], [O{\sc{i}}]$_{63}$ and [O{\sc{iii}}]$_{88}$ are derived for a selection of different galaxy populations in Section \ref{Extend.sec}.
In Section \ref{Conclusions.sec}, we draw together our conclusions. Appendix \ref{DGScompare} discusses the applicability of different unobscured (Section \ref{compareSFR}) and obscured (Section \ref{compareSFRobsc}) indicators as reference SFR tracers for the low-metallicity DGS sample. Appendix \ref{Compare.sec} presents a comparison between \textit{Herschel} and ISO spectroscopy. Tables with source information, measurements of reference SFR calibrators and FIR lines are presented in Appendix \ref{tables} for the DGS sample, the literature sample of galaxies with starburst, composite or active galactic nucleus classifications and high-redshift sources.

\begin{table*}
\caption{Excitation conditions of the fine-structure lines [C{\sc{ii}}], [O{\sc{i}}]$_{63}$ and [O{\sc{iii}}]$_{88}$, with column 2 providing the ionization potential (IP) to create the species. Columns 3, 4 and 5 give the energy of the upper state and critical density for line excitation through collisions with hydrogen atoms and electrons, respectively, calculated for a kinetic temperature $T_{\text{kin}}$ $=$ 8000K. Column 6 provides the reference to the literature work reporting the excitation conditions of each fine-structure line. Column 7 specifies the ISM phase(s) from which the lines might originate.}
\label{critical}
\centering
\begin{tabular}{|l|c|c|c|c|c|c|}
\hline 
Line & IP\tablefootmark{a} & $E_{\text{u}}/k$ & $n_{\text{crit,H}}$ & $n_{\text{crit,e}}$ & Ref\tablefootmark{b} & Origin? \\
\hline 
& [eV] & [K] & [cm$^{-3}$] & [cm$^{-3}$] & & \\
\hline 
$[{\rm CII}]$ & 11.3 & 91 & 1.6 $\times$ 10$^{3}$ & 44 & 1 & PDRs, diffuse H{\sc{i}} clouds, diffuse ionized gas, H{\sc{ii}} regions  \\
$[{\rm OI}]_{63}$ & - & 228 & 5 $\times$ 10$^5$ & - & 2 & warm and/or dense PDRs \\
$[{\rm OIII}]_{88}$ & 35.1 & 163 & - & 510 & 3 & low excitation, highly ionized gas \\
\hline 
\end{tabular}
\tablefoot{
\tablefoottext{a}{\footnotesize The ionization potential refers to the ionization energy of an atom to create the species in Column 1. For example, the ionization potential of 35.1 eV for [O{\sc{iii}}]$_{88}$ indicates the energy required to remove another electron from O$^{+}$ and, thus, ionize O$^{+}$ in order to create O$^{++}$.}\\
\tablefoottext{b}{\footnotesize References: (1) \citet{2012ApJS..203...13G}; (2) \citet{1985ApJ...291..722T}; (3) \citet{1999ApJS..123..311A}.}} 
\end{table*}

\section{Dwarf Galaxy Survey}
\label{data}
\subsection{Sample characteristics}
The Dwarf Galaxy Survey (DGS, \citealt{2013PASP..125..600M}) is a \textit{Herschel} Guaranteed Time Key Program, gathering the PACS \citep{2010A&A...518L...2P} and SPIRE \citep{2010A&A...518L...3G} photometry and PACS spectroscopy of 50 dwarf galaxies in 230 hours.
The sample was selected to cover a wide range in metallicities from 12+$\log$(O/H) = 8.43 (He\,2-10, 0.55\,$Z_{\odot}$) down to 7.14 (I\,Zw18, 0.03\,$Z_{\odot}$)\footnote{Oxygen abundances, which are used here to constrain metallicities, are calculated from optical line intensities following the prescriptions in \citet{2005ApJ...631..231P}, assuming a solar oxygen abundance O/H$_{\odot}$ = 4.9 $\times$ 10$^{-4}$, or 12+$\log$(O/H)$_{\odot}$=8.7 \citep{2009ARA&A..47..481A}.}.
The sample selection was furthermore optimized to maximize the availability of ancillary data. 
With distances ranging from several kpc to 191 Mpc, the Dwarf Galaxy Survey observes the line emission of more distant galaxies within a single beam (PACS beams have full-width at half maximum (FWHM) of 11.5$\arcsec$, 9.5$\arcsec$ and 9.5$\arcsec$ for [C{\sc{ii}}], [O{\sc{i}}]$_{63}$ and [O{\sc{iii}}]$_{88}$), while the extended far-infrared line emission from the brightest star-forming regions is mapped in the most nearby galaxies by the \textit{Herschel} Space Observatory.
More details about the sample selection as well as a description of the scientific goals of the survey are outlined in \citet{2013PASP..125..600M}.

\subsection{Herschel data}
Spectroscopic mapping of the [C{\sc{ii}}] 158 $\mu$m line was performed for 48 galaxies\footnote{The DGS galaxies UGCA\,20 and Tol\,0618-402 were not observed with the PACS spectrometers onboard \textit{Herschel}.}, among which most galaxies were also covered in [O{\sc{iii}}] 88 $\mu$m (43 out of 48 galaxies) and [O{\sc{i}}] 63 \,$\mu$m (38 out of 48 galaxies).
Far-infrared fine-structure lines [O{\sc{i}}] 145\,$\mu$m, [N{\sc{ii}}] 122\,$\mu$m, [N{\sc{ii}}]  205 $\mu$m and [N{\sc{iii}}] 57 $\mu$m were probed in a subsample of the brightest galaxies (see the histogram in Figure 4 of \citealt{2013PASP..125..600M}). An overview of the spectroscopy observations, data reduction, line flux measurements and line ratios is provided in Cormier et al. (in prep).

\subsection{Reference SFR diagnostic}
\label{Ref.sec}
\begin{figure}[!ht]
   \centering
    \includegraphics[width=8.5cm]{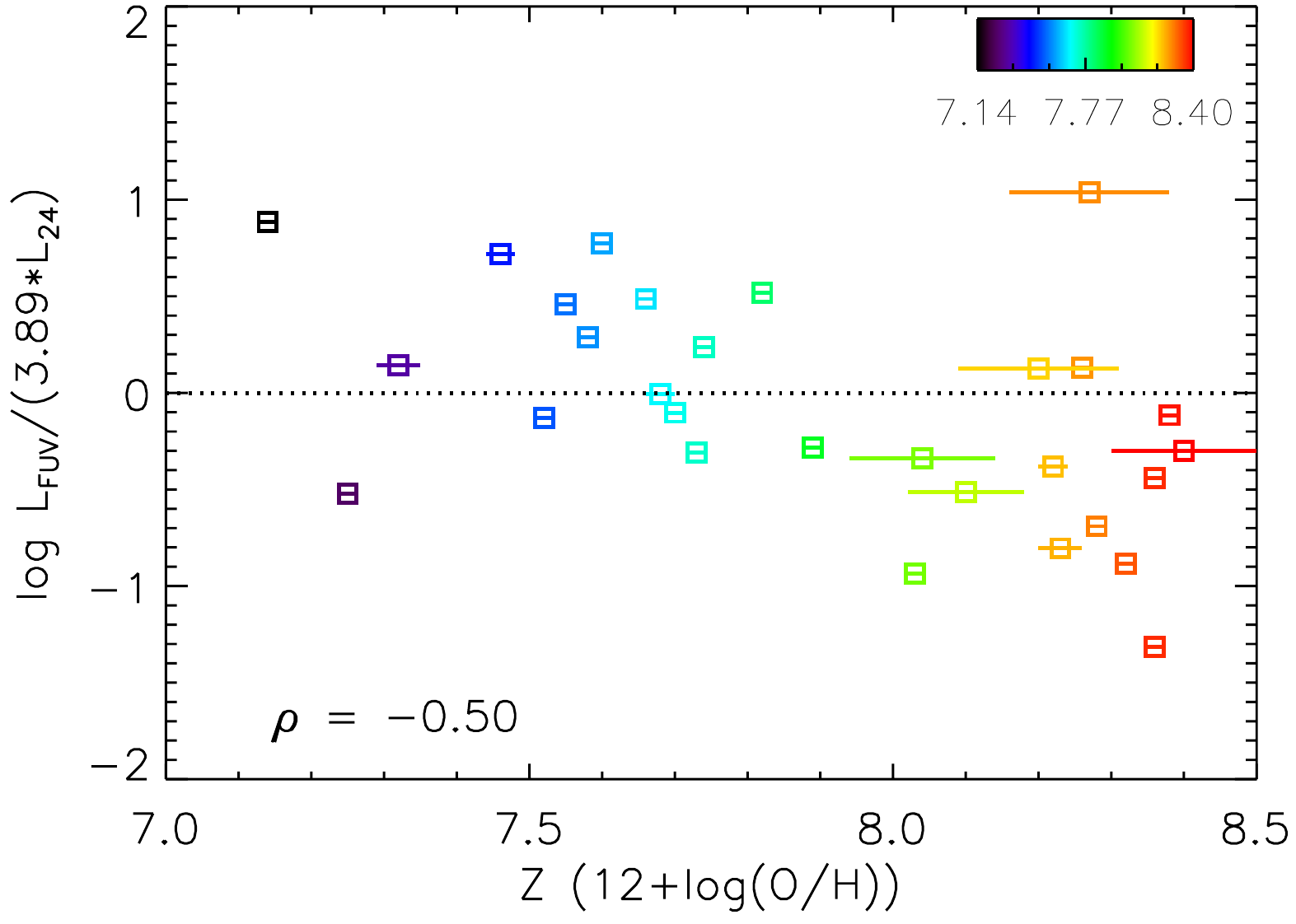}    
   \caption{The ratio of the unobscured ($L_{\text{FUV}}$) versus obscured star formation (3.89*$L_{\text{24}}$), as a function of oxygen abundance, 12+$\log$(O/H), for the DGS sample. The multiplicative factor of 3.89 in the denominator arises from the calibration coefficient that corrects the observed $FUV$ emission for extinction following \citet{2011ApJ...741..124H} (i.e. $L_{\text{FUV}}$(corr) = $L_{\text{FUV}}$(obs) + 3.89 $\times$ $L_{\text{24}}$). Galaxies are color-coded according to metallicity with increasing oxygen abundances going from black over blue, green and yellow to red colors. Data points correspond to global galaxy fluxes.}
              \label{plot_Z_FUVvs24}
    \end{figure}

To establish the applicability of FIR lines to trace the SFR, we need to specify a reference star formation rate tracer.
Typically, combinations of SF diagnostics are used to trace the unobscured and obscured fraction of star formation.
Figure \ref{plot_Z_FUVvs24} shows the ratio of unobscured versus obscured star formation, as probed by GALEX $FUV$ and MIPS\,24\,$\mu$m or our reference calibrators of the unobscured and obscured star formation (see later), respectively, as a function of oxygen abundance for the DGS sample.
The Spearman's rank correlation coefficient, $\rho$, is computed from the IDL procedure r$\_$correlate to quantify the degree of correlation between the oxygen abundance and the ratio of unobscured-versus-obscured star formation. Values of $\rho$ close to +1 and -1 are indicative of a strong correlation or anti-correlation, respectively, while values approaching 0 imply the absence of any correlation.
For the low metallicity (12+$\log$(O/H) $\leq$ 7.8-7.9) galaxies of the DGS sample, the fraction of unobscured star formation starts to dominate (see the negative correlation in Figure \ref{plot_Z_FUVvs24} with $\rho$ = -0.50) and we, thus, need tracers of the unobscured and obscured fraction of star formation.
The increased fraction of unobscured star formation towards lower metal abundances is in agreement with the drop in the ratio of the SFR estimated from the WISE band at 12\,$\mu$m versus the SFR from H$\alpha$ emission, $SFR_{\text{W3}}$/$SFR_{H\alpha}$, with decreasing metallicity reported in \citet{2013ApJ...774...62L}, which was attributed to the low dust-to-gas ratios of metal-poor galaxies (e.g. \citealt{2014A&A...563A..31R}), making the reprocessing of UV photons by dust inefficient \citep{2009MNRAS.394.2001S,2012ApJ...752...64H}.

In this paper, we choose GALEX $FUV$ and MIPS\,24\,$\mu$m as reference SFR tracers to probe the unobscured and obscured star formation component, respectively, and we rely on the SFR calibrations presented by \citet{2011ApJ...741..124H} and \citet{2011ApJ...737...67M} (see Table \ref{SFRref}).
In Section \ref{DGScompare} of the appendix, we motivate this choice of reference SFR calibrators by comparing different unobscured (GALEX $FUV$, H$\alpha$) and obscured (IRAC8\,$\mu$m, MIPS\,24\,$\mu$m, $L_{\text{TIR}}$, 1.4\,GHz) SFR tracers for the DGS galaxy sample.
Among the unobscured SFR diagnostics, we argue that the H$\alpha$ line can provide better SFR estimates compared to $FUV$ due to the limited range of ages to which the SFR calibration for H$\alpha$ ($\sim$ 10 Myr) is sensitive compared to FUV (100 Myr). Given the bursty star formation histories and small sizes of dwarf galaxies, the underlying assumption of continuous star formation activity during the age range of the SFR calibrator is not fulfilled, which becomes worse for the $FUV$ emission probing a longer time scale of star formation activity.
The unavailability of H$\alpha$ maps prevents us from determining the H$\alpha$ emission that corresponds to the galaxy regions covered in our \textit{Herschel} observations. The $FUV$ emission is instead used as reference SFR calibrator, but keeping in mind that $FUV$ might underestimate the SFR by 50$\%$ as compared to H$\alpha$ (see Section \ref{compareSFR}). The analysis in Section \ref{compareSFRobsc} shows that IRAC\,8\,$\mu$m, $L_{\text{TIR}}$ and $L_{\text{1.4\,GHz}}$ are unreliable SFR tracers of the obscured SF fraction due to the dependence of PAH abundance on metallicity, the peculiar SED shapes and/or the burstiness of the star formation histories of metal-poor dwarfs. Although the MIPS\,24\,$\mu$m band emission might not be entirely free of metallicity effects and/or variations in (very small) grain abundance, the analysis of Section \ref{compareSFRobsc} shows that MIPS\,24\,$\mu$m is linked more closely to the SFR over wide ranges of metallicity compared to the other obscured SFR diagnostics.

GALEX $FUV$ data are processed in a similar way to that outlined in \citet{2012A&A...544A.101C}. The data are background subtracted and corrected for Galactic extinction according to the recalibrated $A_{V}$ in \citet{2011ApJ...737..103S} from \citet{1998ApJ...500..525S}, as reported on the NASA/IPAC Extragalactic Database, and assuming an extinction law with $R_V$ = 3.1 from \citet{1999PASP..111...63F}. 
\textit{Spitzer} MIPS data have been retrieved from the \textit{Herschel} Database in Marseille (HeDaM\footnote{http://hedam.lam.fr/}). Details of the data reduction of the ancillary MIPS data set can be found in \citet{2012MNRAS.423..197B}, along with aperture photometry techniques and results. 

\section{Spatially resolved SFR-$L_{\text{line}}$ relation}
\label{Res.sec}
\subsection{Convolution and regridding of the maps}
The closest galaxies in the DGS sample were mapped with \textit{Herschel} to cover most of the star-forming regions and, therefore, can be used to provide a spatially resolved interpretation of the SFR calibrations.
For the spatially resolved analysis, we consider all sources at distances $D$ $\leq$ 7.5 Mpc with GALEX $FUV$ and MIPS 24\,$\mu$m observations\footnote{The distance of 7.5 Mpc has been chosen as a fair compromise between the number of spatially resolved galaxies to be analyzed and the spatial resolution of about 100 pc, corresponding to the nominal pixel size of 3.1333$\arcsec$ at 7.5 Mpc.}. 
A visual inspection of the MIPS 24\,$\mu$m images shows that the mid-infrared emission from VII\,Zw\,403, Mrk\,209, UGC\,4483 and NGC\,625 is barely resolved with respect to the MIPS\,24\,$\mu$m beam (FWHM $\sim$ 6$\arcsec$, \citealt{2007PASP..119..994E}) and, therefore, those galaxies are excluded from the sample for the spatially resolved analysis. We, therefore, end up with a subsample of seven well-sampled galaxies with metal abundances varying from 0.10 Z$_{\odot}$ (NGC\,2366) to 0.38 Z$_{\odot}$ (NGC\,1705).

All maps of the resolved subsample have been convolved from their native resolution ([O{\sc{iii}}]$_{88}$: 9.5$\arcsec$, [O{\sc{i}}]$_{63}$: 9.5$\arcsec$; GALEX $FUV$: 6$\arcsec$; MIPS 24\,$\mu$m: 6$\arcsec$) to the 12$\arcsec$ resolution of PACS at 160\,$\mu$m using the kernels presented in \citet{2011PASP..123.1218A}.
Convolved images are rebinned to maps with pixel size corresponding to regions within galaxies of 114$^{2}$ pc$^{2}$, i.e. the size of a nominal pixel of 3.1333$\arcsec$ at 7.5 Mpc. We caution that pixels of this size are not independent due to the shape of the beam (FWHM of 12$\arcsec$) which is spread across several pixels within one galaxy. Since our main interest is the comparison of the behavior in the relations between the SFR and FIR line emission for different galaxies, we argue that a possible dependence of individual pixels for one specific galaxy will not severely affect the interpretation of our results.
Given that the pixels defined this way correspond to a same physical scale within galaxies, the pixels provide a proxy for the surface density of the SFR, $\Sigma_{SFR}$, and the FIR line surface density, $\Sigma_{\text{line}}$. Only pixels attaining surface brightness levels of signal-to-noise ($S/N$) higher than 5 are taken into consideration, neglecting the uncertainty from the calibration. 
Due to the proximity of NGC\,6822 ($D$ = 0.5 Mpc), the rebinning of pixels results in the absence of any region at sufficient $S/N$ level, limiting the spatially resolved galaxy sample to six objects. For NGC\,1705, the [O{\sc{i}}]$_{63}$ line is not detected at sufficient $S/N$ level on spatially resolved scales.

\subsection{Observed trends}
\label{Trends.sec}
\begin{figure}
   \centering
    \includegraphics[width=8.5cm]{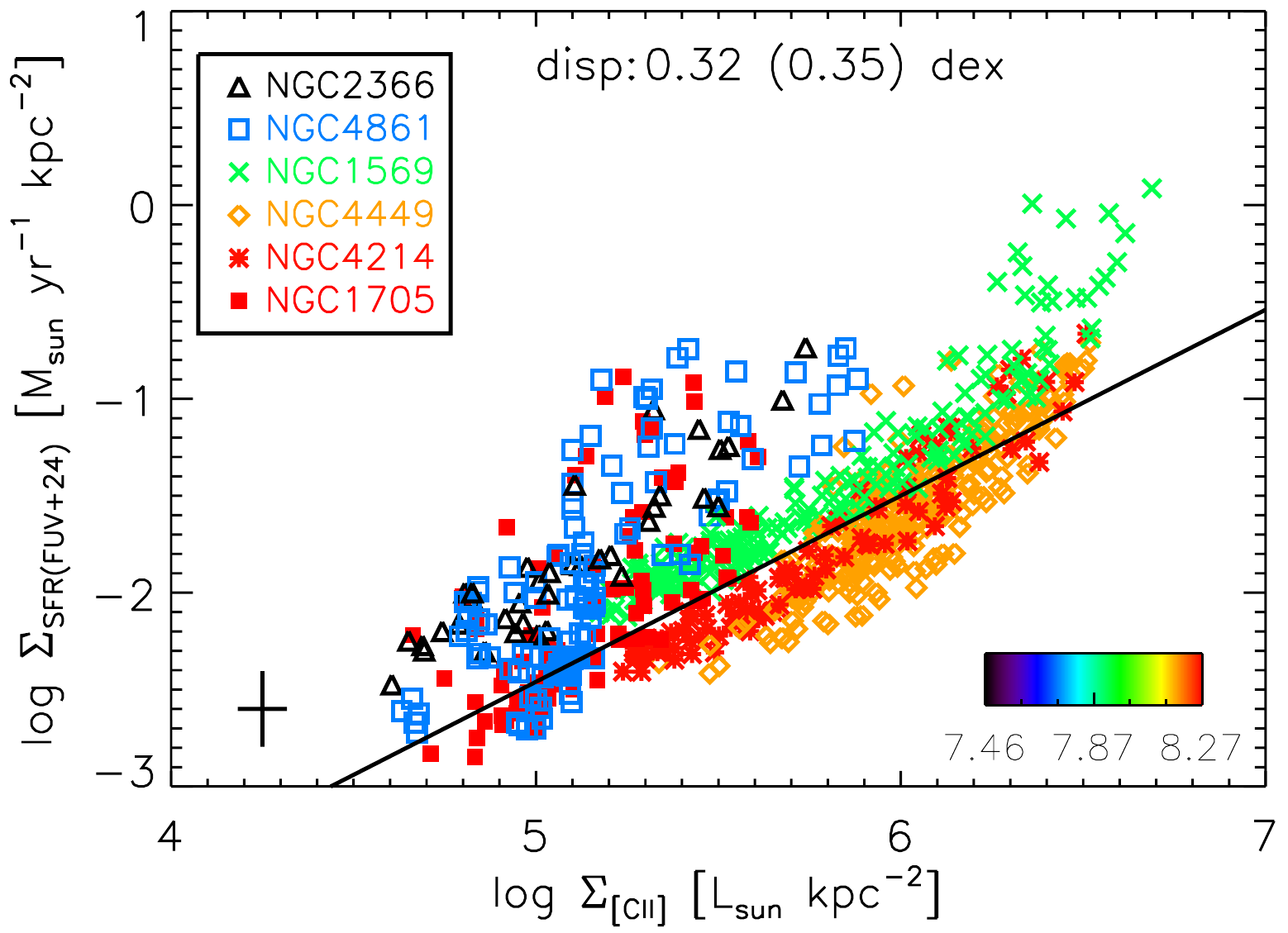}    \\
        \includegraphics[width=8.5cm]{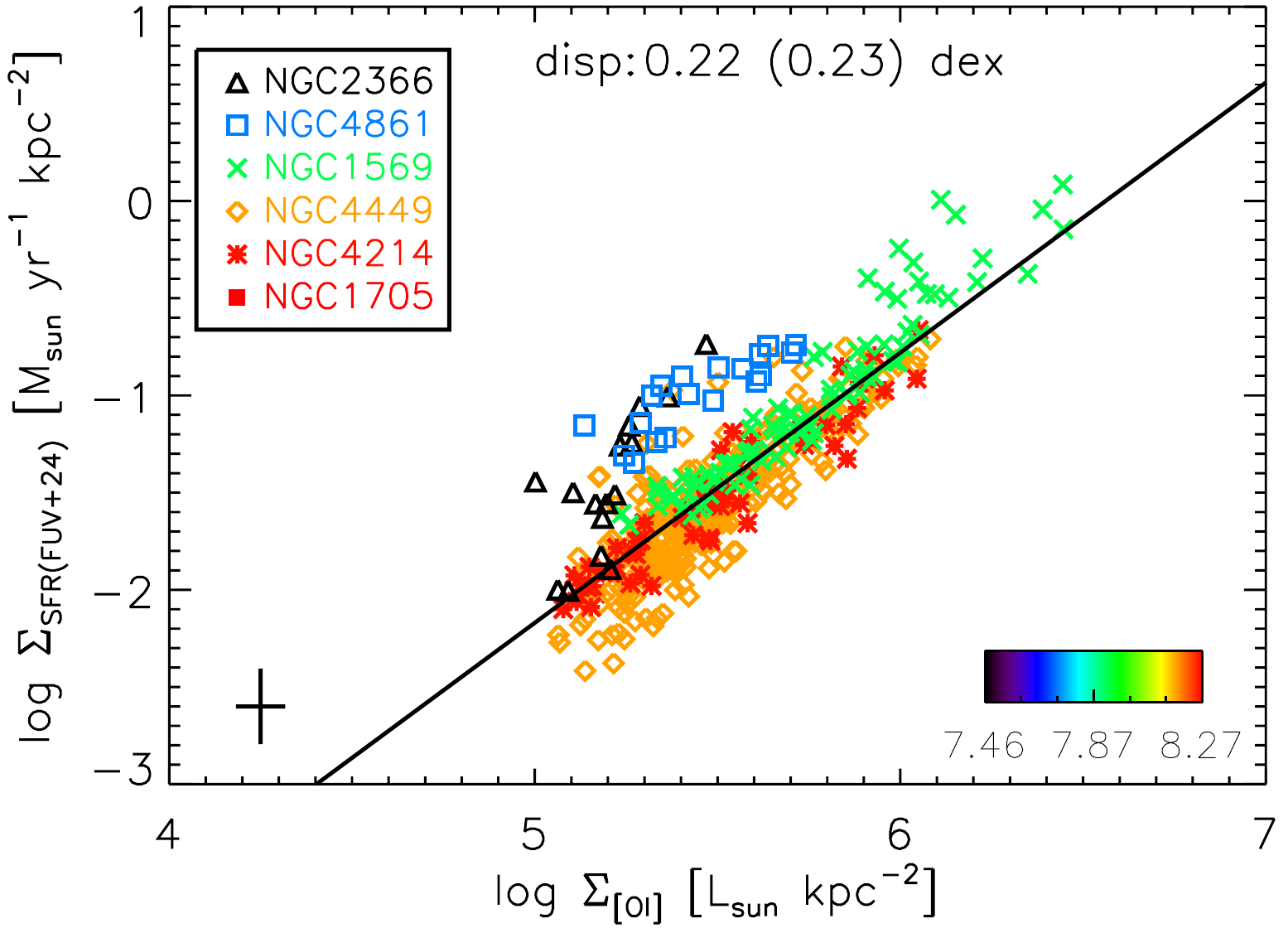}   \\ 
            \includegraphics[width=8.5cm]{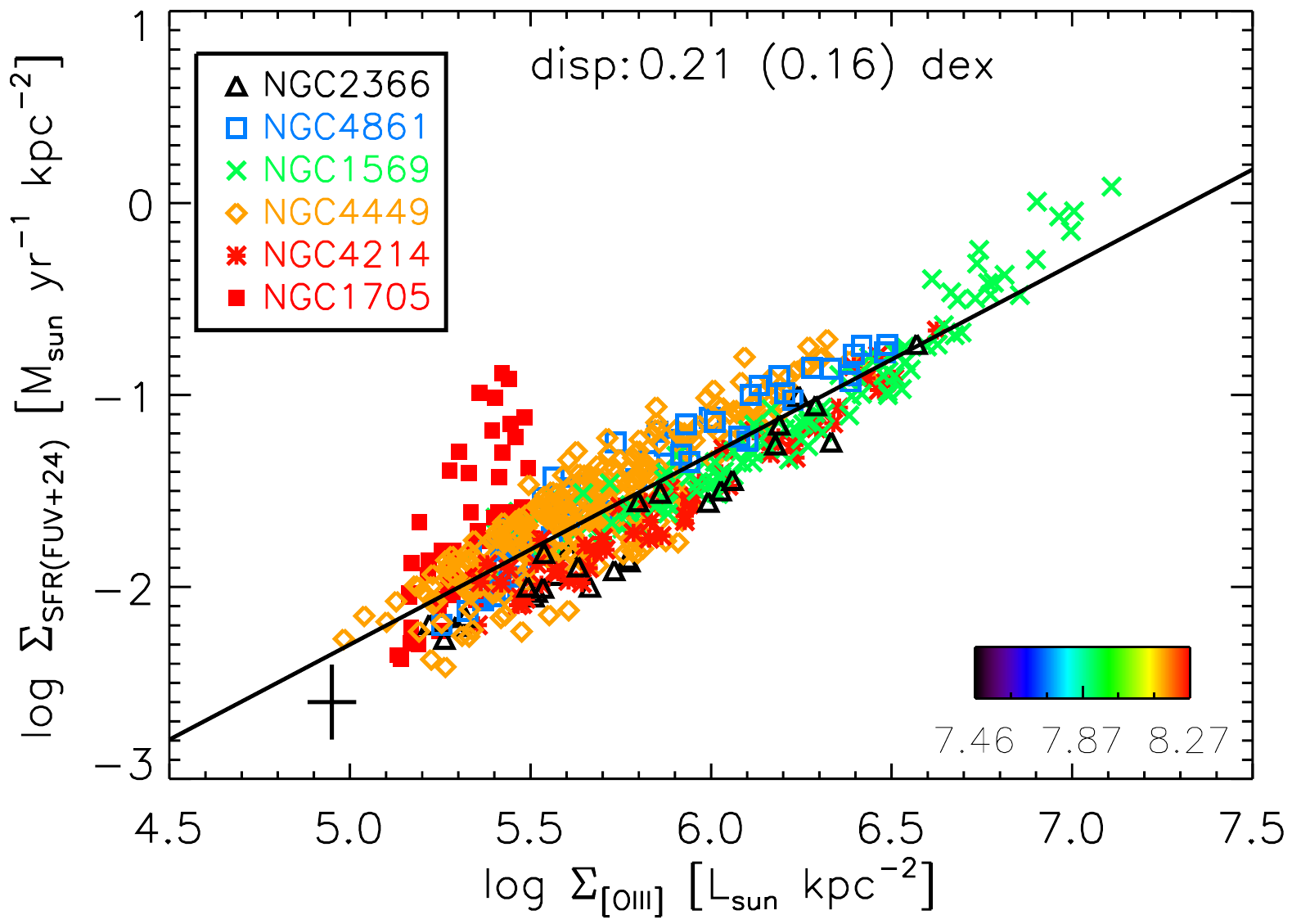}    
   \caption{Spatially resolved galaxy relation between surface densities of the SFR and [C{\sc{ii}}] (top), [O{\sc{i}}]$_{63}$ (middle) and [O{\sc{iii}}]$_{88}$ (bottom) surface densities. The legend explains the symbols used for different galaxies with the color bar indicating the oxygen abundance. Representative error bars are indicated in the lower left corner. Uncertainties on the SFR include the errors on each of the SFR calibrators (GALEX $FUV$, MIPS\,24\,$\mu$m) as well as the average scatter in the calibrations used to convert to the SFR (see Table \ref{SFRref}). Uncertainties on the FIR line surface densities incorporate the errors due to line fitting as well as calibration uncertainties (taken conservatively as $\sim$ 30$\%$). The best fitting SFR calibration is presented as a solid, black line. The dispersion of data points around the SFR calibration is indicated at the top of each panel, with the number in parenthesis indicating the scatter for the complete sample with $>5\sigma$ detections for all three lines.}
              \label{plot_SFR_sd}
    \end{figure}

Figure \ref{plot_SFR_sd} shows the relation between the SFR and [C{\sc{ii}}], [O{\sc{i}}]$_{63}$ and [O{\sc{iii}}]$_{88}$ line emission for the subsample of spatially resolved galaxies. Different galaxies are color-coded according to their oxygen abundance. Since abundances are determined for global galaxies, they may not give representative values for the metal abundance on spatially resolved scales. For example, the inefficient ISM mixing of nebular and neutral gas might cause deviations from this global metallicity value on kiloparsec scales \citep{1995A&A...294..432R,2004A&A...415...55L}. Since the coverage in [C{\sc{ii}}] is larger with respect to the areas mapped in [O{\sc{i}}]$_{63}$ and [O{\sc{iii}}]$_{88}$ and/or the [C{\sc{ii}}] line might attain higher $S/N$ levels in some areas, the number of pixels with $S/N> 5$ is higher for [C{\sc{ii}}] (1274 pixels) than for [O{\sc{i}}]$_{63}$ (602) and [O{\sc{iii}}]$_{88}$ (605) lines.

SFR calibrations of spatially resolved regions are determined from Levenberg-Marquardt least-squares fitting using the IDL procedure \texttt{MPFITFUN}, which is based on the non-linear least-squares fitting package \texttt{MPFIT} \citep{2009ASPC..411..251M}. The MPFITFUN procedure only accounts for the uncertainties on the SFR, which assumes that the line luminosities are error-free. While the line measurements are obviously affected by uncertainties inherent to the calibration and/or line fitting techniques, we prefer to account for the uncertainties on the SFR in the fitting procedure since the precision of the SFR estimate is affected by uncertainties on the reference SFR calibrators as well as the inaccuracy inherent to the applied reference SFR calibrators and, therefore, often larger than the uncertainties on the line measurements.
Table \ref{cali} summarizes the results of the fitted SFR calibrations:
\begin{equation}
\log~\Sigma_{\text{SFR}} = \beta + \alpha*\log \Sigma_{\text{line}}
\end{equation}
where $\Sigma_{\text{line}}$ is the FIR line surface density in units of L$_{\odot}$ kpc$^{-2}$, $\Sigma_{\text{SFR}}$ is the star formation rate in units of M$_{\odot}$ yr$^{-1}$ pc$^{-2}$ and $\alpha$ and $\beta$ represent the slope and intercept of the best fit. We require that the parameter $\alpha$ can be determined at a $>5\sigma$ significance level to determine that two quantities (i.e. SFR and FIR line surface density) are correlated.

The observed trends in Figure \ref{plot_SFR_sd} in combination with the $>5\sigma$ significance of the fitted slope of each trend suggest a correlation between the SFR and [C{\sc{ii}}], [O{\sc{i}}]$_{63}$, [O{\sc{iii}}]$_{88}$ line emission, which persists over almost two orders of magnitude in surface density. The smallest dispersion (0.21 dex) as well as the best constraint on the slope parameter with $S/N \sim 50$ is found for the [O{\sc{iii}}]$_{88}$ line, suggesting that the [O{\sc{iii}}]$_{88}$ line more tightly correlates with the SFR as compared to [O{\sc{i}}]$_{63}$ (0.22 dex) and [C{\sc{ii}}] (0.32 dex) lines on spatially resolved scales of $\sim$ 100 pc. Comparing different galaxies, we observe consistent trends with similar slopes between the FIR line emission and SFR. 
Towards brighter regions, the slope of the SFR-$L_{\text{[CII]}}$ relation appears to get steeper in most galaxies, suggesting that the [C{\sc{ii}}] line is not the dominant coolant in dense star-forming regions, where other cooling lines ([O{\sc{i}}]$_{63}$ or [O{\sc{iii}}]$_{88}$) are favored given the density and ionization state of the gas \citep{2012A&A...548A..91L}.
The steep slope ($\alpha$ = 1.41) in the spatially resolved SFR-$L_{\text{[OI]}}$ relation suggests a sudden drop in the [O{\sc{i}}]$_{63}$ emission for a decrease in star formation activity, while the flatter SFR-$L_{\text{[CII]}}$ ($\alpha$ = 0.93) and SFR-$L_{\text{[OIII]}}$ ($\alpha$ = 1.01) relations indicate that the [C{\sc{ii}}] and [O{\sc{iii}}]$_{88}$ lines remain bright in regions of relatively low SFRs. Given the high upper state energy and critical density for [O{\sc{i}}]$_{63}$ (see Table \ref{critical}), it is not surprising that bright [O{\sc{i}}]$_{63}$ emission only occurs in warm and/or dense star forming regions, where we expect to find the highest level of star formation activity. 

A few galaxies (NGC\,1705, NGC\,2366 and NGC\,4861) show diverging behavior in some of the SFR relations. NGC\,1705 has a peculiar behavior in the SFR-$L_{\text{[CII]}}$ and SFR-$L_{\text{[OIII]}}$ relations with weaker line emission relative to its star formation rate. NGC\,1705 is a dwarf starburst galaxy dominated by a central super star cluster (SSC) straddled by two dusty off-nuclear regions offset by $\sim$ 250 pc that dominate the H$\alpha$, mid- and far-infrared emission of the galaxy \citep{2006ApJ...652.1170C}. The chemistry and heating of gas in the off-nuclear positions do not seem to be regulated directly by the central SSC \citep{2006ApJ...652.1170C}, but rather exposed to the emission of young, massive stars, produced during a second starburst about 3 Myr ago which was presumably induced by the expanding shell after the first central starburst \citep{2009AJ....138..169A}. The weak [O{\sc{iii}}]$_{88}$ emission originates from the eastern dust complex which shows bright [C{\sc{ii}}] and PAH emission, which suggests that the deviation for the eastern regions of NGC\,1705 could be due to a radiation field not strong enough to excite [O{\sc{iii}}]$_{88}$ (requiring massive O6 and earlier-type stars). The western dust region does show bright [O{\sc{iii}}]$_{88}$ emission, but has weak [C{\sc{ii}}] and PAH emission. We, therefore, argue that the western dust complex is exposed to a harder radiation field, capable of destroying PAHs and ionizing the majority of the gas, which makes [O{\sc{iii}}]$_{88}$ a more efficient coolant. The non-detection of [O{\sc{i}}]$_{63}$ (which is an efficient coolant of dense PDRs, see Table \ref{critical}) might suggest rather diffuse ISM regions in the western and eastern dust emission complexes. Alternatively, optical depth effects might play a role in the two dusty off-nuclear regions.
The behavior of NGC\,2366 and NGC\,4861 is mainly divergent in the SFR-$L_{\text{[CII]}}$ and SFR-$L_{\text{[OI]}}$ relations, whereas the [O{\sc{iii}}]$_{88}$ emission correlates remarkably well with the SFR, suggesting that the gas around massive star clusters is mostly ionized and that PDRs only occupy a limited volume of the ISM.

The different galaxies cover a wider range in the SFR-$\Sigma_{\text{line}}$ relations than is observed within one single object, suggesting that the dispersion in the SFR-$\Sigma_{\text{line}}$ relations is driven by the diversity on global galaxy scales rather than variations in the ISM conditions within individual objects.
The SFR-$L_{\text{[CII]}}$ relation is most affected by this different behavior of galaxies with a dispersion of 0.32 dex around the fitted SFR calibration. More metal-rich galaxies show lower star formation rates as traced by $FUV$+MIPS\,24\,$\mu$m than those predicted by our SFR calibration given their [C{\sc{ii}}] emission. Sources with lower metal-abundances, on the other hand, preferentially populate the part of the plot representative of weaker [C{\sc{ii}}] emission and/or higher levels of star formation. Since the SFR based on $FUV$+MIPS\,24\,$\mu$m might be underestimated in more metal-poor dwarfs relative to the SFR calibrators $H\alpha$+MIPS\,24\,$\mu$m (see Section \ref{compareSFR}), the offset of metal-poor dwarf galaxies in the SFR-$\Sigma_{\text{line}}$ relations might even be more pronounced.
With a dispersion of 0.22 and 0.21 dex in the SFR-$L_{\text{[OI]}}$ and SFR-$L_{\text{[OIII]}}$ relations, respectively, the fine-structure lines [O{\sc{i}}]$_{63}$ and [O{\sc{iii}}]$_{88}$ seem to depend less on the ISM conditions in galaxies and might, thus, be potentially better SFR indicators compared to [C{\sc{ii}}]. 

Given that we probe different ISM phases, even on spatially resolved scales of $\sim$ 100 pc, we try to better approximate the overall cooling budget through FIR lines by combining the surface densities of [C{\sc{ii}}], [O{\sc{i}}]$_{63}$ and [O{\sc{iii}}]$_{88}$.
For the combination of several FIR lines, we attempt to fit a SFR function of the form:
\begin{equation}
\label{eqfit}
\log~\Sigma_{\text{SFR}} = \beta + \log(\Sigma_{\text{[CII]}}^{\alpha_{1}} + \Sigma_{\text{[OI]}}^{\alpha_{2}} + \Sigma_{\text{[OIII]}}^{\alpha_{3}})
\end{equation}
where $\Sigma_{\text{line}}$ is the FIR line surface density in units of L$_{\odot}$ kpc$^{-2}$, $\Sigma_{\text{SFR}}$ is the star formation rate in units of M$_{\odot}$ yr$^{-1}$ pc$^{-2}$ and ($\alpha_{1}$, $\alpha_{2}$, $\alpha_{3}$) and $\beta$ represent the slopes and intercept of the best fit.

\begin{figure}
   \centering
    \includegraphics[width=8.5cm]{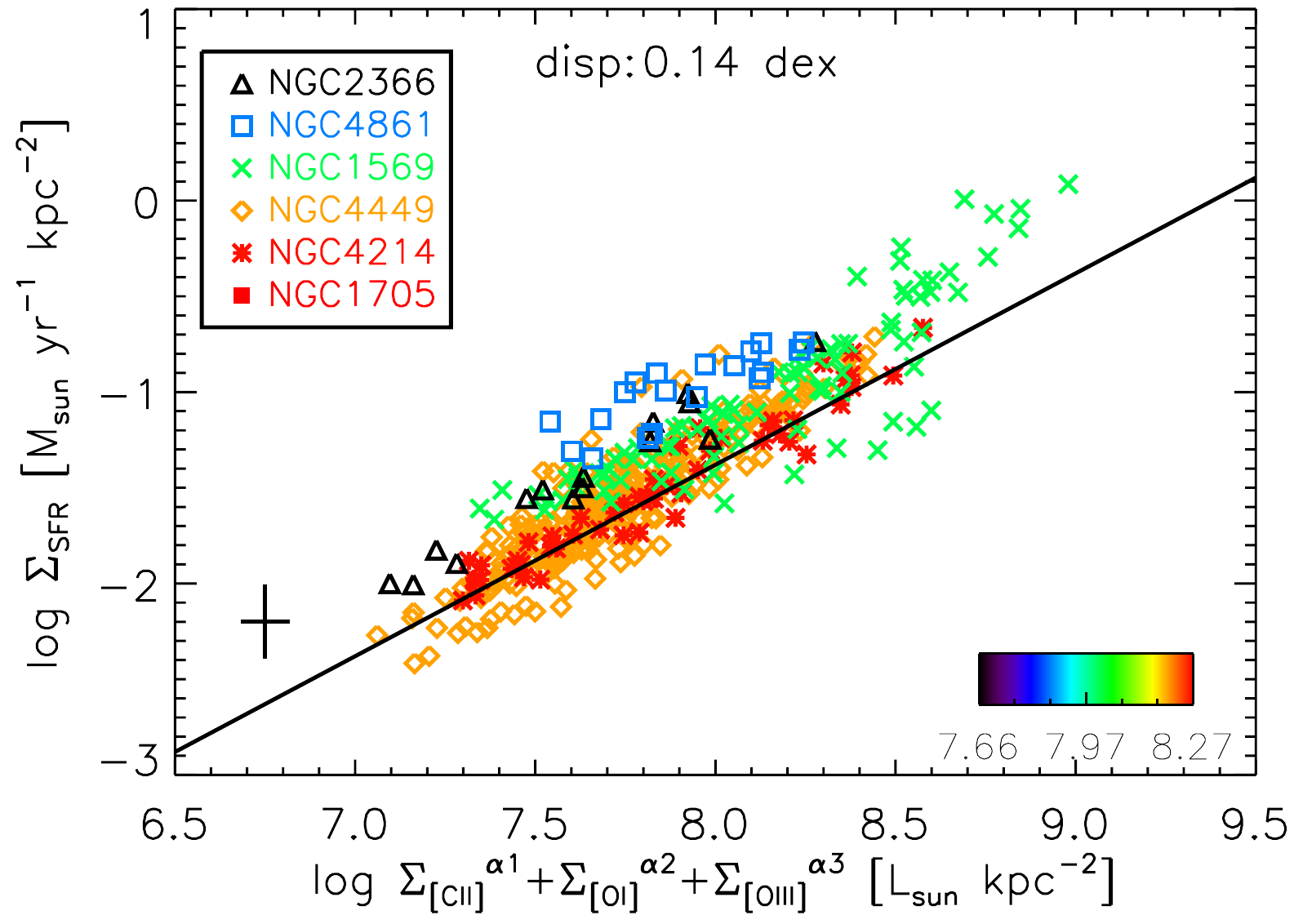}     
   \caption{Spatially resolved galaxy relation between surface densities of the SFR and a combination of [C{\sc{ii}}], [O{\sc{i}}]$_{63}$ and [O{\sc{iii}}]$_{88}$ surface densities. The image format is the same as explained in Fig \ref{plot_SFR_sd}. }
              \label{plot_SFR_CIIOIOIII_res}
    \end{figure}
    
In the fitting procedure, the different data points are equally weighted and the parameters ($\alpha_{1}$, $\alpha_{2}$, $\alpha_{3}$) of the slopes are constrained to positive numbers. 
Similar functions are defined to fit combinations of two FIR lines. 
Best fitting parameters (including calibration coefficients and dispersion) are presented in Table \ref{cali} only for line combinations which improved on the dispersion in the SFR calibrations for single FIR lines.
By combining a number of FIR lines, we are able to significantly reduce the scatter (see Figure \ref{plot_SFR_CIIOIOIII_res}), confirming the hypothesis that the total gas cooling balances the gas heating under the condition of local thermal equilibrium. This, furthermore, suggests that the specific processes that regulate the cooling in the different gas phases are of minor importance. Although the primary heating mechanisms are very different in the neutral gas phase (photo-electric effect and a variable contribution from cosmic rays and soft X-ray heating), as compared to ionized gas media (photo-ionization processes), the main goal is to get access to the total heating due to young stellar emission -irrespective of the dominant heating mechanism in different gas phases- by probing the total cooling budget in galaxies.
In particular, the combinations [C{\sc{ii}}]+[O{\sc{iii}}]$_{88}$, [O{\sc{i}}]$_{63}$+[O{\sc{iii}}]$_{88}$ and  [C{\sc{ii}}]+[O{\sc{i}}]$_{63}$+[O{\sc{iii}}]$_{88}$ provide accurate estimates of the SFR, which suggests that the cooling in the neutral as well as ionized media needs to be probed to approximate the overall cooling budget in metal-poor galaxies and, thus, trace the star formation activity.
However, heating mechanisms not directly linked to the recent star formation activity (e.g soft X-ray heating, \citealt{1969ApJ...158..185S}, which might become substantial in extremely low-metallicity galaxies such as I\,Zw\,18, \citealt{2008A&A...478..371P} and Lebouteiller et al. in prep.) might disperse the link between the emission of cooling lines and the star formation rate (see also Sections \ref{scat_res.sec} and \ref{Trends2.sec}).
 
    \begin{table*}
\caption{Overview of the calibration coefficients for SFR calibrations based on the spatially resolved (top) and global galaxy (bottom) DGS sample. The first column indicates the FIR fine-structure line(s) used in the SFR calibration, with the number of data points, slope, intercept and dispersion of the best fitting line presented in columns 2, 3, 4 and 5, respectively. In parentheses, we show the dispersion for the complete galaxy sample, i.e. galaxy regions or global galaxies which have $>5\sigma$ detections for all three lines [C{\sc{ii}}], [O{\sc{i}}]$_{63}$ and [O{\sc{iii}}]$_{88}$.}
\label{cali}
\centering
\begin{tabular}{|l|c|c|c|c|}
\hline 
SFR calibrator &  Number of data points & Slope & Intercept & 1$\sigma$ dispersion [dex] \\
\hline 
\multicolumn{5}{|c|}{SFR calibration: spatially resolved DGS sample} \\
\hline 
$[{\rm CII}]$  & 1274 & 0.93 $\pm$ 0.06 & -6.99 $\pm$ 0.35 & 0.32 (0.35) \\
$[{\rm OI}]_{63}$ & 602 & 1.41 $\pm$ 0.04 & -9.19 $\pm$ 0.23 & 0.22 (0.23) \\
$[{\rm OIII}]_{88}$  & 605 & 1.01 $\pm$ 0.02 & -7.33 $\pm$ 0.12 & 0.21 (0.16)  \\
$[{\rm CII}]$ +  $[{\rm OIII}]_{88}$  & 605 &  (0.94 $\pm$ 0.03, 1.08 $\pm$ 0.02)  &  -7.82 $\pm$ 0.14 & 0.20 (0.15)     \\
$[{\rm OI}]_{63}$ + $[{\rm OIII}]_{88}$  & 441 &  (1.30 $\pm$ 0.03, 1.25 $\pm$ 0.02)  &  -9.01 $\pm$ 0.15 & 0.14     \\
$[{\rm CII}]$ +  $[{\rm OI}]_{63}$ + $[{\rm OIII}]_{88}$ & 441 & (1.23 $\pm$ 0.03, 1.31 $\pm$ 0.03, 1.25 $\pm$ 0.07)  &  -9.38 $\pm$ 0.16 & 0.14     \\
\hline 
\multicolumn{5}{|c|}{SFR calibration: global galaxy DGS sample} \\
\hline 
$[{\rm CII}]$ & 32 & 0.84 $\pm$ 0.06 & -5.29 $\pm$ 0.34 & 0.38 (0.40)  \\
$[{\rm OI}]_{63}$ & 26 &   0.94 $\pm$ 0.05 & -6.37 $\pm$ 0.29 & 0.25  \\
$[{\rm OIII}]_{88}$   & 28 & 0.92 $\pm$ 0.05 & -6.71 $\pm$ 0.33 & 0.30 (0.30)  \\
\hline 
\end{tabular}
\end{table*}    
    
 \subsection{Scatter in the SFR-$L_{\text{line}}$ relation}
\label{scat_res.sec}
In this section, we want to analyze what is driving the dispersion in the SFR-$L_{\text{line}}$ relations on spatially resolved scales within galaxies by analyzing the trends with several diagnostics for the physical conditions of the interstellar medium. We quantify the dispersion in the respective SFR-$L_{\text{line}}$ relations as the logarithmic distance between the SFR as estimated from the reference SFR calibration based on $FUV$ and 24\,$\mu$m emission and the best fitting line for the spatially resolved SFR-$L_{\text{line}}$ relation.

First of all, we analyze a possible link between the scatter and FIR color (as probed by the PACS\,100\,$\mu$m/PACS\,160\,$\mu$m flux density ratio obtained from dust continuum observations), considered a proxy of the dust temperature (or grain charging) under the assumption that the emission in both bands is heated by the same radiation field. 
Although different radiation fields -originating from star-forming regions or the diffuse interstellar radiation- have been shown to contribute to the emission in far-infrared and submillimeter wavebands (e.g. \citealt{2012MNRAS.419.1833B}), the dust emission in dwarf galaxies seems to be due to heating primarily by young stars, even at wavelengths $>$ 160\,$\mu$m \citep{2010A&A...518L..55G,2012MNRAS.419.1833B}.

\begin{figure}[!ht]
   \centering 
     \includegraphics[width=8.5cm]{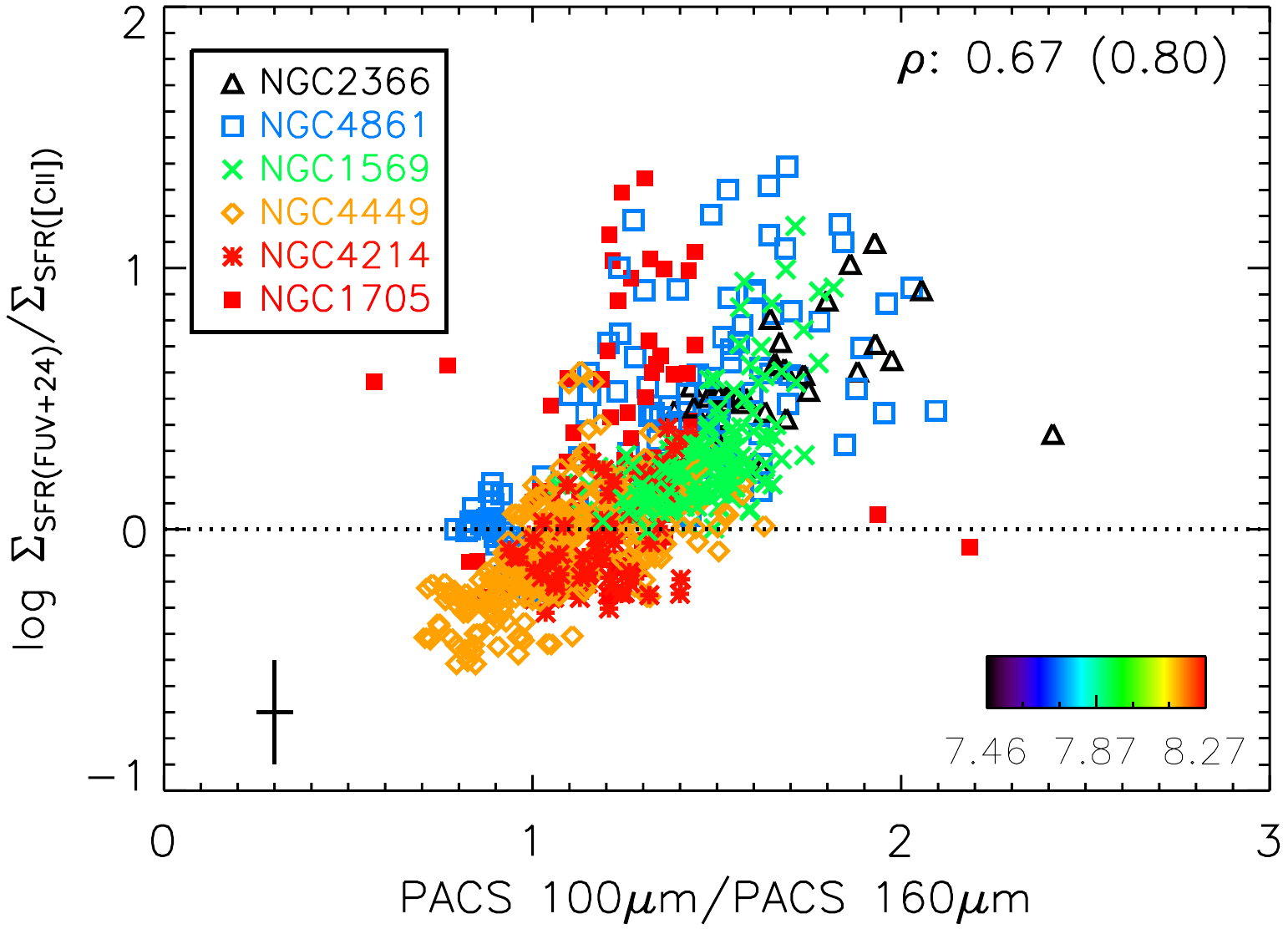} \\
          \includegraphics[width=8.5cm]{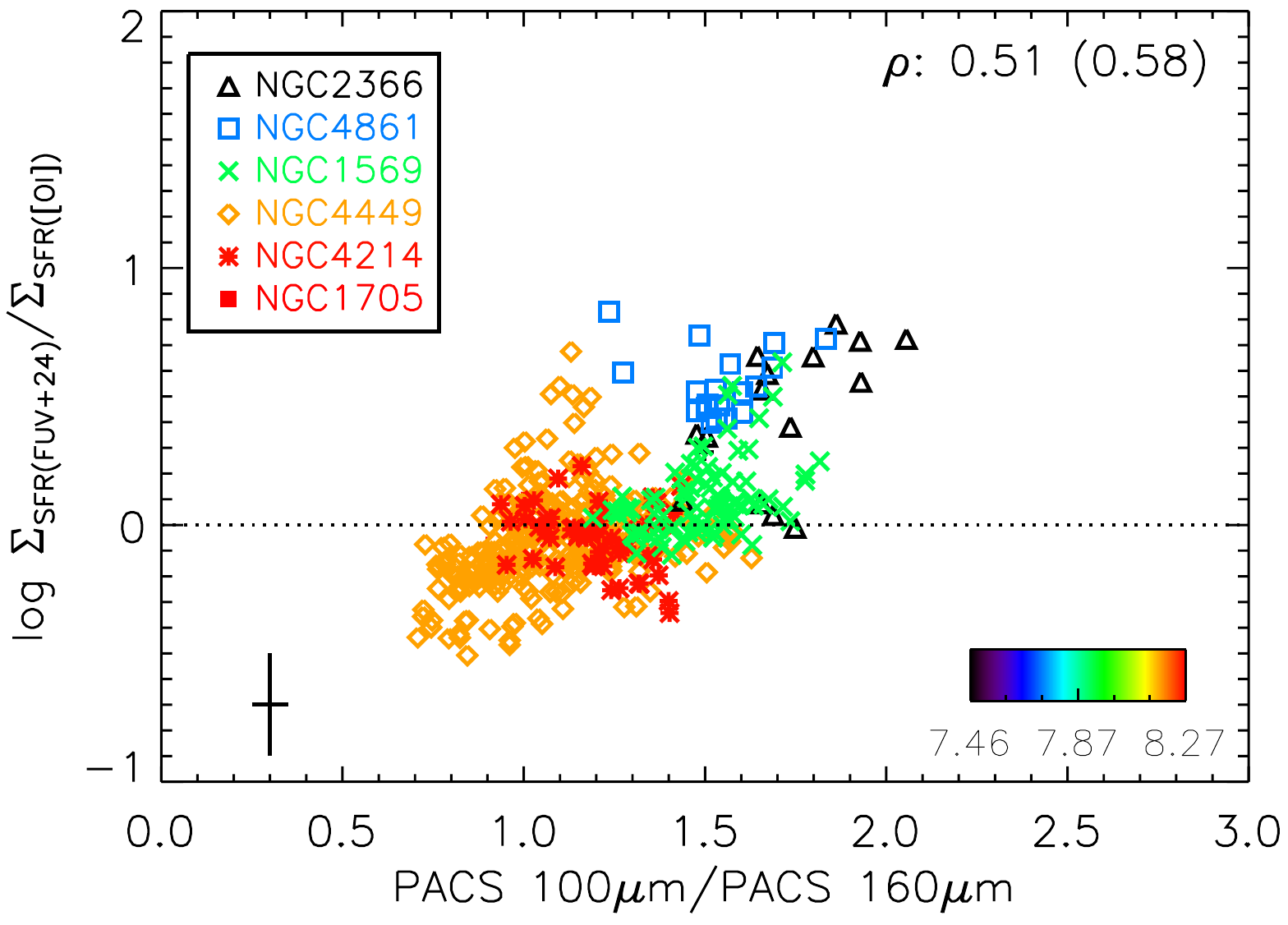}    \\
   \caption{Spatially resolved galaxy relation between the dispersion from the SFR calibrations for [C{\sc{ii}}] (top) and [O{\sc{i}}]$_{63}$ (bottom) as a function of FIR color, i.e. PACS 100\,$\mu$m/PACS 160\,$\mu$m. The legend explains the symbols used for different galaxies with the color bar indicating the oxygen abundance. Representative error bars are indicated in the lower left corner. Representative error bars are indicated in the lower left corner. Uncertainties on the SFR include the errors on each of the SFR calibrators (GALEX $FUV$, MIPS\,24\,$\mu$m) as well as the average scatter in the reference calibration (see Table \ref{SFRref}). Uncertainties on the PACS line ratios incorporate the errors due to map making as well as calibration uncertainties (5$\%$). The Spearman's rank correlation coefficients are presented in the top right corner. In parentheses, we show the dispersion for the complete galaxy sample, i.e. galaxy regions which have $>5\sigma$ detections for all three lines [C{\sc{ii}}], [O{\sc{i}}]$_{63}$ and [O{\sc{iii}}]$_{88}$.}
              \label{plot_PACS70160_sd}%
    \end{figure}

Figure \ref{plot_PACS70160_sd} displays the observed trends between FIR color and dispersion in the SFR-$L_{\text{line}}$ relations for [C{\sc{ii}}] (top panel) and [O{\sc{i}}] (bottom panel). The data reduction of PACS\,100 and 160\,$\mu$m photometry maps is described in \citet{2013A&A...557A..95R}. The dispersion in the SFR-$L_{\text{[CII]}}$ relation ($\rho$ = 0.67) clearly correlates with the FIR colors of galaxy regions in the sense that the fine-structure line [C{\sc{ii}}] does not seem well-suited as a SFR indicator towards warm far-infrared colors. Low-metallicity galaxies often have warm FIR colors (e.g. \citealt{1999ApJ...516..783T,2004ApJS..154..211H,2009A&A...508..645G,2011A&A...532A..56G,2013A&A...557A..95R}), which could explain why galaxies such as NGC\,2366 and NGC\,4861 show an offset in the SFR-$L_{\text{[CII]}}$ relation compared to more metal-rich galaxies.  For increasing dust temperatures, the grain charging parameter increases and, therefore, the photoelectric heating efficiency decreases, diminishing the line cooling.
We observe a similar but more moderate correlation with FIR color for [O{\sc{i}}]$_{63}$ (0.51 dex). 
Similar plots for [O{\sc{iii}}]$_{88}$ are not shown here, since the line emission arises from the ionized gas component, where photo-ionization processes dominate the heating and no physical motivation exists for a correlation with FIR color.

\begin{figure}[!ht]
   \centering
    \includegraphics[width=8.5cm]{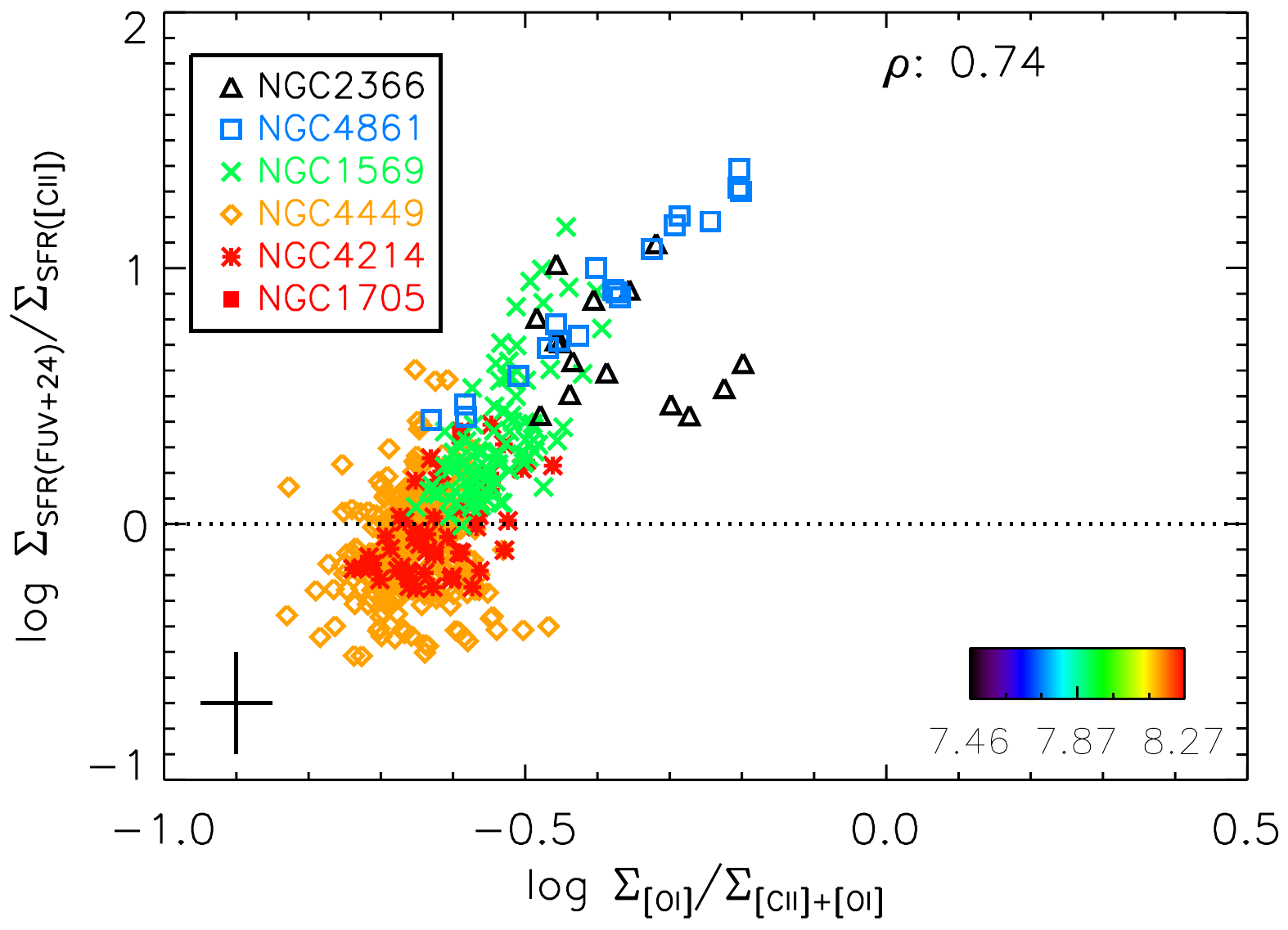}    \\
        \includegraphics[width=8.5cm]{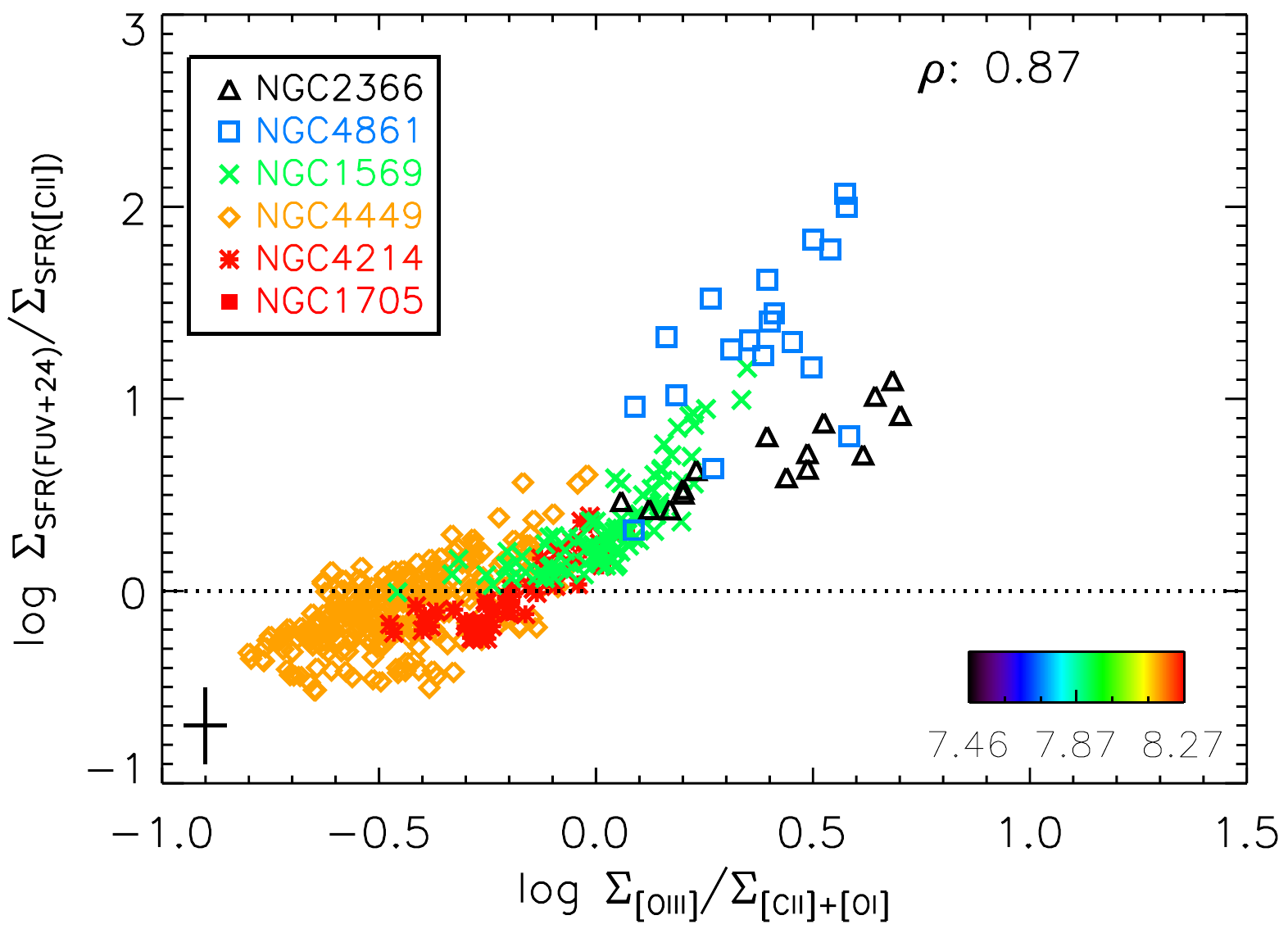}  
           \caption{Spatially resolved galaxy relation between the dispersion from the SFR calibrations for [C{\sc{ii}}] as a function of [O{\sc{i}}]$_{63}$/[C{\sc{ii}}]+[O{\sc{i}}]$_{63}$ (top) and [O{\sc{iii}}]$_{88}$/[C{\sc{ii}}]+[O{\sc{i}}]$_{63}$ (bottom) line ratios. The image format is the same as explained in Fig \ref{plot_PACS70160_sd}.}
              \label{plot_CIIOI_sd}%
    \end{figure}

Figure \ref{plot_CIIOI_sd} shows the behavior of the scatter in the SFR-$L_{\text{[CII]}}$ relation as a function of FIR line ratios [O{\sc{i}}]$_{63}$/[C{\sc{ii}}]+[O{\sc{i}}]$_{63}$ (top) and [O{\sc{iii}}]$_{88}$/[C{\sc{ii}}]+[O{\sc{i}}]$_{63}$ (bottom).
With an upper state energy $E_{\text{u}}/k$ $\sim$ 228 K and critical density $n_{\text{crit,H}}$ $\sim$ 5 $\times$ 10$^{5}$ cm$^{-3}$ for [O{\sc{i}}]$_{63}$ as compared to $T_{\text{exc}}$ $\sim$ 91 K and $n_{\text{crit,H}}$ $\sim$ 1.6 $\times$ 10$^{3}$ cm$^{-3}$ for [C{\sc{ii}}], the [O{\sc{i}}]$_{63}$/[C{\sc{ii}}]+[O{\sc{i}}]$_{63}$ ratio can be interpreted as a proxy for the relative fraction of warm and/or dense gas which increases towards higher values of [O{\sc{i}}]$_{63}$/[C{\sc{ii}}]+[O{\sc{i}}]$_{63}$. Indeed, PDR models have shown that this ratio increases towards higher gas density and radiation field strength \citep{2006ApJ...644..283K}. 
With [O{\sc{iii}}]$_{88}$ emission originating from highly-ionized regions near young O stars, the [O{\sc{iii}}]$_{88}$/[C{\sc{ii}}]+[O{\sc{i}}]$_{63}$ ratio can be interpreted as a proxy for the relative contribution of the ionized gas phase with higher values implying large filling factors of diffuse highly-ionized gas with respect to neutral media. The latter interpretations of line ratios are based on the assumption that most of the [C{\sc{ii}}] emission arises from PDRs rather than diffuse ionized gas media and that also [O{\sc{i}}]$_{63}$ emission can be identified merely with PDRs. However, the interpretation of this line ratio might differ for low-metallicity galaxies, where the filling factors of PDRs are considered to be low based on the weak emission of several PDR tracers (e.g. PAH, CO). High [O{\sc{i}}]$_{63}$/[C{\sc{ii}}]+[O{\sc{i}}]$_{63}$ line ratios in dust deficient objects might, thus, have a different origin than warm and/or dense PDRs. \citet{2008A&A...478..371P} have shown that low-ionization line emission in neutral gas media can be produced by a pseudo-PDR with similar lines as PDRs but with soft X-rays as the dominant heating mechanism (see also Lebouteiller et al. in prep).

There is a clear correlation ($\rho$=0.74) between the [O{\sc{i}}]$_{63}$/[C{\sc{ii}}]+[O{\sc{i}}]$_{63}$ line ratios and the observed scatter in the SFR-$L_{\text{[CII]}}$ relation indicating that regions with higher values of [O{\sc{i}}]$_{63}$/[C{\sc{ii}}]+[O{\sc{i}}]$_{63}$ are offset in the SFR relation towards weaker [C{\sc{ii}}] emission for a certain SFR. This implies that [C{\sc{ii}}] is not the most appropriate SFR indicator in those regions. Higher values of [O{\sc{i}}]$_{63}$/[C{\sc{ii}}]+[O{\sc{i}}]$_{63}$ occur preferentially in galaxies of lower metal abundance (e.g. NGC\,2366, NGC\,4861), which might suggest that the importance of the photoelectric effect diminishes in dust deficient environments and other heatings mechanisms (e.g. soft X-ray heating) become more efficient.

With the [O{\sc{iii}}]$_{88}$/[C{\sc{ii}}]+[O{\sc{i}}]$_{63}$ ratio (see bottom panel of Figure \ref{plot_CIIOI_sd}) covering almost two orders of magnitude, we sample very distinct ISM conditions on hectoparsec scales within spatially resolved galaxies. The clear correlation of [O{\sc{iii}}]$_{88}$/[C{\sc{ii}}]+[O{\sc{i}}]$_{63}$ with the scatter in the SFR-$L_{\text{[CII]}}$ relation ($\rho$ = 0.87) implies that the [C{\sc{ii}}] line is not a good tracer of the star formation activity in regions where the ionized gas phase occupies an important part of the ISM volume. Although the [C{\sc{ii}}] line could also regulate the cooling in ionized gas media besides being the dominant coolant in neutral PDRs, the harder radiation field at lower metallicities will produce hard photons capable of ionizing O$^{+}$ (IP = 35.1 eV). Since the ionization potential of C$^{+}$ is only 24.4 eV, carbon might thus easily become doubly ionized in diffuse ionized gas media resulting in most of the carbon being locked in C$^{++}$ rather than C$^{+}$.
In particular, metal-poor regions seem affected by large filling factors of highly ionized media, which questions the ability of [C{\sc{ii}}] to trace the SFR in those environments. 
    
\section{Global galaxy SFR-$L_{\text{line}}$ relation}
\label{Int.sec}
In the previous section, we analyzed the observed trends and scatter in the SFR-$L_{\text{line}}$ relations for a subsample of spatially resolved galaxies down to hectoparsec scales. We verify whether the trends and scatter remain present on global galaxy scales, when averaged out over the different ISM phases. For this analysis, we consider all galaxies with GALEX $FUV$ and MIPS\,24\,$\mu$m observations (32 out of 48 galaxies), including the resolved sources from Section \ref{Res.sec} with their line and continuum flux measurements reduced to one data point.
The extension of the subsample of spatially resolved galaxies to the entire DGS sample broadens the range covered in metallicity from 0.03 Z$_{\odot}$ (I\,Zw\,18) to 0.5 Z$_{\odot}$ (HS\,2352+2733) and in SFR from 0.001 M$_{\odot}$ yr$^{-1}$ (UGC\,4483) to 43 M$_{\odot}$ yr$^{-1}$ (Haro\,11) as traced by $FUV$+24\,$\mu$m.

\subsection{Global fluxes}
\label{global.sec}
For global galaxy fluxes of FIR lines, we rely on the aperture photometry results for fine-structure lines reported in Cormier et al. (in prep.), where FIR line fluxes within apertures covering the brightest fine-structure line emission are computed. We assume a 30$\%$ calibration error on top of the uncertainties that result from line fitting. 
In some cases, the [C{\sc{ii}}] emission is more extended with respect to the [O{\sc{i}}]$_{63}$ and [O{\sc{iii}}]$_{88}$ emission or simply observed across a wider field, in which cases, the [C{\sc{ii}}] apertures are bigger to include the total region mapped. Here, we measure the [C{\sc{ii}}] flux within the same apertures as the [O{\sc{i}}]$_{63}$ and [O{\sc{iii}}]$_{88}$ emission for 6 galaxies using the same techniques as Cormier et al. (in prep.).
For NGC\,6822, we only include the FIR line measurements from the H{\sc{ii}} region Hubble V, since it is the only area in NGC\,6822 covered in all three lines.

Corresponding GALEX $FUV$ and MIPS\,24\,$\mu$m fluxes are obtained from aperture photometry using the central positions and aperture sizes applied to the FIR fine-structure lines. Table \ref{table2} gives an overview of the aperture photometry results for GALEX $FUV$ and MIPS\,24\,$\mu$m bands.
For point sources, GALEX $FUV$ and MIPS\,24\,$\mu$m measurements usually correspond to total galaxy fluxes. 
Total MIPS\,24\,$\mu$m fluxes for point sources are adopted from \citet{2012MNRAS.423..197B} and are indicated with an asterisk in column 5 of Table \ref{table2}. In some cases, the $FUV$ data show an extended tail of emission (HS\,1442+4250, UGC\,04483, UM\,133) with no counterpart in MIPS 24\,$\mu$m nor PACS maps. Rather than measuring the global $FUV$ emission for these galaxies with a cometary structure, we rely on aperture photometry within apertures that encompass the brightest 24\,$\mu$m emission features. In this manner, we avoid overestimating the total SFR for these galaxies by neglecting the $FUV$ emission that was either not covered in our \textit{Herschel} observations or did not show any dust emission, suggesting that little dust is present in those areas.
Table \ref{table2} reports the $FUV$ flux densities within those apertures, but also provides in parentheses the photometry results for apertures encompassing the total $FUV$ emission.

\subsection{Observed trends}
\label{Trends2.sec}  
    
\begin{figure}[!ht]
   \centering
    \includegraphics[width=8.5cm]{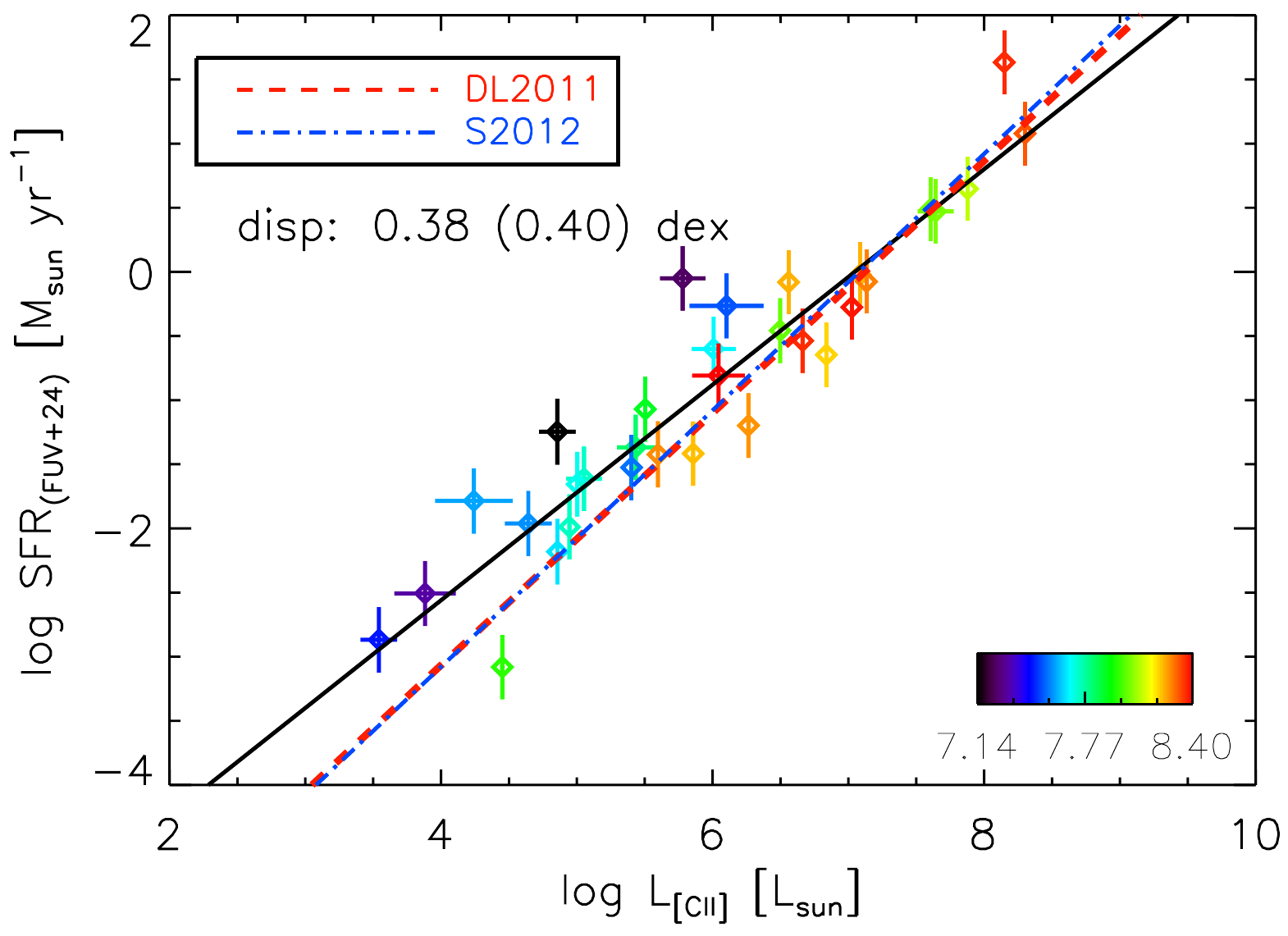}    \\
        \includegraphics[width=8.5cm]{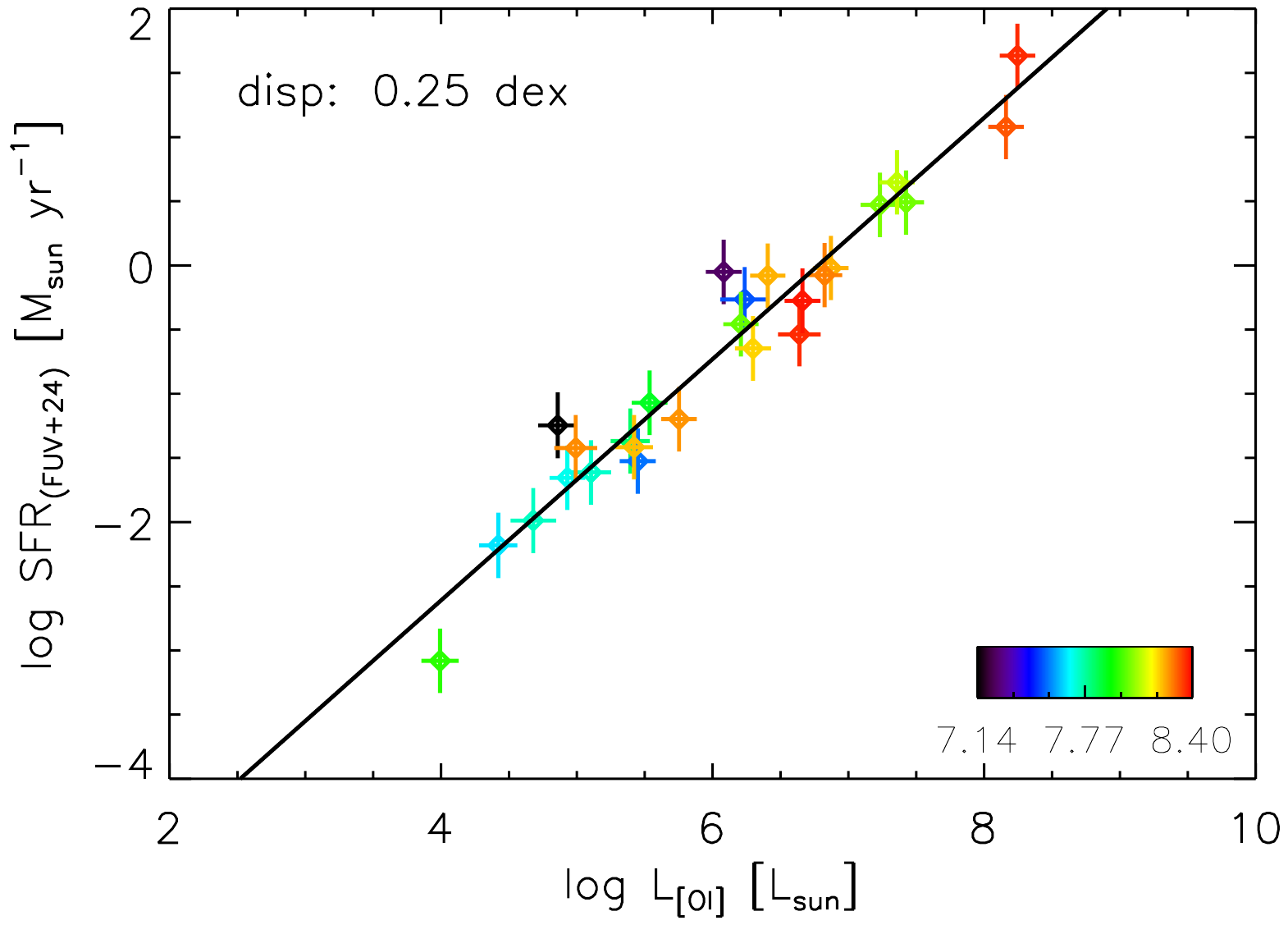}   \\ 
            \includegraphics[width=8.5cm]{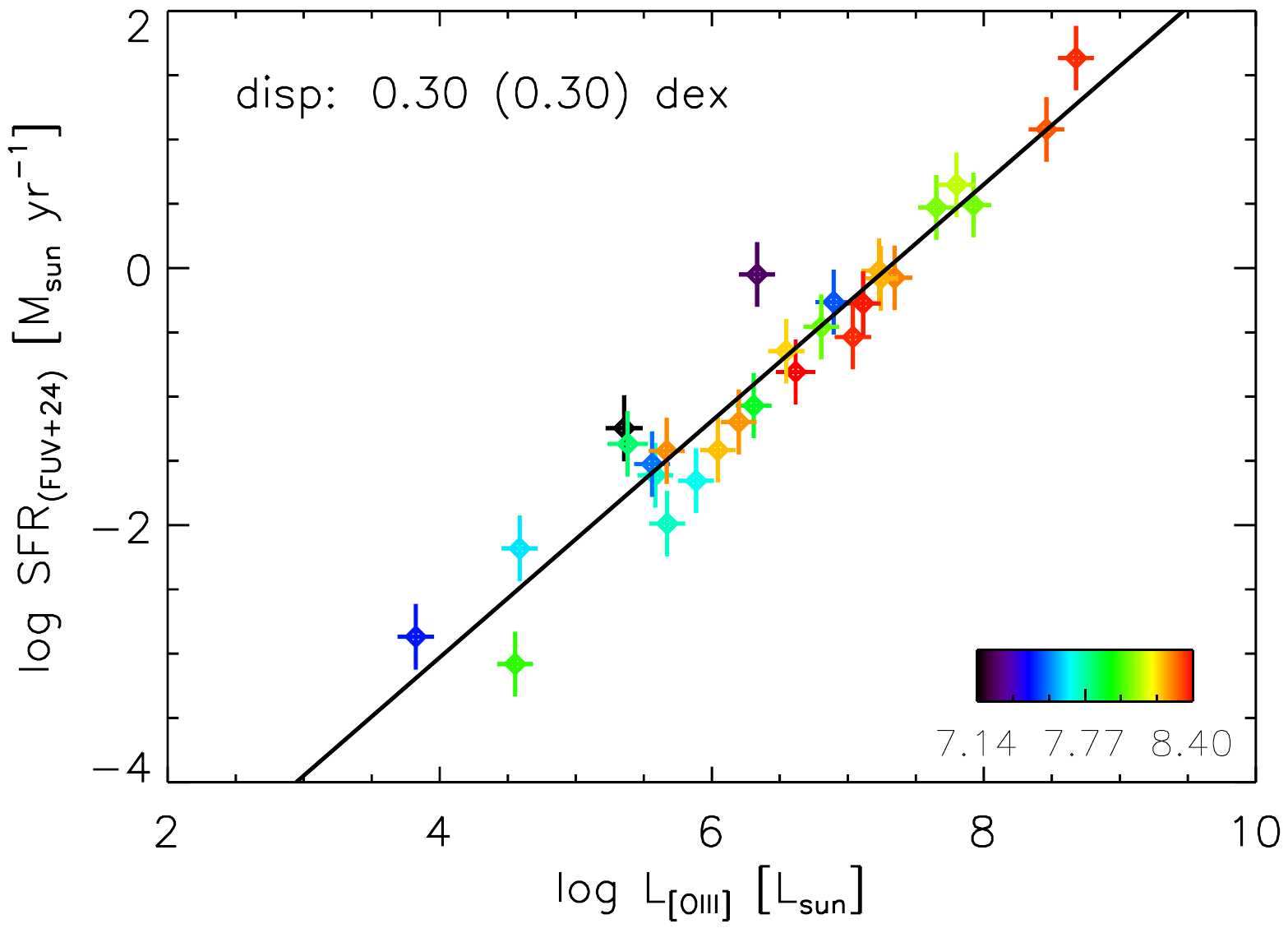}    
      \caption{Relation between the SFR and [C{\sc{ii}}] (top), [O{\sc{i}}]$_{63}$ (middle) and [O{\sc{iii}}]$_{88}$ (bottom) luminosities on global galaxy scales. Galaxies are color-coded according to metallicity with increasing oxygen abundances going from black over blue, green and yellow to red colors. The best fitting SFR calibration is presented as a solid, black line. The dispersion of data points around the SFR calibration is indicated at the top of each panel, with the number in parenthesis indicating the scatter for the complete sample with $>5\sigma$ detections for all three lines.}
              \label{plot_CII_int}%
    \end{figure}

Figure \ref{plot_CII_int} presents the SFR-$L_{\text{line}}$ relations on global galaxy scales for DGS sources with GALEX $FUV$ and MIPS 24\,$\mu$m observations. 
Based on the observed trends, SFR calibrations are derived from linear regression fits:
\begin{equation}
\label{func1}
\log~\text{SFR} = \beta + \alpha*\log L_{\text{line}}.
\end{equation}
where $L_{\text{line}}$ is the FIR line luminosity in units of L$_{\odot}$, SFR is the star formation rate in units of M$_{\odot}$ yr$^{-1}$ and $\alpha$ and $\beta$ represent the slope and intercept of the best fit. 
Table \ref{cali} (see bottom part) summarizes the calibration coefficients (slope, intercept) retrieved from the fits and the dispersion of data points around the best fit. 

With the slopes of all best fitting lines determined with at least 5$\sigma$ significance, we are confident that the SFR also correlates with the [C{\sc{ii}}], [O{\sc{i}}]$_{63}$ and [O{\sc{iii}}]$_{88}$ line emission on global scales. The smallest dispersion (0.25 dex) and strongest constraint on the slope (S/N $\sim$ 19) could be obtained for the [O{\sc{i}}]$_{63}$ line, from which the SFR can be estimated with an uncertainty factor of 1.8. The [O{\sc{iii}}]$_{88}$ line probes the SFR within an uncertainty factor of 2, while the link between the SFR and [C{\sc{ii}}] line is more dispersed and results in an estimated SFR uncertain by a factor of $\sim$ 2.4. 
The top panel of Figure \ref{plot_CII_int} includes previous SFR calibrations reported in \citet{2011MNRAS.416.2712D}\footnote{The SFR calibration in \citet{2011MNRAS.416.2712D} was derived based on the reference SFR tracers GALEX $FUV$ and MIPS 24\,$\mu$m and the scaling factor $\alpha$ = 6.31 as derived by \citet{2008ApJ...686..155Z}. Recalibrating their relation with the scaling factor ($\alpha$ = 3.89) applied in this paper would only shift their relation by 0.2 dex at most.} (red dashed line) and \citet{2012ApJ...755..171S} (blue dashed-dotted line). 
The SFR-$L_{\text{[C{\sc{ii}}]}}$ calibration derived for the DGS sample\footnote{Although the [C{\sc{ii}}] luminosity range in \citet{2011MNRAS.416.2712D} (5.7 $\leq$ $\log$ $L_{\text{[CII]}}$ [L$_{\odot}$] $\leq$ 9.1) does not extend to the faintest [C{\sc{ii}}] luminosities for galaxies, it largely overlaps with the $L_{\text{[CII]}}$ range covered by DGS sources while the SFR relation in \citet{2012ApJ...755..171S} was calibrated for higher [C{\sc{ii}}] luminosities (7 $\leq$ $\log$ $L_{\text{[CII]}}$ [L$_{\odot}$] $\leq$ 9).} has a shallower slope ($\alpha$ = 0.84) compared to the nearly one-to-one correlation obtained in \citet{2011MNRAS.416.2712D} and \citet{2012ApJ...755..171S}, which can be attributed to a decreasing [C{\sc{ii}}] emission towards lower metal abundances. For [C{\sc{ii}}] and [O{\sc{iii}}]$_{88}$, the SFR seems qualitatively linked in the same way to the line emission on global galaxy and spatially resolved scales. On spatially resolved scales, the slope of the SFR calibrations for [O{\sc{i}}]$_{63}$ ($\alpha$ = 1.41) differs from the correlation observed on global galaxy scales ($\alpha$ = 0.94).

Compared to the dispersion in the spatially resolved SFR-$L_{\text{line}}$ relations, the averaging over the different ISM phases on global galaxy scales does not reduce the scatter in the observed SFR-$L_{\text{line}}$ trends. This again shows that the dispersion in the SFR relations is not driven by variations within one single galaxy but rather originates from the diversity of ISM conditions in a large sample of galaxies covering wide ranges in metallicity. 
The dispersion is largest in the SFR-$L_{\text{[CII]}}$ trend (0.38 dex) as compared to the SFR-$L_{\text{[OIII]}}$ (0.30 dex) and SFR-$L_{\text{[OI]}}$ (0.25 dex) relations. 
For a sample of similar size (24 galaxies), the SFR calibration for [C{\sc{ii}}] presented in \citet{2011MNRAS.416.2712D} reports a 1$\sigma$ dispersion of only 0.27 dex. Part of the increased scatter observed for the DGS sample might be attributed to the uncertainties on the reference SFR tracer, which was shown to be sensitive to the star formation history and, possibly, the grain properties of metal-poor dwarf galaxies (see Sections \ref{compareSFR} and \ref{compareSFRobsc}).
We argue, however, that the significant scatter in the SFR-$L_{\text{[C{\sc{ii}}]}}$ relation indicates the large variety of ISM conditions (i.e. gas density, radiation field, filling factors of neutral and ionized gas, excitation conditions) probed in the DGS galaxy sample (see Section \ref{scat_int.sec}). This diversity might not be surprising given the different morphological classifications (e.g. blue compact dwarfs, low-surface brightness objects, Luminous Infrared Galaxies, interacting galaxies) of the dwarf galaxies in the DGS sample.

On global galaxy scales, the [O{\sc{i}}]$_{63}$ line is considered a better intrinsic tracer of the SFR for the DGS sample compared to [C{\sc{ii}}] and [O{\sc{iii}}]$_{88}$, which suggests that the fraction of gas heating in warm and/or dense PDRs is a good approximation of the level of star formation activity across a wide range of metallicities. We do need to caution that the [O{\sc{i}}]$_{63}$ emission in extremely metal deficient objects is not necessarily linked to the classical PDRs, but might rather be powered by soft X-rays (e.g. \citealt{2008A&A...478..371P}, Lebouteiller et al. in prep.). 

To better approximate the overall gas cooling budget in galaxies, and, hereby, the heating through star formation under the assumption of local thermal equilibrium, we attempt to fit SFR calibrations with different combinations of FIR lines of the form:
\begin{equation}
\label{eqfit4}
\log~\text{SFR} = \beta + \log(L_{\text{[CII]}}^{\alpha_{1}} + L_{\text{[OI]}}^{\alpha_{2}} + L_{\text{[OIII]}}^{\alpha_{3}}).
\end{equation}
By combining the emission of two or three FIR lines (C{\sc{ii}}], [O{\sc{i}}]$_{63}$ and [O{\sc{iii}}]$_{88}$), we do not improve on the scatter in the SFR calibrations on global galaxy scales.
The combination of the brightest FIR lines on spatially resolved scales of about 100 pc did diminish the scatter in the SFR-$L_{\text{line}}$ relations, suggesting that other cooling lines (potentially in the optical wavelength domain) and/or gas heating mechanisms (unrelated to star formation) become important on global galaxy scales.

\subsection{Scatter in the SFR-$L_{\text{line}}$ relation}    
\label{scat_int.sec}        
\begin{figure}[!ht]
   \centering
    \includegraphics[width=8.5cm]{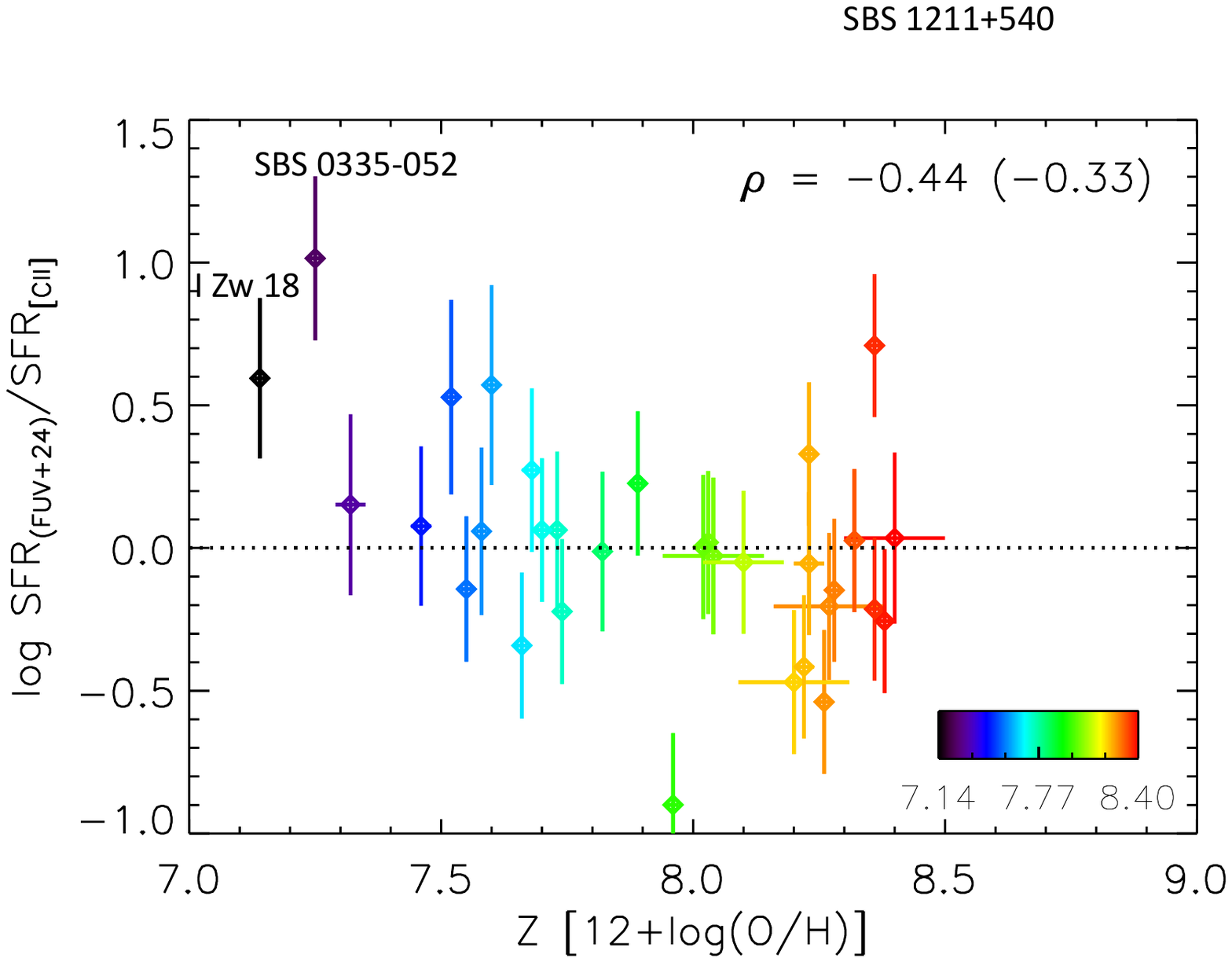}    \\
                    \includegraphics[width=8.5cm]{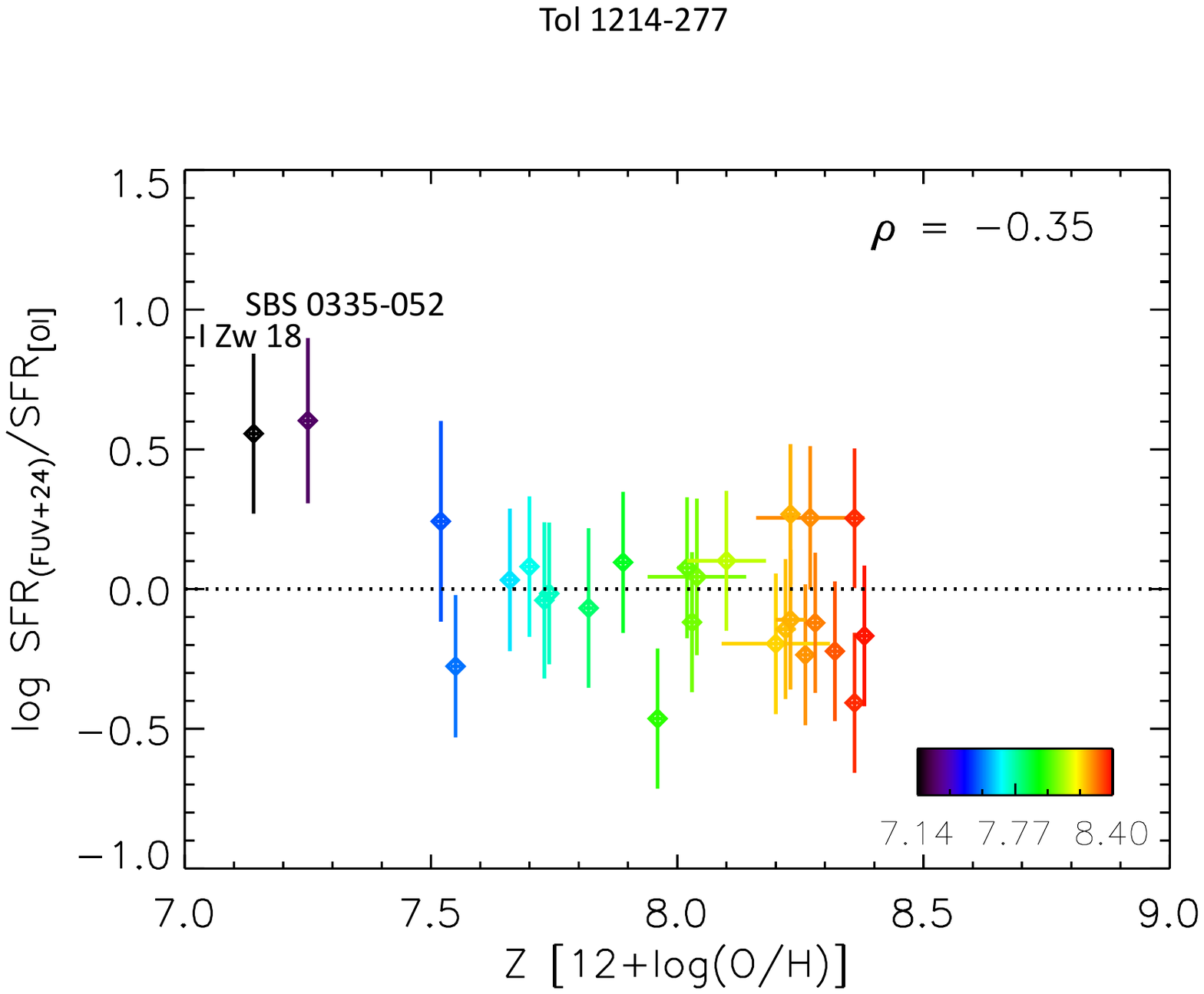}    \\
        \includegraphics[width=8.5cm]{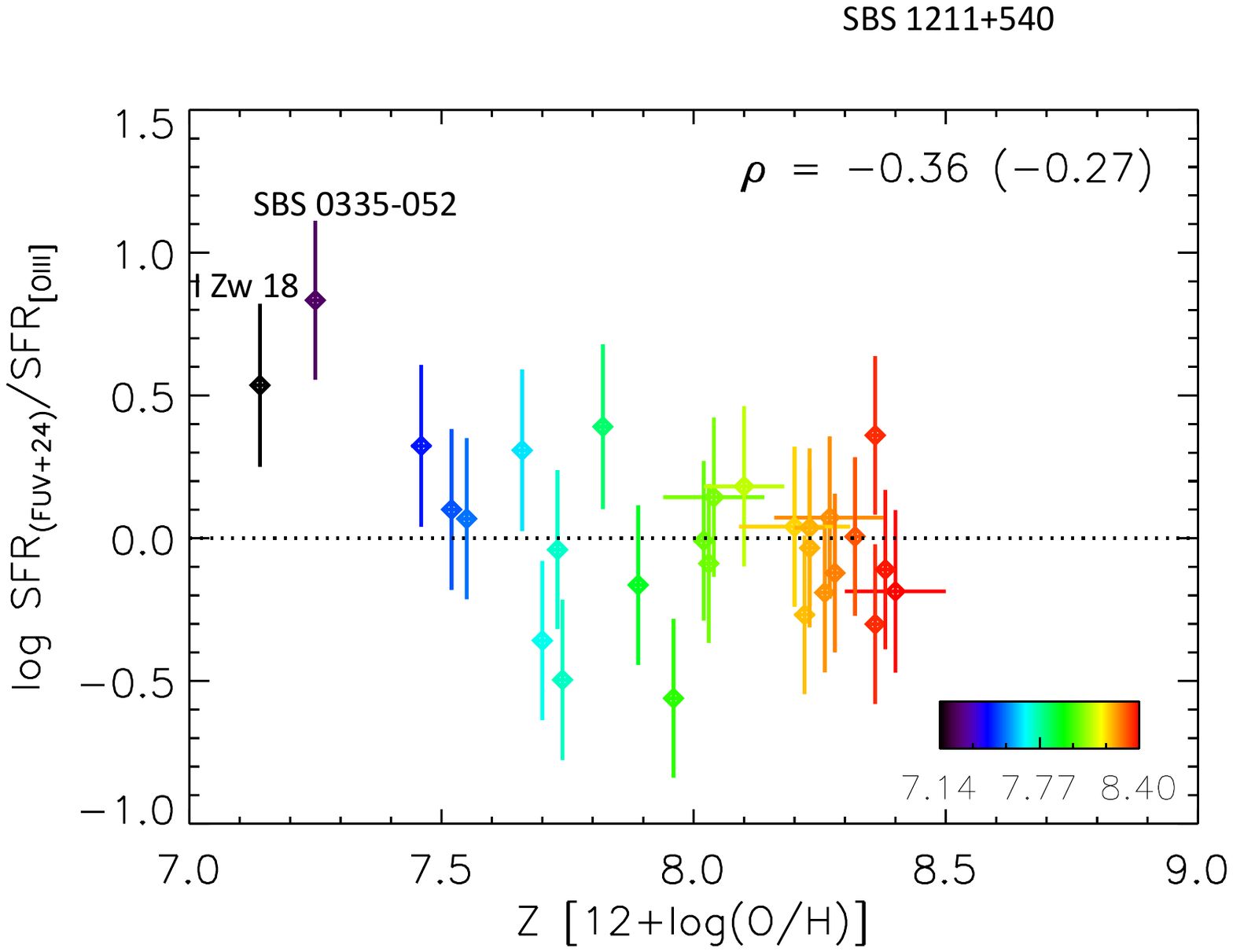}     
      \caption{Relation between the dispersion from the SFR calibrations for [C{\sc{ii}}] (top), [O{\sc{i}}]$_{63}$ (middle), [O{\sc{iii}}]$_{88}$ (bottom) as a function of oxygen abundance, 12+$\log$(O/H), on global galaxy scales. Galaxies are color-coded according to metallicity with increasing oxygen abundances going from black over blue, green and yellow to red colors. The Spearman's rank correlation coefficients are indicated in the top right corner. In parentheses, we show the dispersion for the complete galaxy sample, i.e. global galaxies which have $>5\sigma$ detections for all three lines [C{\sc{ii}}], [O{\sc{i}}]$_{63}$ and [O{\sc{iii}}]$_{88}$.}
              \label{plot_SFRdisp_Z}
    \end{figure}

\begin{figure}[!ht]
   \centering 
     \includegraphics[width=8.5cm]{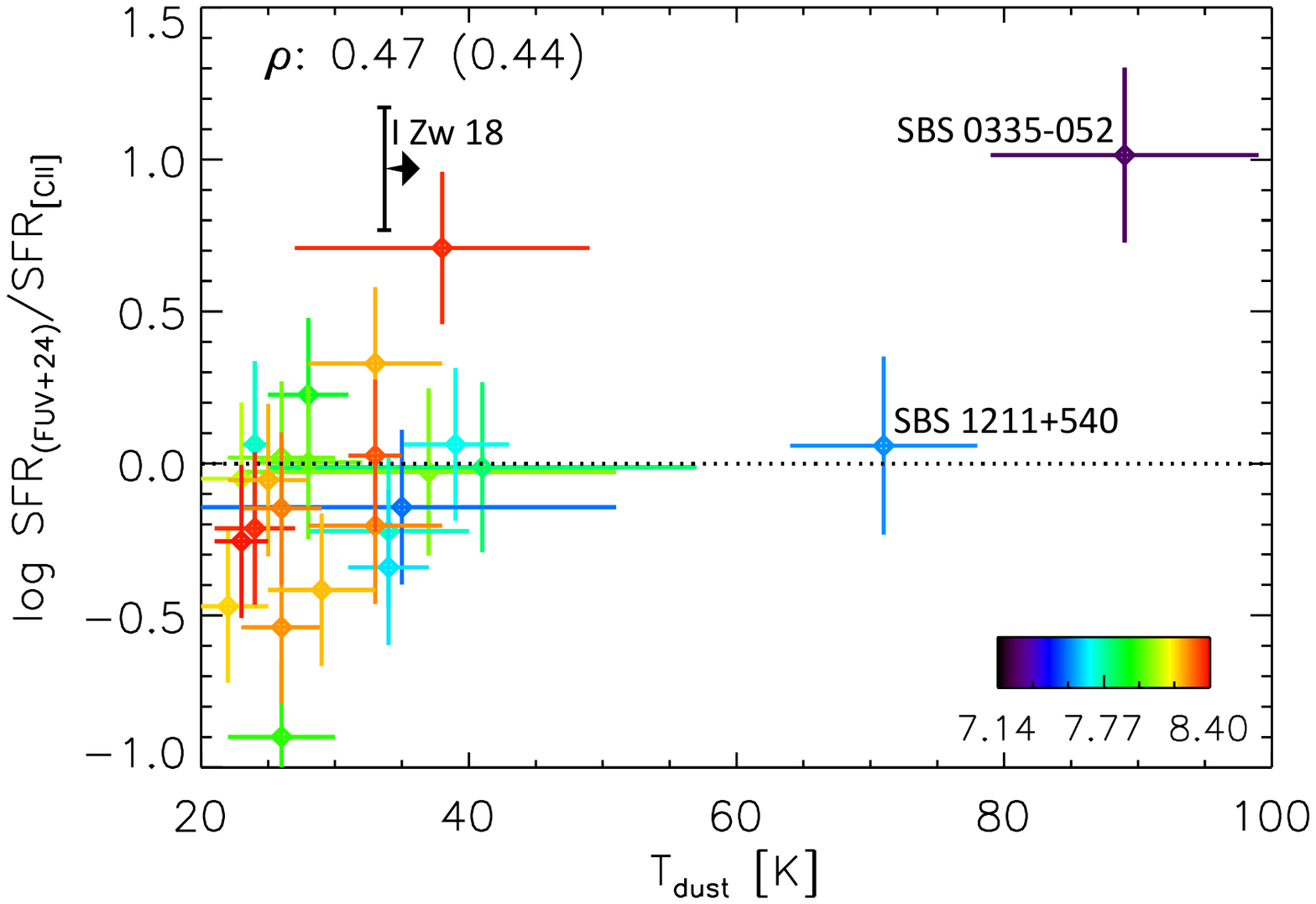} \\
          \includegraphics[width=8.5cm]{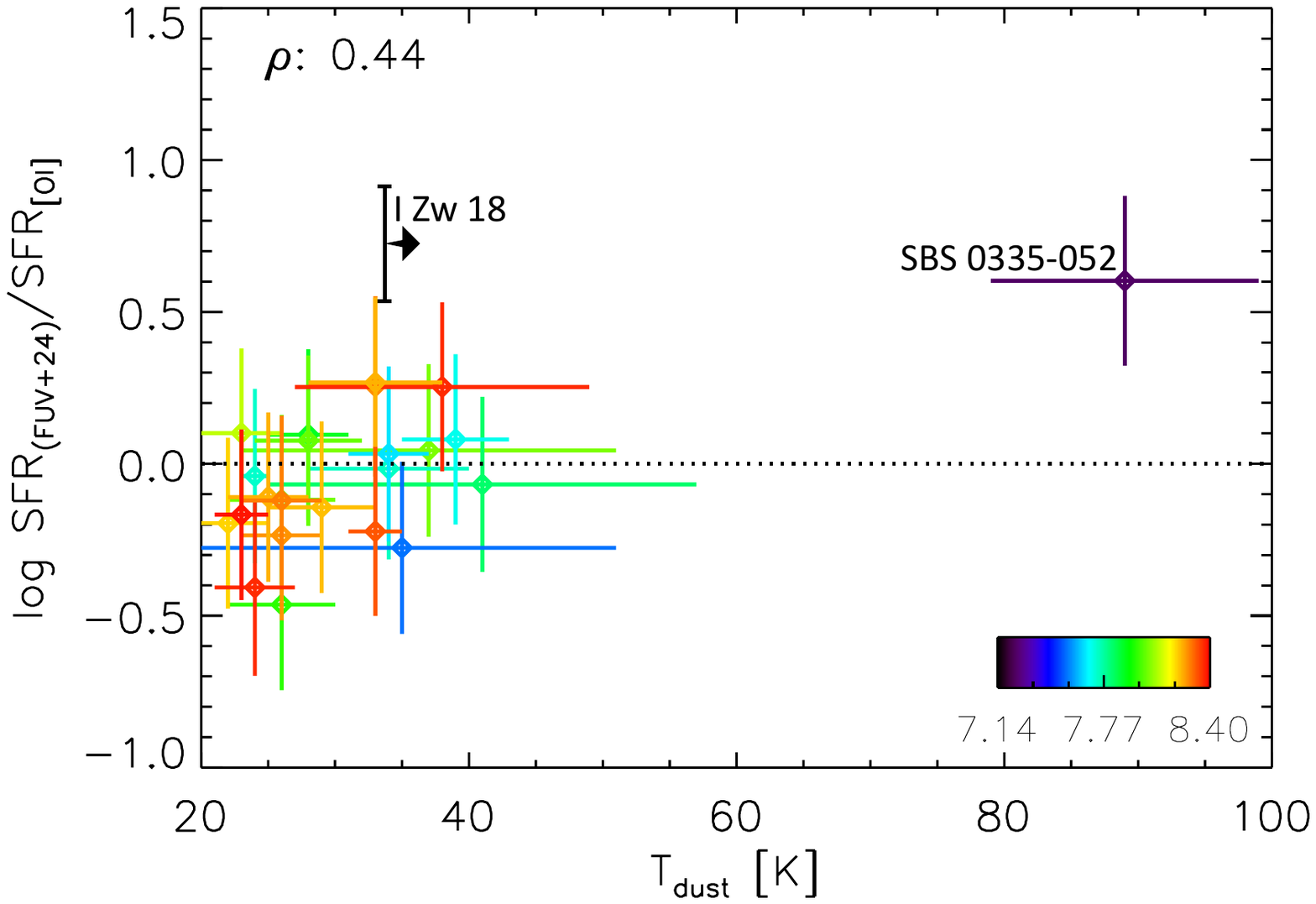} \\  
   \caption{Relation between the dispersion from the SFR calibrations for [C{\sc{ii}}] (top) and [O{\sc{i}}]$_{63}$ (bottom) as a function of dust temperature, $T_{\text{dust}}$, on global galaxy scales. The galaxy I\,Zw\,18 was not detected at PACS\,160\,$\mu$m wavelengths and, therefore, the fitting procedure was not attempted in \citet{2013A&A...557A..95R}. \citet{2012ApJ...752..112H} estimated a lower limit for the dust temperature $T_{\text{d}}$ $\geq$ 33.7 K based on the PACS\,70\,$\mu$m flux and PACS\,160\,$\mu$m upper limit, which is used to indicate the position of I\,Zw\,18 in the plots of Figure \ref{plot_PACS70160_sd}. The image format is the same as explained in Fig \ref{plot_SFRdisp_Z}.}
              \label{plot_CIIdisp_PACS70160_int}
    \end{figure}

\begin{figure}[!ht]
   \centering
    \includegraphics[width=8.5cm]{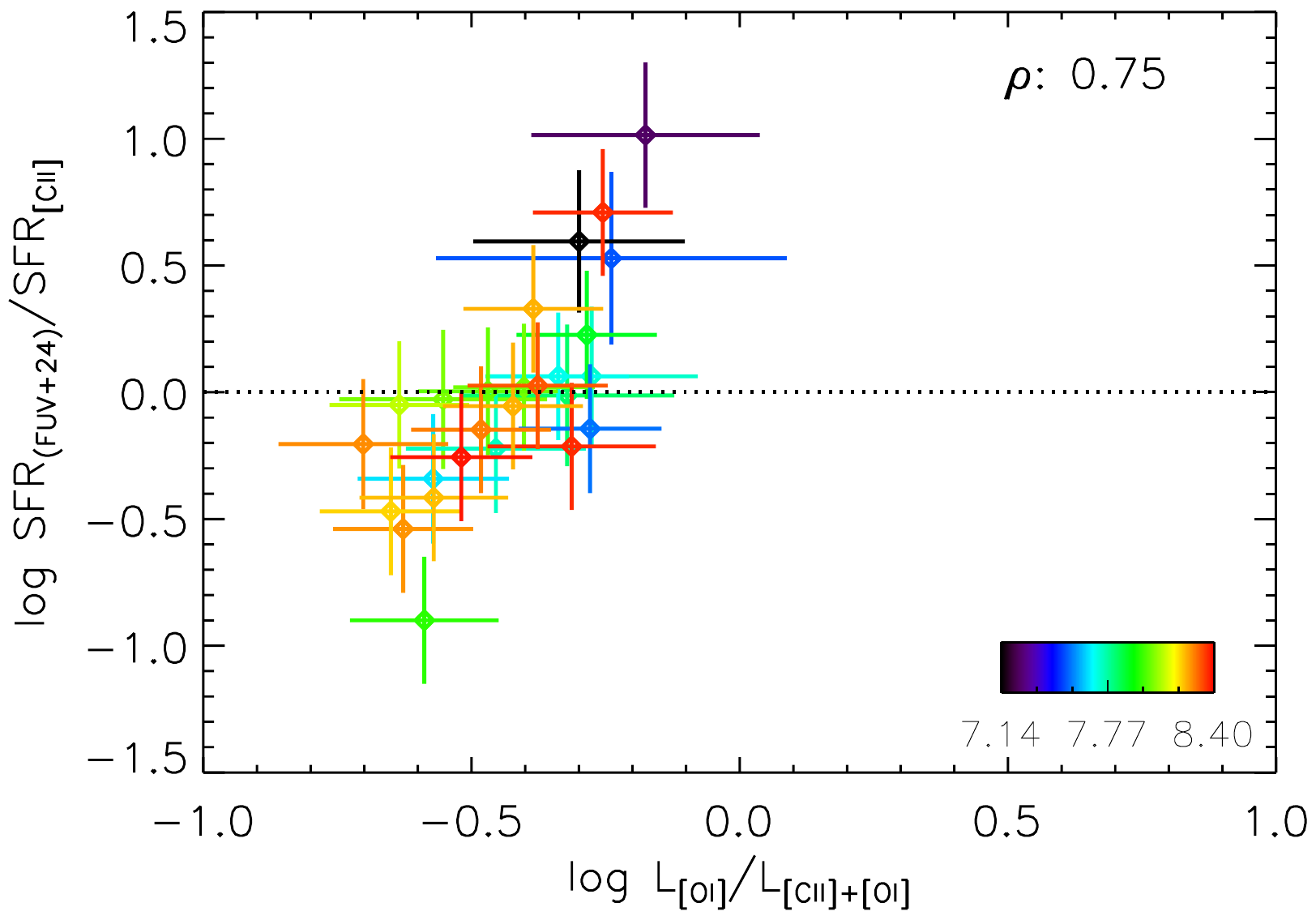}    \\
        \includegraphics[width=8.5cm]{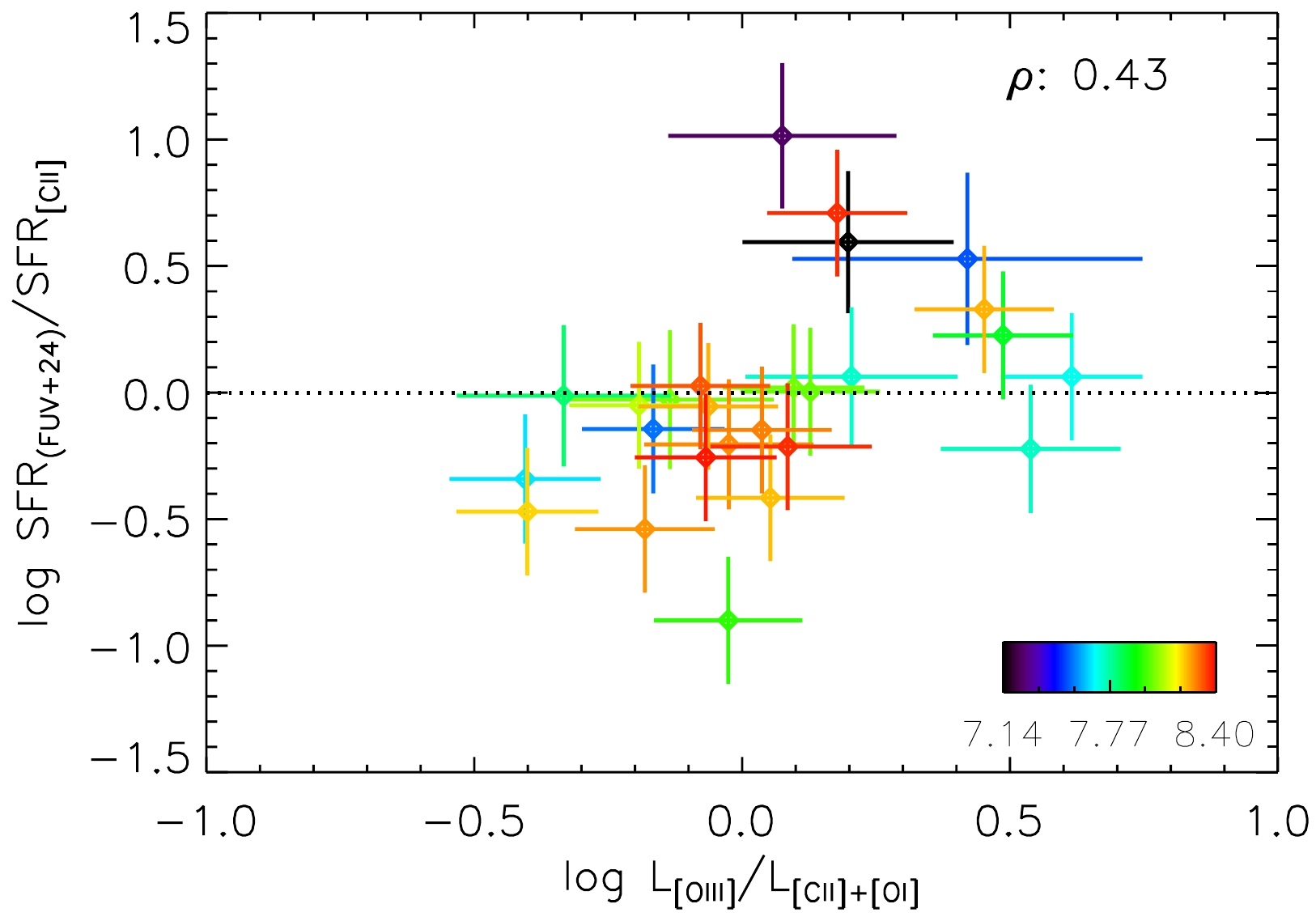}      
   \caption{Relation between the dispersion from the SFR calibrations for [C{\sc{ii}}] as a function of [O{\sc{i}}]/[C{\sc{ii}}]+[O{\sc{i}}]$_{63}$ (top) and [O{\sc{iii}}]$_{88}$/[C{\sc{ii}}]+[O{\sc{i}}]$_{63}$ (bottom) line ratios, on global galaxy scales. The image format is the same as explained in Fig \ref{plot_SFRdisp_Z}.} 
              \label{plot_CIIdisp_CIIOI}%
    \end{figure}

In this section, we focus on identifying the parameters that drive the dispersion in the SFR-$L_{\text{line}}$ relations. Figures \ref{plot_SFRdisp_Z}, \ref{plot_CIIdisp_PACS70160_int} and \ref{plot_CIIdisp_CIIOI} show the observed trends between the scatter in the $SFR-L_{\text{line}}$ relations and the metal abundance, dust temperature and ISM structure (as probed through the line ratios [O{\sc{i}}]$_{63}$/[C{\sc{ii}}]+[O{\sc{i}}]$_{63}$ and [O{\sc{iii}}]$_{88}$/[C{\sc{ii}}]+[O{\sc{i}}]$_{63}$), respectively.
The dust temperatures of 25 out of 32 galaxies are constrained by the results from a modified blackbody fitting routine with variable dust emissivity index $\beta$ presented in \citet{2013A&A...557A..95R}.

The shallow slope in the SFR calibrations for the DGS sample (as compared to the slopes for the literature sample in Section \ref{Extend.sec}) and the weak correlations in Figure \ref{plot_SFRdisp_Z} ($\rho =$ -0.44 for [C{\sc{ii}}], $\rho =$ -0.35 for [O{\sc{i}}]$_{63}$ and $\rho$ $=$ -0.36 for [O{\sc{iii}}]$_{88}$) suggest that the metal abundance has an effect on the quantitative link between the SFR and FIR line emission (in particular for [C{\sc{ii}}]). The true offset of the lowest abundance dwarfs might be even higher due to an underestimation of their SFR based on $FUV$ emission (see discussion in Section \ref{compareSFR}). The weaker [C{\sc{ii}}] emission towards lower metal abundances is consistent with the drop in [C{\sc{ii}}] surface brightness in PDR models by a factor of about 5 from metallicities of 12+$\log$(O/H) $\sim$ 8.5 down to 12+$\log$(O/H) $\sim$ 7.5 \citep{2006A&A...451..917R} (i.e. drop of a factor of 10 in metallicity) for a single cloud with density $n_{\text{H}}$ $\sim$ 10$^{3}$ cm$^{-3}$ (see their Figure 6). The weak trends for [O{\sc{i}}]$_{63}$ and [O{\sc{iii}}]$_{88}$ are mainly driven by two galaxies of extreme low metal abundance, I\,Zw\,18 and SBS 0335-052, in which the [O{\sc{i}}]$_{63}$ and [O{\sc{iii}}]$_{88}$ lines do not seem to add significantly to the overall gas cooling. Since the SFR is unlikely to be overestimated for these galaxies based on $FUV$+MIPS\,24\,$\mu$m (see comparison with other SFR tracers in Section \ref{DGScompare} of the Appendix), the offset of these extremely dust deficient galaxies might suggest that other lines dominate the cooling processes (e.g. Ly $\alpha$). The gas heating might, alternatively, be dominated by heating mechanisms other than the photoelectric effect (e.g. soft X-ray heating, mechanical heating, cosmic rays), which could disperse the link between the emission of cooling lines and the SFR. For I\,Zw\,18, the heating has indeed been shown to be dominated by soft X-ray heating (\citealt{2008A&A...478..371P} and Lebouteiller et al. in prep.), which is likely to also affect the SFR-$L_{\text{line}}$ relations.

The dust temperatures of galaxies seem to correlate (weakly) with the dispersion in the SFR calibration for [C{\sc{ii}}] ($\rho$ = 0.47) and [O{\sc{i}}]$_{63}$ ($\rho$ = 0.44). The global galaxy analysis, hereby, confirms the trends observed in Figure \ref{plot_PACS70160_sd} on spatially resolved galaxy scales. With the DGS sources showing a trend of increasing dust temperature with decreasing metal abundance \citep{2013A&A...557A..95R}, the correlation of the dispersion in the SFR-$L_{\text{[CII]}}$ and SFR-$L_{\text{[OI]}}$ relations with dust temperature seems -at least in part- related to the metallicity of DGS sources.

The dispersion in the SFR-$L_{\text{[CII]}}$ relation clearly correlates with the [O{\sc{i}}]$_{63}$/[C{\sc{ii}}]+[O{\sc{i}}]$_{63}$ line ratio ($\rho$ = 0.75), while a trend is also present -although less obvious- for [O{\sc{iii}}]$_{88}$/[C{\sc{ii}}]+[O{\sc{i}}]$_{63}$ ($\rho$ = 0.43). Making similar plots for the dispersion in the SFR relation for [O{\sc{i}}]$_{63}$ as a function of [O{\sc{i}}]$_{63}$/[C{\sc{ii}}]+[O{\sc{i}}]$_{63}$ ($\rho$ = 0.35) does not reveal a clear trend (graph is not shown here), suggesting that the [O{\sc{i}}]$_{63}$ line is capable of tracing the SFR in a consistent way irrespective of changes in the ISM structure.

To understand the observed trends between the scatter in the SFR-$L_{\text{line}}$ relations and several physical parameters, we try to link the low abundance to the warm dust temperatures and different ISM structure observed in low-metallicity galaxies. In low abundance galaxies, the fraction of metals is lower in the solid as well as gas phase. The lower abundance of grains, however, does not directly cause a decrease of the photo-electric heating efficiency, since it is, at least partially, compensated by a higher heating rate in dwarf galaxies, exhibited by their hotter average temperature (e.g. \citealt{2013A&A...557A..95R}). Similarly, the lower abundances of C and O in the gas phase will be balanced by higher line cooling rates. Deficits of species like C and N could, however, occur relative to the O/H abundance (used here to measure metallicity via the relative O abundance) which could result in relatively less cooling provided by the C lines.

The photon escape fraction might become more important with decreasing metallicity due to the porosity of the ISM, which lowers the energy input for the heating of dust and gas through the photo-electric effect. Other than higher photon escape fractions, the hard radiation fields in low-metallicity environments can also enhance grain charging, making grains less efficient for the photo-electric effect (e.g. \citealt{1985ApJ...291..722T,1997ApJ...491L..27M,2001A&A...375..566N,2012ApJ...747...81C,2013ApJ...776...38F}).

Indeed, stars at lower metallicities have higher effective temperatures due to line blanketing effects. For a given stellar age and mass, they will produce more hard photons than at solar luminosities.
Due to the longer mean free path lengths of UV photons in dust deficient media, the ionization of gas and participation in the gas/dust heating persist over large distances from the ionizing sources in star-forming complexes, which furthermore increases the dust temperatures.
As a consequence of the longer distances traversed by ionizing photons, the $C^{+}$-emitting zone in galaxies can be enlarged compared to higher metallicity environments due to the deeper penetration of $FUV$ photons capable of photo-dissociating CO molecules (e.g. \citealt{1995ApJ...454..293P,1996ApJ...465..738I,1997ApJ...483..200M,2011A&A...531A..19I}). Also the filling factor of ionized gas phases will enlarge due to the hardness of the radiation field and the transparency of the ISM in low metallicity objects. 

Grain properties might, furthermore, vary in objects of lower metal abundance. PAH emission has been shown to decrease below 12 + $\log$(O/H) $\sim$ 8.1 \citep{2004A&A...428..409B,2005ApJ...628L..29E,2006ApJ...646..192J,2006A&A...446..877M,2007ApJ...663..866D,2008ApJ...678..804E,2008ApJ...672..214G}, while the abundance of very small grains grows drastically relative to the large grain population due to the fragmentation of those large grains through shocks experienced in the turbulent ISM \citep{1996ApJ...469..740J,2002A&A...382..860L,2003A&A...407..159G,2005A&A...434..867G}. Knowing that PAHs and very small grains are the main contributors to the photo-electric effect, the outcome on the gas heating efficiency and the subsequent gas cooling remains a puzzle.

In summary, we argue that the best SFR tracer varies for different environments depending on the density and ionization state of the gas. Due to the hardness of the radiation field and longer mean free path lengths in metal-poor galaxies, the filling factors of ionized gas media are bound to grow drastically, resulting in [C{\sc{ii}}] and [O{\sc{i}}]$_{63}$ being poor SFR diagnostics. In such highly ionized regions, we expect most of the carbon and oxygen to be locked in elements with higher ionization potentials. The [C{\sc{ii}}] and [O{\sc{i}}]$_{63}$ line emission might, furthermore, be affected by a decrease in the photo-electric efficiency due to higher photon escape fractions and/or grain charging. The reliability of [O{\sc{iii}}]$_{88}$ as a SFR indicator mainly relies on the filling factor of diffuse, highly ionized gas. The large range covered in [O{\sc{iii}}]$_{88}$/[C{\sc{ii}}]+[O{\sc{i}}]$_{63}$ (from -0.4 to 0.6) suggests that the relative filling factors of PDRs and ionized media can differ significantly from one galaxy to another, depending on the hardness of the radiation field, excitation conditions and filling factor of low-density gas relative to compact gas clumps.
The choice of a reference SFR tracer would, thus, benefit from knowledge on the ionization state and density of the gas. Without any precursory constraints on the ISM conditions, the [O{\sc{i}}]$_{63}$ line is considered to be the most reliable SFR indicator for galaxies covering a wide range in metal abundances. 
    
\section{Prescriptions for extending the SFR calibrations to other galaxy samples}
\label{Extend.sec}

In this section, we derive SFR calibrations for different galaxy populations. Hereto, we gather FIR fine-structure line measurements from the literature for dwarf galaxies, starbursts, Ultra-Luminous InfraRed Galaxies (ULIRGs), galaxies harboring an active galactic nucleus (AGN) and high-redshift galaxies (ranging from $z = 0.5$ to $6.6$). The entire galaxy sample constitutes of 530, 150 and 102 galaxies with [C{\sc{ii}}], [O{\sc{i}}]$_{63}$ and [O{\sc{iii}}]$_{88}$ detections, respectively. 

\subsection{Literature sample}
The literature sample of the local Universe ($z < 0.5$) was assembled from FIR line measurements published based on ISO observations in \citet{2008ApJS..178..280B} (83 galaxies) and \textit{Herschel} data in \citet{2013ApJ...776...65P} (1 galaxy), \citet{2012ApJ...755..171S} (101 galaxies), \citet{2013ApJ...774...68D} (206 galaxies), \citet{2013ApJ...776...38F} (24 galaxies), \citet{2011ApJ...728L...7G} and Graci{\'a}-Carpio et al. (in prep.) (56 galaxies). Where duplications exist between ISO and \textit{Herschel} spectroscopy, we choose the \textit{Herschel} data (see Section \ref{Compare.sec} for a comparison between \textit{Herschel} and ISO spectroscopy measurements.).
For the \citet{2008ApJS..178..280B} sample of ISO observations, we consider all galaxies with emission unresolved with respect to the ISO beam. While \citet{2011MNRAS.416.2712D} only considered galaxies with GALEX $FUV$ and MIPS\,24\,$\mu$m observations, we extend the literature sample to 84 galaxies from the \citet{2008ApJS..178..280B} sample with IRAS\,12, 25, 60 and 100\,$\mu$m flux measurements, from which the TIR luminosity and, thus, the SFR can be computed. Although some of the galaxies from the \citet{2008ApJS..178..280B} sample have been observed with \textit{Herschel}, the lack of their published FIR line fluxes led to the usage of the ISO flux measurements.
All the literature works of \textit{Herschel} observations present the FIR line measurements of all three lines of interest ([C{\sc{ii}}], [O{\sc{i}}]$_{63}$, [O{\sc{iii}}]$_{88}$), with the exception of \citet{2012ApJ...755..171S} and \citet{2013ApJ...774...68D}\footnote{Other FIR lines have been observed for the GOALS sample, but have not yet been published.} reporting only [C{\sc{ii}}] measurements. 
The Great Observatories All-sky LIRG Survey (GOALS) sample \citep{2013ApJ...774...68D} was complemented with data from other \textit{Herschel} programs. We exclude sources already presented in \citet{2012ApJ...755..171S} and Graci{\'a}-Carpio et al. (in prep.) (and, thus, already part of our literature sample), resulting in 206 sources of the original GOALS sample.

FIR line detections and upper limits for high-redshift galaxies\footnote{We refer to high-redshift galaxies starting from redshifts $z \geq 0.5$.} are gathered for fine-structure lines [C{\sc{ii}}], [O{\sc{i}}]$_{63}$ and [O{\sc{iii}}]$_{88}$ based on observations with a large variety of ground-based facilities and the \textit{Herschel} Space Observatory.
To convert redshifts to luminosity distances, we use the NED cosmology calculator \citep{2006PASP..118.1711W} assuming a spatially flat cosmology with $H_{\text{0}}$ = 67.3 km s$^{-1}$ Mpc$^{-1}$, $\Omega_{\lambda}$ = 0.685 and $\Omega_{\text{m}}$ = 0.315 \citep{2013arXiv1303.5076P}.

\subsection{Source classification}

We classify galaxies as dwarfs if the criterion $L_{\text{H}} < 10^{9.6} L_{\text{H},\odot}$ is fulfilled, similar to the selection procedure applied in \citet{2008ApJ...674..742B}. We do not distinguish between the different classifications of dwarf galaxies (e.g. blue compact dwarfs, late-type spirals, Magellanic irregulars).

For the more massive galaxy populations, we make a distinction between the dominant power source for infrared emission, i.e. star formation or AGN activity.
To homogenize the classification of starburst, composite and AGN sources for the different literature datasets, we adapt the source classification of \citet{2012ApJ...755..171S} to the selection criteria used in \citet{2013ApJ...774...68D} based on the EW of the mid-infrared PAH feature at 6.2\,$\mu$m. More specifically, galaxies are considered to be AGN-dominated if EW (PAH 6.2\,$\mu$m) $\lesssim$ 0.3 and classified as pure starburst if EW (PAH 6.2\,$\mu$m) $\gtrsim$ 0.5. Galaxies characterized by intermediate equivalent width values are considered composite sources, i.e. with starburst and AGN contributions to the mid-infrared features.
Applying these selection criteria results in the classification of 94 composite/AGN sources and 7 starburst galaxies from the galaxy sample presented in \citet{2012ApJ...755..171S}, among which 19 can be classified as ULIRGs. The GOALS sample consists of 129 starburst galaxies and 77 AGN or composite sources, among which 2 can be assigned ULIRGs.
Based on the optical source classification of \citet{2013ApJ...776...38F}, we identify 6 H{\sc{ii}}-dominated/starburst galaxies and 18 LINER/Seyfert galaxies.
For the \citet{2008ApJS..178..280B} sample, 
we use an optical classification similar to \citet{2011MNRAS.416.2712D} to distinguish between purely star-forming objects (37 H{\sc{ii}}/starburst galaxies) and objects with power sources other than star formation (36 transition/LINER/Seyfert galaxies). Galaxies with no or an uncertain object classification on the NASA Extragalactic Database (NED) were omitted from our sample. The \citet{2008ApJS..178..280B} sample, furthermore, includes 10 dwarf galaxies.
Based on the optical source classification for the SHINING sample (Survey with \textit{Herschel} of the Interstellar Medium in Nearby Infrared Galaxies, \citealt{2010A&A...518L..41F,2010A&A...518L..36S,2011ApJ...728L...7G}, Graci{\'a}-Carpio et al. in prep.), we identify 20 starburst galaxies and 36 composite or AGN sources, among which 21 objects fulfill the criterion for ULIRGs ($L_{\text{IR}}$ $>$ $10^{12}$ L$_{\odot}$).
We classify the central region of M51 as H{\sc{ii}}-dominated, since \citet{2013ApJ...776...65P} argue that the AGN in M51 does not significantly affect the excitation of gas.

To verify that the optical source classification is consistent with the classification based on the equivalent width (EW) of the mid-infrared PAH feature at 6.2\,$\mu$m, we compare the results for a subsample of 19 galaxies from the SHINING sample with measurements of EW (PAH 6.2\,$\mu$m) reported in \citet{2013ApJS..206....1S} for all objects of the GOALS sample. The optical classification coincides with the limits in EW (PAH 6.2\,$\mu$m) to distinguish between pure starbursts and composite/AGN sources, except for 5 ULIRGs. The high level of obscuration in ULIRGs impedes the classification, but since we treat ULIRGs as a separate population with $L_{\text{IR}}$ $>$ $10^{12}$ L$_{\odot}$ distinct from starburst and composite/AGN galaxies with lower infrared luminosities, we are confident that the different methods applied for the source classification are consistent for galaxies with $L_{\text{IR}}$ $<$ $10^{12}$ L$_{\odot}$.

\subsection{Reference SFR calibrator}

For dwarf galaxies, we estimate the SFR from the same combination of (un)obscured SFR tracers (GALEX\,FUV) used for the DGS sample (see Section \ref{Ref.sec}). 
We use the GALEX $FUV$ and MIPS 24\,$\mu$m flux measurements reported in \citet{2011MNRAS.416.2712D}, when available. For the remaining sources, we retrieve GALEX $FUV$ fluxes from the GALEX catalog\footnote{http://galex.stsci.edu/GR6/}.
Catalog $FUV$ measurements have been corrected for Galactic extinction according to the recalibrated $A_{V}$ in \citet{2011ApJ...737..103S} from \citet{1998ApJ...500..525S}, as reported on the NASA/IPAC Extragalactic Database, and assuming an extinction law with $R_V$ = 3.1 derived in \citet{1999PASP..111...63F}. Relying on the conclusions drawn in \citet{2009ApJ...703.1672K} for the SINGS galaxy sample, we assume that the emission from MIPS\,24\,$\mu$m and IRAS\,25\,$\mu$m can be used interchangeably. We collect IRAS\,25\,$\mu$m flux densities from the IRAS Revised Bright Galaxy Sample \citep{2003AJ....126.1607S} or, alternatively, from the IRAS Faint Source Catalog \citep{1990IRASF.C......0M}.

For all other galaxies in the literature sample, we estimate the star formation rates based on the TIR luminosity ($L_{\text{TIR}}$\footnote{Often FIR luminosities (42.5-122.5\,$\mu$m) are reported, while the SFR calibration requires the total infrared luminosity, $L_{\text{TIR}}$ (see Table \ref{SFRref}). We use a common conversion factor of 1.75 to translate the quoted far-infrared into total infrared luminosities, following \citet{2000ApJ...533..682C}. Some authors apply the convention that $L_{\text{FIR}}$ is the luminosity in the wavelength range 40-500\,$\mu$m. We convert the latter FIR luminosities to $L_{\text{TIR}}$ using a conversion factor of 1.167, following the factor of 1.75 \citep{2000ApJ...533..682C} to convert from FIR (42.5-122.5\,$\mu$m) to TIR (8-1000\,$\mu$m) and the conversion factor $L_{\text{40-500}\mu m}$ = 1.5$\times L_{\text{42.5-122.5}\mu m}$ \citep{2003AJ....126.1607S}.}, 8 - 1000 $\mu$m) and the SFR calibration reported in \citet{2011ApJ...741..124H,2011ApJ...737...67M} (see also Table \ref{SFRref}). 
Total-infrared luminosities are reported in \citet{2012ApJ...755..171S} and \citet{2013ApJ...776...38F}. For the SHINING sample, we compute FIR (42.5-122.5\,$\mu$m) luminosities from the \textit{Herschel} continuum flux densities at 63 and 122\,$\mu$m, which could be determined based on a proper continuum estimation from the [O{\sc{i}}]$_{63}$ and [N{\sc{ii}}]$_{122}$ line observations (see Graci{\'a}-Carpio et al. in prep.). Constraining the FIR luminosities in this manner (rather than relying on the total IRAS flux densities to compute FIR) allows us to determine the infrared emission within the same regions as the PACS line observations, preventing any overestimation of the SFR for galaxies only partly covered by \textit{Herschel} spectroscopy observations. 
For the \citet{2008ApJS..178..280B} sample, we use the IRAS flux densities at 12, 25, 60 and 100\,$\mu$m to compute the TIR luminosity based on the formulas from \citet{1996ARA&A..34..749S}. Similarly, the IRAS 60 and 100\,$\mu$m flux densities are used to compute $L_{\text{FIR}}$ and converted to $L_{\text{TIR}}$ using a correction factor of 1.75 for the GOALS sample.
The SFR in M51 is estimated from total-infrared luminosity in the central 80$\arcsec$ region, reported in \citet{2013ApJ...776...65P}.

Table \ref{table3} summarizes the FIR line measurements obtained from the literature and quotes $L_{\text{TIR}}$ and SFR of high-redshift sources derived in this manner. Since the uncertainties on the line and FIR luminosities (and thus SFR estimates) for high-redshift galaxies depend strongly on the uncertainty of their assumed distance -relying on a specific cosmological model with underlying uncertainties- as well as a possible magnification factor for lensed sources, we safely assume a conservative uncertainty of 50$\%$ on both the line luminosity and SFR estimate.
    
Tables \ref{table2amin}, \ref{table2a} and  \ref{table2b} give an overview of the galaxies classified as dwarfs, H{\sc{ii}}/starburst and AGN, respectively, and indicate their name, luminosity distance $D_{\text{L}}$, reference for their FIR line measurements, total infrared luminosity (8-1000\,$\mu$m) and SFR. In both tables, ULIRGs are indicated with an asterisk behind their name.

\subsection{SFR-$L_{\text{line}}$ calibrations}

        \begin{figure*}
   \centering
    \includegraphics[width=8.45cm]{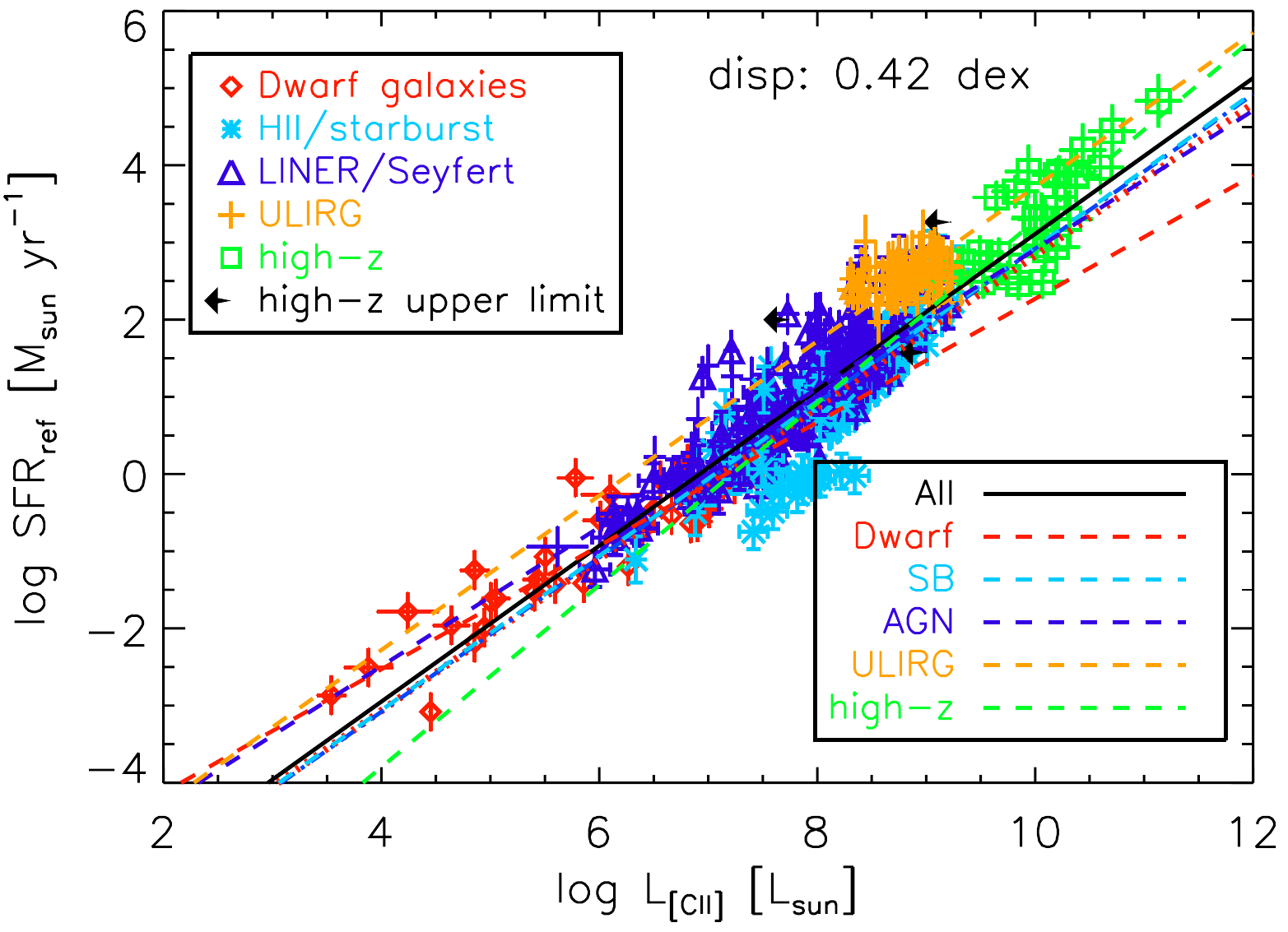}    
         \includegraphics[width=8.45cm]{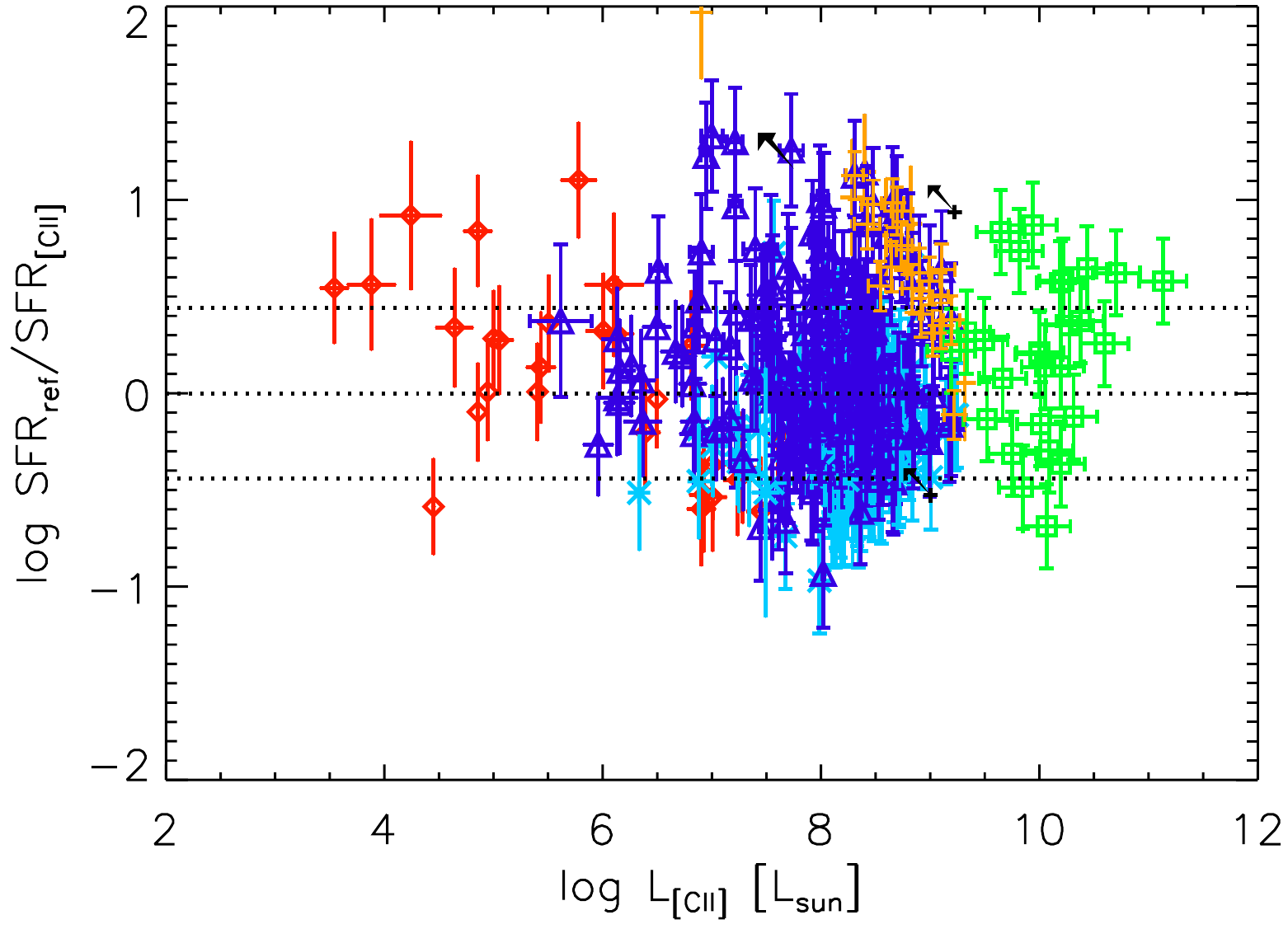}    \\
    \includegraphics[width=8.45cm]{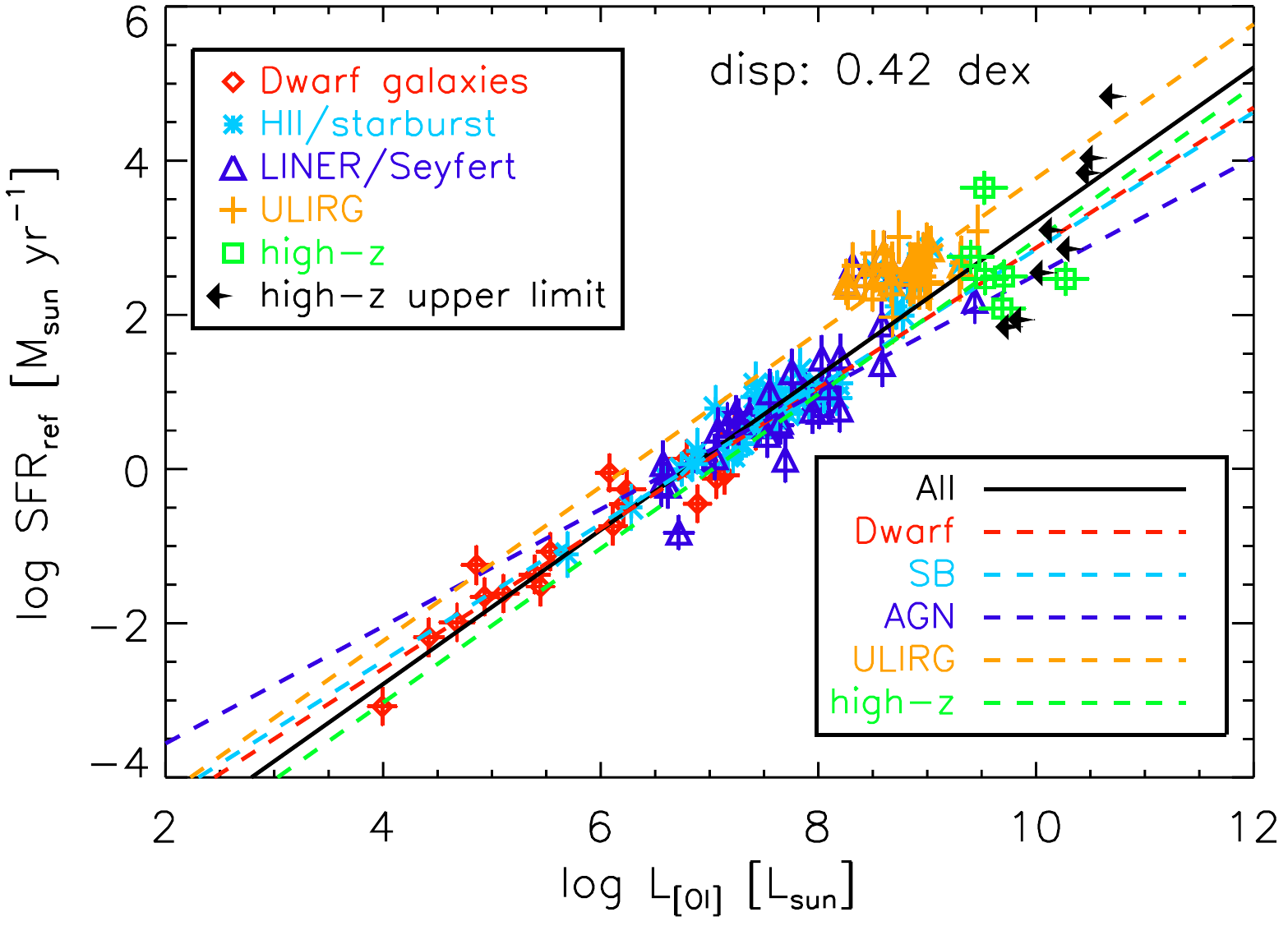}      
        \includegraphics[width=8.45cm]{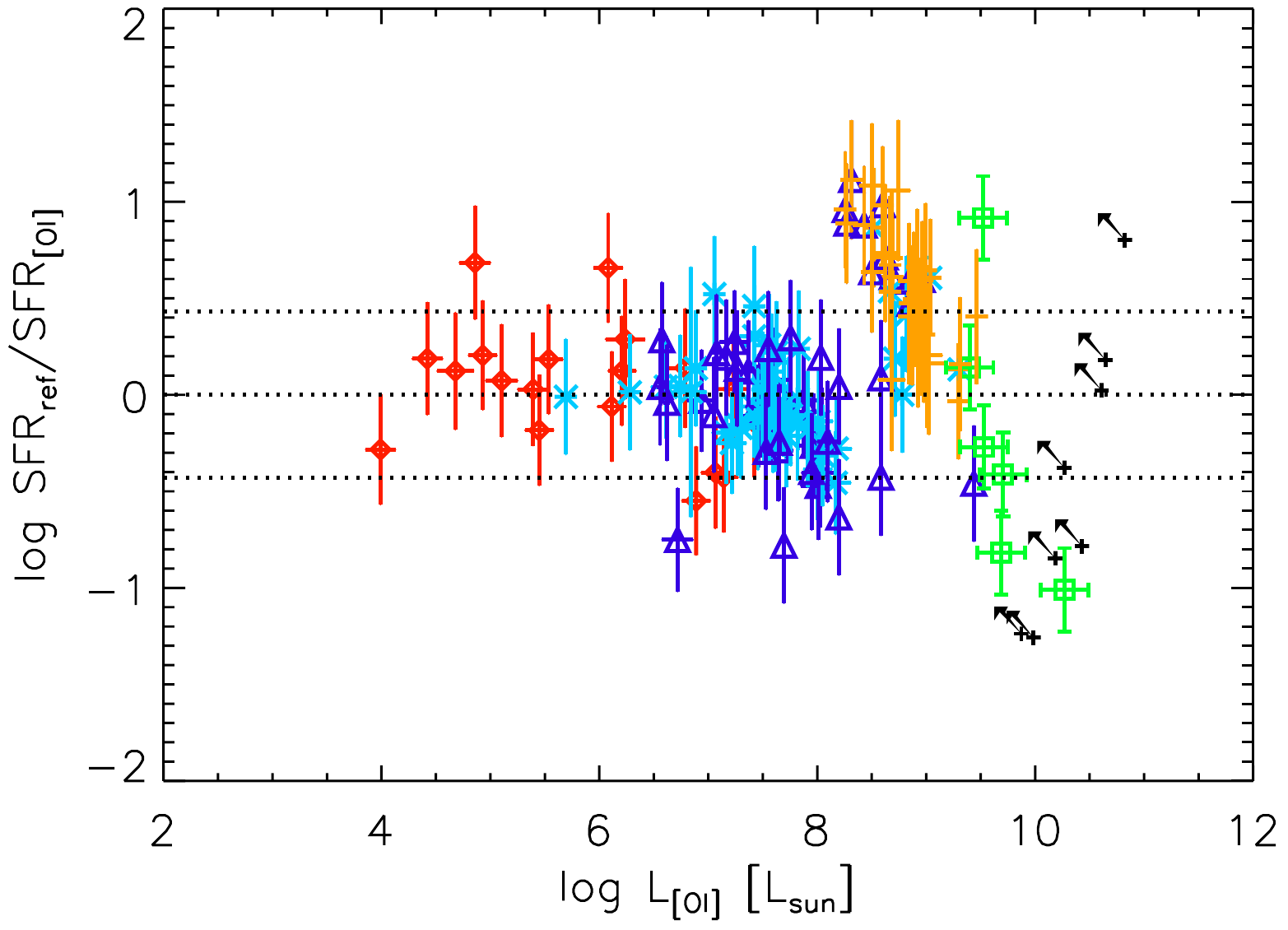}    \\ 
    \includegraphics[width=8.45cm]{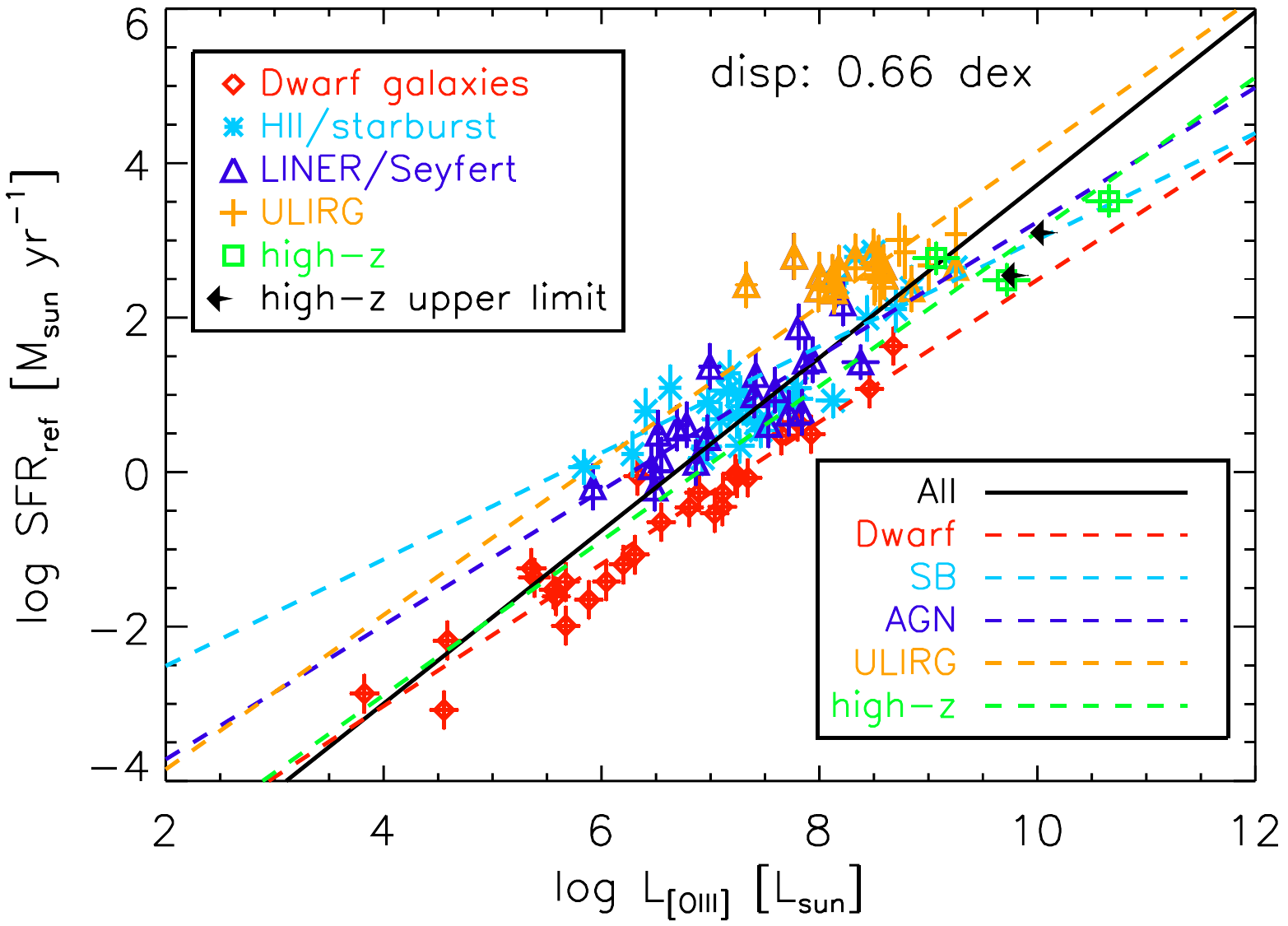}       
    \includegraphics[width=8.45cm]{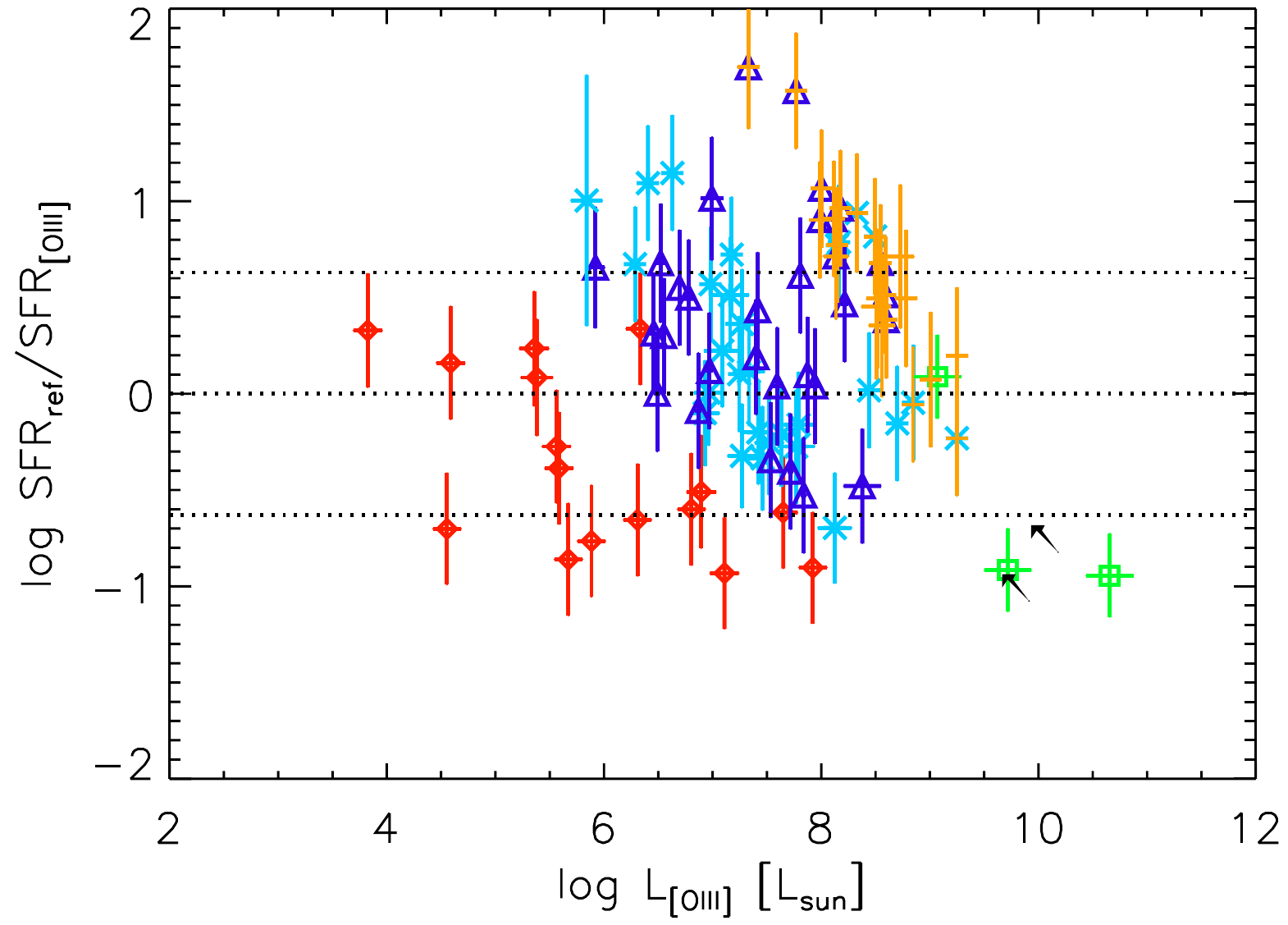}    \\
   \caption{Left: SFR calibrations based on a literature sample of different galaxy populations for FIR fine-structure lines ([C{\sc{ii}}] (top), [O{\sc{i}}]$_{63}$ (middle), [O{\sc{iii}}]$_{88}$ (bottom). DGS dwarf galaxies, H{\sc{ii}}/starburst galaxies and composite, LINER or AGN sources are presented as red diamonds, blue asterisks and purple triangles, respectively. ULIRGs with $L_{\text{FIR}}$ $>$ 10$^{12}$ L$_{\odot}$ are indicated as orange crosses. High-redshift sources can be identified as green squares while upper limits for high-redshift objects are shown as black arrows. The black solid line shows the best fitting relation for the entire galaxy sample, while the red, cyan, purple, orange and green dashed curves represent the SFR calibrations derived for separate galaxy populations, i.e. dwarf galaxies, starbursts, AGNs, ULIRGs and high-redshift sources, respectively. The 1$\sigma$ dispersion of the entire galaxy sample around the best fit is indicated in the bottom right corner of each panel. Right: dispersion plots indicating the logarithmic distance between the SFR estimates obtained from the reference SFR tracer and the FIR line emission. The 1$\sigma$ dispersion of the entire literature sample is indicated as black dashed lines. The top panel also includes previous SFR calibrations reported in \citet{2011MNRAS.416.2712D} (red dotted line) and \citet{2012ApJ...755..171S} (blue dashed-dotted line).} 
                 \label{extend}%
    \end{figure*}

For the sample of literature data, we derive SFR calibrations for each of the fine-structure lines [C{\sc{ii}}], [O{\sc{i}}]$_{63}$ and [O{\sc{iii}}]$_{88}$, as well as combinations of all lines based on the IDL procedure \texttt{MPFITFUN} and similar functions as defined in Eq. \ref{func1} and \ref{eqfit4}. To identify a correlation between the SFR and FIR line luminosities, we again require that the parameter $\alpha$ is determined at the $>$5$\sigma$ significance level.
The best fitting SFR calibrations are presented in Table \ref{calilit} for each of the different galaxy populations along with the number of galaxies used for the calibration, the slope and intercept of the best fitting line and the dispersion (or the uncertainty on the SFR estimate in parentheses).
The SFR calibrations for the entire galaxy sample allow us to compare the different FIR lines and their applicability to trace the star formation across a large sample of galaxy populations. Although all correlations are significant, 
the large dispersion in the SFR-$L_{\text{line}}$ relations (ranging from 0.42 to 0.66 dex) immediately tells us that [C{\sc{ii}}], [O{\sc{i}}]$_{63}$ and [O{\sc{iii}}]$_{88}$ are fairly unreliable SFR tracers when calibrated for the entire literature sample. In particular, the link of the [O{\sc{iii}}]$_{88}$ emission with the SFR appears to depend strongly on galaxy type.

To improve the applicability of each of the FIR lines as a SFR diagnostic, we derive separate SFR calibrations for each of the different galaxy populations in the literature sample, i.e. dwarf galaxies, H{\sc{ii}}/starburst galaxies, composite/AGN sources, ULIRGs and high-redshift galaxies. All galaxy subpopulations are exclusive, i.e. the ULIRG population consists of starburst, composite and AGN galaxies, but the starburst and AGN samples do not contain ULIRGs to prevent SFR calibrations biased by the line deficits observed in ULIRGs. 
For every galaxy population, the fitting of a combination of FIR emission lines to probe the SFR was attempted (according to formula \ref{eqfit4}) with the aim of decreasing the scatter in the SFR calibrations. The majority of line combinations did not result in an improvement of the scatter, suggesting that other FIR cooling lines are necessary to supplement the [C{\sc{ii}}], [O{\sc{i}}]$_{63}$ and [O{\sc{iii}}]$_{88}$ emission and, thus, trace a more complete cooling budget. While the three fine-structure lines ([C{\sc{ii}}], [O{\sc{i}}]$_{63}$, [O{\sc{iii}}]$_{88}$) under investigation in this paper are the brightest FIR lines in our sample of metal-poor dwarfs, other composite line tracers might be more appropriate for high-energy sources like AGNs, starbursts and ULIRGs (e.g. [N{\sc{ii}}]$_{122,205}$, [N{\sc{iii}}]$_{57}$, higher-J CO lines,...). \citet{2013ApJ...776...38F} find that [O{\sc{i}}]$_{63,145}$ and [N{\sc{ii}}]$_{122}$ are the most reliable SFR tracers for a sample of ULIRGs, while \citet{2013ApJ...765L..13Z} have shown that [N{\sc{ii}}]$_{205}$ is a potentially powerful SFR indicator in local Luminous InfraRed Galaxies (LIRGs) as well as in the more distant Universe.

\begin{table*}
\caption{Prescriptions to estimate the SFR from the relation log $SFR$ [M$_{\odot}$ yr$^{-1}$] = $\beta$ + $\alpha$ $\times$ $\log$ $L_{\text{line}}$ [L$_{\odot}$] depending on galaxy type, i.e. metal-poor dwarf galaxies, H{\sc{ii}}/starburst galaxies, composite or AGN sources, ULIRGs and high-redshift galaxies.
The first and second column indicate the FIR fine-structure line(s) and number of galaxies used in the SFR calibration, with the slope(s) $\alpha$, intercept $\beta$ and dispersion of the best fitting line for the different samples presented in columns 3, 4 and 5.
Between parenthesis, we note the uncertainty factor on the derived SFR estimates corresponding to the dispersion in the relation.}
\label{calilit}
\centering
\begin{tabular}{|l|c|c|c|c|}
\hline 
SFR calibrator & Number of galaxies & Slope & Intercept & 1$\sigma$ dispersion [dex] \\
\hline 
\multicolumn{5}{|c|}{SFR calibration: entire literature sample} \\
\hline 
$[{\rm CII}]$ & 530 & 1.01 $\pm$ 0.02 & -6.99 $\pm$ 0.14 & 0.42 (2.6) \\
$[{\rm OI}]_{63}$ & 150 & 1.00 $\pm$ 0.03 & -6.79 $\pm$ 0.22  & 0.42 (2.6) \\
$[{\rm OIII}]_{88}$\tablefootmark{a}  & 83 & 1.12 $\pm$ 0.06 & -7.48 $\pm$ 0.42 & 0.66 (4.6)  \\
\hline 
\multicolumn{5}{|c|}{SFR calibration: metal-poor dwarf galaxies} \\
\hline 
$[{\rm CII}]$ & 42 & 0.80 $\pm$ 0.05 & -5.73 $\pm$ 0.32 & 0.37 (2.3) \\
$[{\rm OI}]_{63}$ & 31 &  0.91 $\pm$ 0.05 & -6.23 $\pm$ 0.30 & 0.27 (1.9) \\
$[{\rm OIII}]_{88}$\tablefootmark{a} & 28 & 0.92 $\pm$ 0.05 & -6.71 $\pm$ 0.33 & 0.30 (2.0)  \\
\hline 
\multicolumn{5}{|c|}{SFR calibration: HII/starburst galaxies} \\
\hline 
$[{\rm CII}]$ & 184 & 1.00 $\pm$ 0.04 & -7.06 $\pm$ 0.33 & 0.27 (1.9) \\
$[{\rm OI}]_{63}$ & 41 & 0.89 $\pm$ 0.06 & -6.05 $\pm$ 0.44  & 0.20 (1.6) \\
$[{\rm OIII}]_{88}$\tablefootmark{a} & 9 & 0.69 $\pm$ 0.09  & -3.89 $\pm$ 0.63 & 0.23 (1.7)  \\
\hline 
\multicolumn{5}{|c|}{SFR calibration: composite/AGN sources} \\
\hline 
$[{\rm CII}]$ & 212 & 0.90 $\pm$ 0.04 & -6.09 $\pm$ 0.29 & 0.37 (2.3) \\
$[{\rm OI}]_{63}$ & 37 & 0.76 $\pm$ 0.09 & -5.08 $\pm$ 0.73 & 0.35 (2.2) \\
$[{\rm OIII}]_{88}$\tablefootmark{a}  & 20 & 0.87 $\pm$ 0.14 & -5.46 $\pm$ 0.98 & 0.35 (2.2)  \\
\hline 
\multicolumn{5}{|c|}{SFR calibration: ULIRGs} \\
\hline 
$[{\rm CII}]$ & 65 & 1.0\tablefootmark{b} & -6.28 $\pm$ 0.04 & 0.31 (2.0) \\
$[{\rm OI}]_{63}$ & 35 & 1.0\tablefootmark{b} & -6.23 $\pm$ 0.06 & 0.33 (2.1) \\
$[{\rm OIII}]_{88}$  & 23 & 1.0\tablefootmark{b}  & -5.80 $\pm$ 0.09 & 0.40 (2.5)  \\
\hline 
\multicolumn{5}{|c|}{SFR calibration: high-redshift ($z>0.5$)} \\
\hline 
$[{\rm CII}]$ & 27 & 1.18 $\pm$ 0.19 & -8.52 $\pm$ 1.92 & 0.40 (2.5) \\
$[{\rm OI}]_{63}$ & 6 & 1.0\tablefootmark{b} & -7.03 $\pm$ 0.29 & 0.64 (4.4) \\
$[{\rm OIII}]_{88}$  & 3 & 1.0\tablefootmark{b} & -6.89 $\pm$ 0.30 & 0.42 (2.6)  \\
\hline 
\end{tabular}
\tablefoot{
\tablefoottext{a}{We only rely on \textit{Herschel} observations for the SFR calibrations for [O{\sc{iii}}]$_{88}$ because of the significant difference found between the calibration of the \textit{Herschel} and ISO instruments (see Section \ref{Compare.sec}).}
\tablefoottext{b}{The number of galaxies and/or FIR line luminosity range was insufficient to constrain the slope and intercept of the best fitting line. 
Therefore, the fitting procedure was performed for a fixed slope of 1.}}
\end{table*}  

\subsection{Prescriptions for different galaxy populations}
Compared to the scatter in the SFR calibrations for the entire literature sample, the dispersion for each of the separate galaxy populations is significantly reduced (see Table \ref{calilit}).
Given that the dispersion in the SFR relations also differs significantly among galaxy populations and for the different FIR lines, the correlation between the SFR and FIR lines is clearly dependent on galaxy type. 
As a guideline, we briefly summarize the reliability of the three FIR fine-structure lines [C{\sc{ii}}], [O{\sc{i}}]$_{63}$ and [O{\sc{iii}}]$_{88}$ to trace the SFR in each of the following galaxy populations.
In case knowledge on the source classification is lacking, the calibrations derived for the [C{\sc{ii}}] and [O{\sc{i}}]$_{63}$ lines for the entire source sample (see top part of Table \ref{calilit}) will provide the most reliable SFR estimates with an uncertainty of factor 2.6. 

\subsubsection{Metal-poor dwarf galaxies}
The most reliable estimate of the SFR in metal-poor dwarf galaxies can be derived from the [O{\sc{i}}]$_{63}$ luminosity following the calibration:
\begin{equation}
\log SFR~=  -6.23 + 0.91 \times \log L_{\text{[OI]}},
\end{equation}
with an uncertainty factor of $\sim$ 1.9.
The star formation activity can be traced with an uncertainty of factor $\sim$ 2 and $\sim$ 2.3 from the [O{\sc{iii}}]$_{88}$ and [C{\sc{ii}}] lines, respectively, based on:
\begin{equation}
\log SFR~=  -6.71 + 0.92 \times \log L_{\text{[OIII]}},
\end{equation}
and
\begin{equation}
\log SFR~=  -5.73 + 0.80 \times \log L_{\text{[CII]}}.
\end{equation}
All SFR calibrations for metal-poor dwarf galaxies have shallower slopes compared to the entire literature sample, due to their decreasing FIR line luminosity towards lower metal abundances (see Section \ref{scat_int.sec}).

\subsubsection{Starburst galaxies}
The [C{\sc{ii}}] and [O{\sc{i}}]$_{63}$ lines can estimate the SFR in starburst galaxies within uncertainty factors of 1.9 and 1.6, respectively, following the calibrations:
\begin{equation}
\log SFR~=  -7.06 + 1.00 \times \log L_{\text{[CII]}}
\end{equation} 
and
\begin{equation}
\log SFR~=  -6.05+ 0.89 \times \log L_{\text{[OI]}}.
\end{equation}
The SFR calibration for [C{\sc{ii}}] is not very different from previous calibrations obtained by \citet{2011MNRAS.416.2712D} and \citet{2012ApJ...755..171S} for normal-star forming galaxies and starbursts, respectively (see Fig. \ref{extend}), which suggests that the [C{\sc{ii}}] line is linked to star formation in all galaxies extending from low levels of star formation activity (SFR $\sim$ 0.1 M$_{\odot}$ $yr^{-1}$) to extremely active starbursts (SFR $\sim$ 100 M$_{\odot}$ $yr^{-1}$).

For [O{\sc{iii}}]$_{88}$, we only have 9 [O{\sc{iii}}]$_{88}$ line fluxes from \textit{Herschel} after excluding the ISO measurements, resulting in the following SFR calibration with an uncertainty factor of $\sim$ 1.7 on the estimated SFR:
\begin{equation}
\log SFR~=  -3.89 + 0.69 \times \log L_{\text{[OIII]}}.
\end{equation}
Since [O{\sc{iii}}]$_{88}$ emission requires highly ionized gas of low density, it is not surprising that the [O{\sc{iii}}]$_{88}$ emission is weaker in starburst galaxies (with an average [O{\sc{iii}}]$_{88}$/[O{\sc{i}}]$_{63}$ line ratio of 0.4 in starburst as compared to 3 in dwarfs), where gas densities are also higher and mean free path lengths shorter. Although the hard radiation to ionize O$^{+}$ is likely present in local starbursts, the radiation is produced in compact, dusty regions, prohibiting the high-energy photons to reach the lower density gas surrounding dense cores (e.g. \citealt{2009ApJ...701.1147A}).

\subsubsection{Composite/AGN sources}
The SFR calibrations are more dispersed for composite and AGN sources compared to starburst galaxies. The substantial scatter might be due to a possible contribution from dust heated by the AGN to the total infrared luminosity (e.g. \citealt{2012ApJ...755..171S}). Alternatively, some AGNs appear to show line deficits similar to ULIRGs caused by highly charged dust grains which limit the photoelectric heating efficiency \citep{1985ApJ...291..722T,1997ApJ...491L..27M,2001A&A...375..566N,2012ApJ...747...81C,2013ApJ...776...38F} and/or high dust-to-gas opacities due to an increased average ionization parameter (e.g. \citealt{2011ApJ...728L...7G,2013ApJ...774...68D,2013ApJ...776...38F}).  
Part of the dispersion for the [O{\sc{i}}]$_{63}$ line might be caused by self-absorption and optical depth effects as well as the excitation through shocks\footnote{The literature data for AGNs and ULIRGs mostly correspond to total galaxy values, whereas the link between the SFR and [O{\sc{i}}]$_{63}$ line might be more dispersed zooming in into the central regions of galaxies hosting AGNs.}.

The star formation activity in AGNs can be constrained up to a factor of $\sim$ 2.3 based on all three lines:
\begin{equation}
\log SFR~=  -6.09 + 0.90 \times \log L_{\text{[CII]}},
\end{equation}
\begin{equation}
\log SFR~=  -5.08 + 0.76 \times \log L_{\text{[OI]}},
\end{equation}
\begin{equation}
\log SFR~=  -5.46 + 0.87 \times \log L_{\text{[OIII]}}.
\end{equation}
Several combinations of FIR lines, in particular for [O{\sc{iii}}]$_{88}$, result in SFR calibrations with reduced scatter. With the [O{\sc{iii}}]$_{88}$ line being on average $\sim$ 5 times fainter than [C{\sc{ii}}], we believe the results are an artifact of the fitting procedure and do not have any physical interpretation.

\subsubsection{Ultra-Luminous InfraRed galaxies}
Since the ULIRG sample does not cover a sufficient range in luminosity to constrain the slope of the SFR calibration, we fix the slope to a value of 1 (similar to the slope for the entire literature sample) and determine the intercept from the fitting procedure. The SFR calibrations for ULIRGs are offset by about 0.5 to 1.0 dex from starbursts and AGNs due to line deficits relative to their total-infrared luminosity, which are caused either by the compactness of the size of starburst regions (e.g. \citealt{2011ApJ...728L...7G,2013ApJ...774...68D,2013ApJ...776...38F}) and/or enhanced grain charging in regions with high $G_{\text{0}}$/$n_{\text{H}}$ values \citep{1985ApJ...291..722T,1997ApJ...491L..27M,2001A&A...375..566N,2012ApJ...747...81C,2013ApJ...776...38F}. Interestingly, the occurrence of line deficits has been shown to coincide with the transition between two different modes of star formation \citep{2011ApJ...728L...7G}, i.e. the star-forming disk galaxies populating the main sequence in the gas-star formation diagrams and ultra-luminous gas-rich mergers with elevated levels of star formation for the same gas fractions \citep{2010ApJ...714L.118D,2010MNRAS.407.2091G}.

With a fixed slope of 1, the SFR can be determined from the [C{\sc{ii}}], [O{\sc{i}}]$_{63}$ and [O{\sc{iii}}]$_{88}$ luminosities within uncertainty factors of 2, 2.1 and 2.5, respectively, and using the calibrations:
\begin{equation}
\log SFR~=  -6.28 + 1.0 \times \log L_{\text{[CII]}},
\end{equation}
\begin{equation}
\log SFR~=  -6.23 + 1.0 \times \log L_{\text{[OI]}},
\end{equation}
\begin{equation}
\log SFR~=  -5.80 + 1.0 \times \log L_{\text{[OIII]}}.
\end{equation}

The SFR calibrations derived from our sample of ULIRGs are offset from the SFR calibrations reported by \citet{2013ApJ...776...38F} in the sense that our SFR estimates are 2 to 4 times higher. 
Given that our literature sample contains the same ULIRGs presented in \citet{2013ApJ...776...38F}, we believe the difference in the SFR estimate can be attributed to the reference SFR tracer that was used to calibrate the SFR relations. While we rely on the TIR luminosity and the SFR(TIR) calibration presented in \citet{2011ApJ...741..124H}, \citet{2013ApJ...776...38F} use the PAH luminosity and the SFR(PAH) relation presented in \citet{2007ApJ...667..149F}.

\subsubsection{High-redshift galaxies} 
As [C{\sc{ii}}] observations in high-redshift galaxies have been more popular than other FIR fine-structure lines, we can report a relatively reliable SFR calibration for the [C{\sc{ii}}] line, based on:
\begin{equation}
\log SFR~ = -8.52 + 1.18 \times \log L_{\text{[CII]}}.
\end{equation}
Most high-redshift sources follow the trend of local starbursts and AGNs but with significant dispersion (0.40 dex), which results in an uncertainty factor on the SFR estimate of about 2.5. 
The large scatter can be attributed to some high-redshift galaxies, revealing similar [C{\sc{ii}}] deficits as ULIRGs. Relying on the warmer temperatures inferred for high-redshift sources (e.g. \citealt{2012ApJ...760....6M}), it might not be surprising that [C{\sc{ii}}] is incapable of tracing the SFR accurately due to the presence of strong radiation fields.

For the [O{\sc{i}}]$_{63}$ and [O{\sc{iii}}]$_{88}$ lines, the literature high-redshift sample did not contain a sufficient number of objects to constrain the slope and intercept in our fitting procedure. Therefore, the slope was fixed to a value of 1, which is similar to the slope in the SFR calibrations for the entire literature sample.
The SFR calibrations for [O{\sc{i}}]$_{63}$ and [O{\sc{iii}}]$_{88}$ determined in this way are:
\begin{equation}
\log SFR~=  -7.03 + 1.0 \times \log L_{\text{[OI]}}
\end{equation}
and
\begin{equation}
\log SFR~=  -6.89 + 1.0 \times \log L_{\text{[OIII]}}.
\end{equation}

The scatter in the SFR relations for [O{\sc{i}}]$_{63}$ quickly increases with only 6 high-redshift detections resulting in an uncertainty factor of $\sim$4.4 on the SFR estimate.
The [O{\sc{i}}]$_{63}$ detections from \citet{2010A&A...518L..36S}, \citet{2012MNRAS.427..520C} and \citet{2014ApJ...780..142F} suggest that the line emission is significantly brighter at high-redshift compared to the SFR calibration derived in the local Universe, which could imply that the ISM in those early Universe objects is warmer and denser compared to average conditions in the local Universe. 
The [O{\sc{i}}]$_{63}$ line might, however, easily become optically thick and could be hampered by other excitation mechanisms (e.g. shocks) in dusty high-redshift galaxies, especially during merger episodes.
We furthermore need to caution that the few [O{\sc{i}}]$_{63}$ detections of high-redshift galaxies might be biased towards hot, dense objects given the difficulty to detect [O{\sc{i}}]$_{63}$ at high redshift.
One exception is the intermediate redshift ($z=0.59$) galaxy IRAS F16413+3954 \citep{2004ApJ...604..565D}, which shows a similar [O{\sc{i}}]$_{63}$ deficit as local ULIRGs.
More [O{\sc{i}}]$_{63}$ detections at high-redshift sources are mandatory to infer the behavior of this line with the star formation activity in early Universe objects.

Based on the high-redshift [O{\sc{iii}}]$_{88}$ detections and upper limits reported in the literature for 5 galaxies \citep{2010A&A...518L..35I,2010ApJ...714L.147F,2011MNRAS.415.3473V}, the [O{\sc{iii}}]$_{88}$ line might be a potentially powerful tracer of the star formation activity in the early Universe in the absence of [C{\sc{ii}}] and with a similar degree of uncertainty on the SFR estimate (factor of $\sim$ 2.6). 
Requiring hard radiation to ionize O$^{+}$, it is not surprising that the [O{\sc{iii}}]$_{88}$ line is bright in high-redshift sources, which are known to harbor strong radiation fields (e.g. \citealt{2012ApJ...760....6M}).
Aside from the compact star-forming regions, high-redshift sources could have low-density components where the chemistry and heating is regulated by the hard radiation field.

\section{Conclusions}
\label{Conclusions.sec}

Based on \textit{Herschel} observations of low-metallicity dwarf galaxies from the Dwarf Galaxy Survey, we have analyzed the applicability of FIR fine-structure lines to reliably trace the star formation activity.
More specifically, we investigated whether three of the brightest cooling lines in the DGS sample ([C{\sc{ii}}], [O{\sc{i}}]$_{63}$, [O{\sc{iii}}]$_{88}$) are linked to the star formation rate as probed through a composite SFR tracer (GALEX $FUV$+MIPS 24\,$\mu$m).
We briefly summarize the results of our analysis:
\begin{itemize}
\item On spatially resolved galaxy scales, the [O{\sc{iii}}]$_{88}$ line shows the tightest correlation with the SFR (0.25 dex), which provides determination of the SFR with an uncertainty factor of 1.6. 
Also [O{\sc{i}}]$_{63}$ is a reasonably good SFR tracer with an uncertainty factor of 1.7 on the SFR estimate. The spatially resolved relation between [C{\sc{ii}}] and the SFR is heavily dispersed and does not allow us to constrain the SFR within a factor of 2.
\item The dispersion in the SFR calibrations results from the diversity in ISM conditions (i.e. density and ionization state of the gas) for the DGS sample covering a wide range in metallicity, rather than from variations within one single galaxy (on spatially resolved scales).
\item On global galaxy scales, the dispersion in the SFR-$L_{\text{[CII]}}$ relation (0.38 dex) is again worse compared to the [O{\sc{i}}]$_{63}$ (0.25 dex) and [O{\sc{iii}}]$_{88}$ (0.30 dex) lines.
The [O{\sc{i}}]$_{63}$ line is the most reliable overall SFR indicator in galaxies of sub-solar metallicity with an uncertainty factor of 1.8 on the SFR estimate, while the SFR derived from [O{\sc{iii}}]$_{88}$ is uncertain by a factor of 2. The [C{\sc{ii}}] line is not considered a reliable SFR tracer in galaxies of low metal abundance.
\item The scatter in the SFR-$L_{\text{[CII]}}$ relation increases towards low metallicities, warm dust temperatures and large filling factors of diffuse, highly ionized gas. Due to the porosity of the ISM and the exposure to hard radiation fields, an increased number of ionizing photons is capable of ionizing gas at large distances from the star-forming regions, which favors line cooling through ionized gas tracers such as [O{\sc{iii}}]$_{88}$. The photo-electric efficiency might, furthermore, reduce in low-metallicity environments due to grain charging and/or increased photon escape fractions.
\item On spatially resolved scales, we can reduce the scatter in the SFR calibration by combining the emission from multiple FIR lines.
Ideally, we want to probe the emission from all cooling lines that constitute the total gas cooling budget.
\end{itemize} 

Based on the assembly of literature data, we furthermore analyze the applicability of fine-structure lines [C{\sc{ii}}], [O{\sc{i}}]$_{63}$ and [O{\sc{iii}}]$_{88}$ to probe the SFR (as traced by the TIR luminosity) in H{\sc{ii}}/starburst galaxies, AGNs, ULIRGs and high-redshift objects:
\begin{itemize}
\item The [C{\sc{ii}}] and [O{\sc{i}}]$_{63}$ lines are considered to be the most reliable SFR tracers to recover the star formation activity in starburst galaxies with uncertainty factors of 1.9 and 1.6, respectively. The [O{\sc{iii}}] line, on the other hand, is weak and the SFR calibration could not be well constrained due to the low number of \textit{Herschel} [O{\sc{}iii}]$_{88}$ detections in starbursts.
\item All three FIR lines can recover the SFR from composite or AGN sources within an uncertainty of factor $\sim$ 2.3. The increased scatter in the SFR calibrations for AGNs (as compared to starbursts) might result from a possible AGN contribution to the total infrared luminosity (used to derive the SFR). Alternatively, some AGNs might show line deficits similar to ULIRGs.
\item ULIRGs are offset from the SFR calibrations for starbursts and AGNs due to line deficits relative to their total-infrared luminosity and, therefore, require separate SFR calibrations. The star formation rate in ULIRGs is preferentially traced through [C{\sc{ii}}] and [O{\sc{i}}]$_{63}$ line emission, providing SFR estimates with uncertainties of factor $\sim$ 2, while the SFR([O{\sc{iii}}]) estimate is uncertain by a factor of 2.5.
\item At high-redshift, we can only reliably determine a SFR calibration for the [C{\sc{ii}}] line (with an uncertainty factor of $\sim$ 2.5 on the SFR estimate), due to the low number of observations for the other lines.
The relatively few detections of [O{\sc{i}}]$_{63}$ and [O{\sc{iii}}]$_{88}$ appear to be bright at high-redshift, suggesting that the [O{\sc{i}}]$_{63}$ and [O{\sc{iii}}]$_{88}$ lines are also potentially powerful tracers of the SFR at high redshift, but more detections are mandatory to acquire conclusive evidence.
\end{itemize}

\begin{acknowledgements}
IDL is a postdoctoral researcher of the FWO-Vlaanderen (Belgium).
V.L. is supported by a CEA/Marie Curie Eurotalents fellowship. 
This research was supported by the Agence Nationale de la Recherche (ANR) through the programme SYMPATICO (Program Blanc Projet ANR-11-BS56-0023).
PACS has been developed by a consortium of institutes
led by MPE (Germany) and including UVIE
(Austria); KU Leuven, CSL, IMEC (Belgium);
CEA, LAM (France); MPIA (Germany); INAFIFSI/
OAA/OAP/OAT, LENS, SISSA (Italy);
IAC (Spain). This development has been supported
by the funding agencies BMVIT (Austria),
ESA-PRODEX (Belgium), CEA/CNES (France),
DLR (Germany), ASI/INAF (Italy), and CICYT/
MCYT (Spain). SPIRE has been developed
by a consortium of institutes led by Cardiff
University (UK) and including Univ. Lethbridge
(Canada); NAOC (China); CEA, LAM
(France); IFSI, Univ. Padua (Italy); IAC (Spain);
Stockholm Observatory (Sweden); Imperial College
London, RAL, UCL-MSSL, UKATC, Univ.
Sussex (UK); and Caltech, JPL, NHSC, Univ.
Colorado (USA). This development has been
supported by national funding agencies: CSA
(Canada); NAOC (China); CEA, CNES, CNRS
(France); ASI (Italy); MCINN (Spain); SNSB
(Sweden); STFC and UKSA (UK); and NASA
(USA).

\end{acknowledgements}

\appendix

\section{Comparison between different SFR tracers for the DGS sample}
\label{DGScompare}
\subsection{Limitations of SFR tracers in metal-poor dwarf galaxies}
In this work, we rely on the most recent calibrations reported in \citet{2012ARA&A..50..531K}, which are calibrated for an initial mass function (IMF) characterized by a broken power law with a slope of -2.35 from 1 to 100 M$_{\odot}$ and -1.3 between 0.1 and 1 M$_{\odot}$ \citep{2003ApJ...598.1076K}.
Table \ref{SFRref} gives an overview of the reference SFR calibrations used in this work. SFR calibrations based on single tracers in the first part of Table \ref{SFRref} predict the SFR in $M_{\odot}$ yr$^{-1}$ following the prescription $\log$ SFR = $\log$ $L_{\text{x}}$ - $\log$ $C_{\text{x}}$, where $L_{\text{x}}$ is the SFR indicator in units of erg s$^{-1}$ and $C_{\text{x}}$ represents the calibration coefficients for a specific SFR diagnostic. The second part of Table \ref{SFRref} provides the calibrations to correct unobscured SFR indicators for extinction. 

\begin{table}
\caption{Overview of the different reference SFR calibrations. In the first part, the SFR calibration coefficients are provided for several SFR indicators following the prescription $\log$ SFR = $\log$ $L_{\text{x}}$ - $\log$ $C_{\text{x}}$. The column "Age range'' specifies the lower age limit, the mean age and the age of stars below which 90$\%$ of the emission is contributed, respectively. The last column refers to the literature work(s) reporting the calibrations. The second part gives the calibration factors for several composite SFR tracers used to correct the unobscured SFR indicators for extinction, with columns 2 and 3 presenting the calibration coefficients $\nu$ and average dispersion in the calibrations, respectively.}
\label{SFRref}
\centering
\begin{tabular}{|l|c|c|c|}
\hline 
\multicolumn{4}{|c|}{SFR calibrations} \\
\hline 
$L_{\text{x}}$ [erg s$^{-1}$] & Age range [Myr] & $C_{\text{x}}$ & Ref\tablefootmark{a}  \\
\hline 
$FUV_{\text{corr}}$ & 0-10-100 & 43.35 & 1,2  \\
H$\alpha_{\text{corr}}$ & 0-3-10 & 41.27 & 1,2  \\
24\,$\mu$m & 0-5-100 & 42.69 & 3  \\
70\,$\mu$m & 0-5-100 & 43.23 & 4  \\
$TIR$ & 0-5-100 & 43.41 & 1,2  \\
\hline 
\multicolumn{4}{|c|}{Composite SFR calibrations} \\
\hline Composite tracer\tablefootmark{b} & $\nu$ &  1$\sigma$ disp [dex] & Ref\tablefootmark{a}  \\
\hline 
$L_{\text{FUV,obs}}$+$\nu$$L_{\text{24}}$\tablefootmark{c,d} & 3.89 & 0.13 & 1 \\
$L_{\text{FUV,obs}}$+$\nu$$L_{\text{1.4 GHz}}$ & 7.2 $\times$ 10$^{14}$ & 0.14 & 1 \\
$L_{\text{FUV,obs}}$+$\nu$$L_{\text{TIR}}$& 0.27 & 0.09 & 1 \\
$L_{\text{H$\alpha$,obs}}$+$\nu$$L_{\text{24}}$\tablefootmark{c} & 0.020  & 0.12 & 5 \\
\hline 
\end{tabular}
\tablefoot{
\tablefoottext{a}{\footnotesize References: (1) \citet{2011ApJ...741..124H}; (2) \citet{2011ApJ...737...67M}; (3) \citet{2009ApJ...692..556R}; (4) \citet{2010ApJ...714.1256C} ; (5) \citet{2009ApJ...703.1672K}. } \\
\tablefoottext{b}{\footnotesize The luminosities of composite tracers are in units of erg s$^{-1}$, except $L_{\text{1.4 GHz}}$ which is in units of erg s$^{-1}$ Hz$^{-1}$.} \\
\tablefoottext{c}{\footnotesize The SFR calibration is valid for the emission from MIPS 24\,$\mu$m or IRAS 25\,$\mu$m filters, due to the excellent correspondence that exists between both bands \citep{2009ApJ...703.1672K}.} \\
\tablefoottext{d}{\footnotesize Two calibrations based on $FUV$ and MIPS 24\,$\mu$m have been reported in the literature.
\citet{2008ApJ...686..155Z} found a calibration coefficient $\nu$ $=$ 6.31 based on a sample of 187 star-forming non-AGN galaxies, while
\citet{2011ApJ...741..124H} found a lower scaling factor $\nu$ $=$ 3.89, calibrated on a sample of 144 galaxies retained from the spectrophotometric survey by \citet{2006ApJS..164...81M} (97 objects) and the SINGS survey (47 objects, \citealt{2003PASP..115..928K}). No clear reason has been found to explain the difference in calibration coefficients derived by \citet{2008ApJ...686..155Z} and \citet{2011ApJ...741..124H}.}}
\end{table}

The empirically derived SFR calibrations from Table \ref{SFRref} assume a constant star formation rate over time scales comparable to or longer than the lifetime of stars to which the SFR tracers are sensitive. 
The age range for the SFR tracers used in this analysis are summarized in the second column of Table \ref{SFRref} and were adopted from \citet{2012ARA&A..50..531K}, with the first, second and third number representing the lower age boundary, the mean age and the age of stars below which 90$\%$ of the emission is contributed, respectively.
For observations covering an entire galaxy, we sample a wide range in age across the different star-forming complexes, sufficient to maintain a constant level of star formation activity when averaging out over different galaxy regions.
With low-mass galaxies being dominated by one or only few H{\sc{ii}} regions and furthermore characterized by bursty star formation histories (e.g. \citealt{1998ARA&A..36..435M,1999A&A...344..393F}), the assumption of constant star formation rate might not be valid due to the insufficient sampling of different ages. The SFR calibrations might, therefore, no longer hold. In particular, on spatially resolved scales (i.e. few hundreds of parsec or the size of a typical H{\sc{ii}} region for nearby DGS sources achieved in \textit{Herschel} observations), the sampling of different ages will not suffice to sample the entire age range to maintain a constant star formation activity (i.e. constant SFR across the entire age range) and the SFR calibrations might break down. Therefore, we caution the interpretation of the SFR in the analysis of metal-poor dwarf galaxies, in particular for the spatially resolved analysis in Section \ref{Res.sec}, in the sense that the unobscured SFR obtained from GALEX $FUV$ might underestimate the "true" level of star formation activity all throughout the galaxy.

Other than the constant SFR, empirical SFR calibrations assume the universality of the initial mass function (IMF). 
\citet{2007ApJ...671.1550P} question this universality of the IMF on global galaxy scales, especially in dwarf galaxies, arguing that the maximum stellar mass in a star cluster is limited by the total mass of the cluster with the latter being constrained by the SFR. Neglecting this effect could result in an underestimation of the SFR by up to 3 orders of magnitude.
Since the data currently at hand do not allow a proper investigation of any possible deviations from an universal IMF, we apply and derive SFR calibrations in this paper under the assumption of universality of the IMF, but keep in mind that possible systematic effects might bias our analysis. 
The extendibility of SFR calibrations to the early Universe might furthermore be affected by deviations from this universal IMF, with indications for a top heavy IMF at high-redshift \citep{1998MNRAS.301..569L,2005MNRAS.356.1191B,2008MNRAS.385..147D,2008MNRAS.385.1155L,2008MNRAS.391..363W}.

\subsection{Comparison of unobscured SFR tracers: GALEX $FUV$ and H$\alpha$}
\label{compareSFR}
$FUV$ and H$\alpha$ are commonly-used tracers of the unobscured star formation. 
The applicability of $FUV$ as a star formation rate tracer might, however, be affected by the recent star formation history in dwarfs, which is often bursty and dominated by few, giant H{\sc{ii}} regions (e.g. \citealt{1998ARA&A..36..435M,1999A&A...344..393F}). With H$\alpha$ being sensitive to the emission of OB stars with mean ages $\sim$ 10$^7$ yr and $FUV$ tracing the emission of massive A stars with ages $\sim$ 10$^8$ yr \citep{1989LNP...333..147L}, H$\alpha$ might be more appropriate as SFR tracer in dwarf galaxies with a particular bursty star formation history. 

\begin{figure}[!ht]
   \centering
    \includegraphics[width=8.5cm]{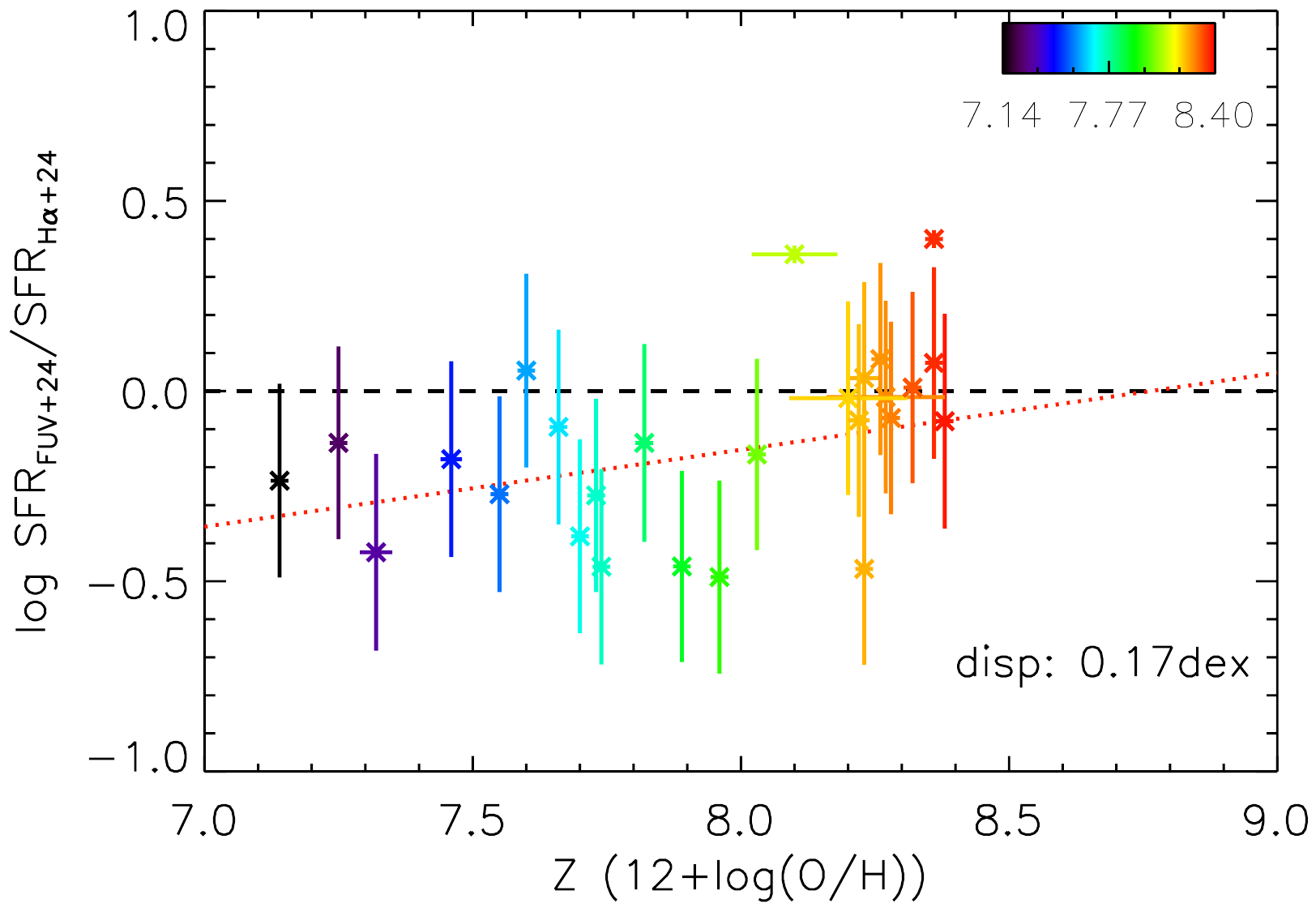}    
   \caption{Comparison between the ratio of the SFR as obtained from the SFR calibrators $FUV$+MIPS\,24\,$\mu$m and H$\alpha$+MIPS\,24\,$\mu$m as a function of oxygen abundance. Galaxies are color-coded according to metallicity with increasing oxygen abundances going from black over blue, green and yellow to red colors. The red, dotted and black, dashed line represents the best fit and a perfect one-to-one correlation, respectively. The dispersion around the best fit is indicated in the bottom right corner.}
              \label{plot_FUVvsHalpha}
    \end{figure}

In Figure \ref{plot_FUVvsHalpha}, we compare the ratio of the SFR as obtained from the SFR calibrators $FUV$+MIPS\,24\,$\mu$m and H$\alpha$+MIPS\,24\,$\mu$m (see SFR relations in Table \ref{SFRref}) as a function of oxygen abundance. Total H$\alpha$ fluxes are taken from \citet{2003ApJS..147...29G,2006ApJS..164...81M,2008ApJS..178..247K,2009AJ....138..923O}.
We determine the best fit to the data points based on linear regression fits using the IDL procedure \texttt{MPFITEXY}, which is based on the non-linear least-squares fitting package \texttt{MPFIT} \citep{2009ASPC..411..251M}. The best fitting line and a perfect one-to-one correlation are indicated as red, dotted and black, dashed lines, respectively.
The dispersion around the best fit is indicated in the bottom right corner. 

For galaxies with oxygen abundances 12+$\log$(O/H) $\geq$ 8.1, the two composite SFR tracers seem to provide consistent estimates of the SFR, while below 12+$\log$(O/H) $<$ 8.1 galaxies start to deviate from a perfect one-to-one correlation. Due to the lower stellar masses of metal-poor dwarfs, their star formation history is dominated by only few giant H{\sc{ii}} regions and, thus, heavily dependent on the time delay since the last burst of star formation. 
With $FUV$ tracing the unobscured star formation over a time-scale of 10 to 100 Myr \citep{1998ARA&A..36..189K,2005ApJ...633..871C,2007ApJS..173..267S},  the SFR derived from $FUV$ emission will be lower compared to H$\alpha$ in galaxies characterized by a single recent burst of star formation ($<$ 10 Myr) over a time scale of 100 Myr. Therefore, H$\alpha$ is considered a better SFR calibrator in low-mass dwarfs which are dominated by single H{\sc{ii}} regions.
The correspondence between SFR estimates from $FUV$ and $H\alpha$ emission for galaxies with 12+$\log$(O/H) $\geq$ 8.1 might indicate an age effect, where the latter objects are rather in a post-starburst phase and the most recent starburst occurred more than 10 Myr ago.
The unavailability of H$\alpha$ maps (only global H$\alpha$ fluxes could be retrieved from the literature for most galaxies) prevents us from using H$\alpha$ as reference SFR tracer, since we are unable to recover the H$\alpha$ emission that corresponds to the areas in galaxies covered by our \textit{Herschel} observations.
Therefore, $FUV$ is used as reference SFR tracer for this analysis, bearing in mind that the conversion to SFR might break down for low-mass metal-poor dwarfs where $FUV$ will on average underestimate the SFR derived from H$\alpha$ by 50$\%$ (see Figure \ref{plot_FUVvsHalpha}).

\subsection{Comparison of obscured SFR tracers: IRAC8\,$\mu$m, MIPS\,24\,$\mu$m, $L_{\text{TIR}}$ and 1.4\,GHz}
\label{compareSFRobsc}
Different monochromatic and multi-band data in the infrared and radio wavelength domain can be used to trace the obscured star formation component.
Here, we compare IRAC 8\,$\mu$m, MIPS 24\,$\mu$m, PACS\,70\,$\mu$m, the total-infrared luminosity and radio continuum emission at 1.4\,GHz.
Total IRAC 8\,$\mu$m flux densities have been adopted from R{\'e}my-Ruyer et al. (in prep.). The DGS sample has been observed by the \textit{Herschel} PACS (70, 100, 160\,$\mu$m) and SPIRE (250, 350, 500\,$\mu$m) photometers in all continuum bands. 
Details about the observing strategy, the applied data reduction techniques and aperture photometry results are presented in \citet{2013A&A...557A..95R}.
Total infrared luminosities $L_{\text{TIR}}$ are taken from \citet{2013PASP..125..600M}, as determined from Spitzer bands using the prescriptions in \citet{2002ApJ...576..159D}. Radio continuum measurements are retrieved from the NRAO VLA Sky Survey (NVSS) catalog (\citealt{1998AJ....115.1693C}), \citet{2004ApJ...610..772C}, \citet{2004AJ....128..617T} and \citet{2005A&A...436..837H}.

The emission from polycyclic aromatic hydrocarbons (PAHs) usually dominates the IRAC 8\,$\mu$m band in metal-rich galaxies. In low-metallicity galaxies, the IRAC\,8\,$\mu$m band might also contain an important contribution from the warm continuum emission from very small grains.
Since the PAH emission has been observed to be under-luminous below 12 + $\log$(O/H) $\sim$ 8.1 \citep{2004A&A...428..409B,2005ApJ...628L..29E,2006ApJ...646..192J,2006A&A...446..877M,2007ApJ...663..866D,2008ApJ...678..804E,2008ApJ...672..214G}, in combination with the uncertainty to quantifying the 8\,$\mu$m band in terms of PAH and VSG contribution, the IRAC\,8\,$\mu$m band is considered an unreliable SFR calibrator for the DGS sample covering a wide range in metallicity. 
\citet{2007ApJ...666..870C} could indeed identify that the sensitivity of the IRAC\,8\,$\mu$m band to metallicity is about one order of magnitude worse compared to MIPS\,24\,$\mu$m. The weak PAH emission towards lower metallicities is not directly related to the lower metal abundance but rather emanates from the generally strong and hard radiation fields in low-metallicity systems destroying and/or ionizing PAHs (e.g. \citealt{2008ApJ...682..336G,2012ApJ...744...20S}). Other than its dependence on metallicity (or thus radiation field), PAH emission tends to be inhibited in regions of strong star formation activity while it can be several times more luminous compared to other star formation rate tracers in regions with relatively weak or nonexistent star formation (e.g. \citealt{2005ApJ...633..871C,2008MNRAS.389..629B,2008ApJ...682..336G}). Prior to any conversion of IRAC\,8\,$\mu$m emission to SFR, the band emission needs to be corrected for any stellar contribution. Given that the contribution from the stellar continuum could be substantial in low abundance galaxies, this will make the correction using standard recipes (e.g. \citealt{2004ApJS..154..253H}) highly uncertain. 
All together, we argue that the IRAC 8\,$\mu$m is not appropriate as SFR indicator in our sample of dwarf galaxies with widely varying metallicities.

The most common monochromatic tracer of obscured star formation is MIPS 24\,$\mu$m emission, 
which generally originates from a combination of stochastically-heated very small grains (VSGs) and large grains at an equilibrium temperature of $\sim$100 K. 
For a grain size distribution similar to our Galaxy, we expect large equilibrium grains only to start dominating the MIPS 24\,$\mu$m emission above a threshold of $G_{\text{0}}$ $\sim$ 100 (where $G_{\text{0}}$ is the scaling factor of the interstellar radiation field, expressed relative to the $FUV$ interstellar radiation field from 6 to 13.6 eV for the solar neighborhood in units of Habing flux, i.e. 1.6 $\times$ 10$^{-3}$ erg s$^{-1}$ cm$^{-2}$). The emission in the MIPS\,24\,$\mu$m band has been shown to be well-correlated with other star formation rate tracers on both local scales (\citealt{2007ApJ...666..870C,2008AJ....136.2782L}) and global scales (e.g. \citealt{2005ApJ...633..871C,2007ApJ...666..870C,2005ApJ...632L..79W,2006ApJ...650..835A,2006ApJ...648..987P,2008ApJ...686..155Z,2009ApJ...703.1672K,2009ApJ...692..556R,2011ApJ...741..124H}) and directly traces the ongoing star formation over a timescale of $\sim$ 10 Myr \citep{2005ApJ...633..871C,2006ApJ...648..987P,2007ApJ...666..870C}.

Since the grain properties and size distribution has been shown to be sensitive to the metallicity of galaxies \citep{2002A&A...382..860L,2003A&A...407..159G,2005A&A...434..867G}, we need to verify whether MIPS\,24\,$\mu$m is an appropriate SFR tracer for the DGS sample. In low-metallicity objects, small grain sizes ($\lesssim$ 3 nm) start to dominate the 24\,$\mu$m emission compared to larger dust grains  \citep{2002A&A...382..860L,2003A&A...407..159G,2005A&A...434..867G}, due to the fragmentation of these larger dust grains through shocks experienced in the turbulent ISM. The hard radiation field in low metallicity galaxies (see Section \ref{scat_int.sec}) furthermore increases the maximum temperature of stochastically heated grains and shifts the peak of the SED to shorter wavelengths, boosting the MIPS\,24\,$\mu$m flux (e.g. \citealt{1999ApJ...516..783T,2004ApJS..154..211H,2009A&A...508..645G,2011A&A...532A..56G}).
To verify the influence of metallicity on the MIPS\,24\,$\mu$m band emission, we compare the estimated star formation rates obtained from MIPS\,24\,$\mu$m to other SFR indicators which should not be biased (or at least less) by variations in the dust composition across galaxies with different metal abundances. 

\begin{figure}[!ht]
   \centering
    \includegraphics[width=8.5cm]{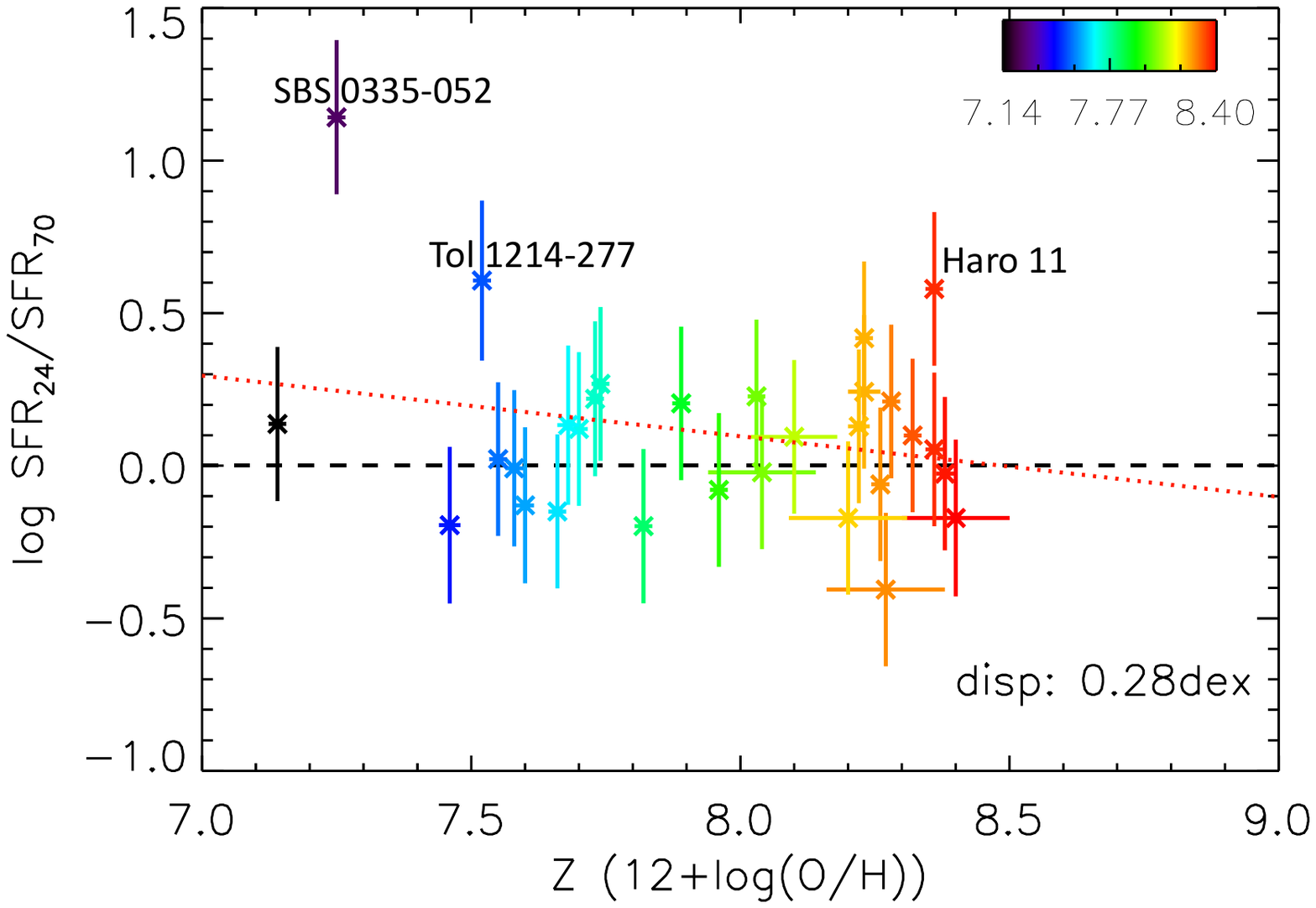}    \\
        \includegraphics[width=8.5cm]{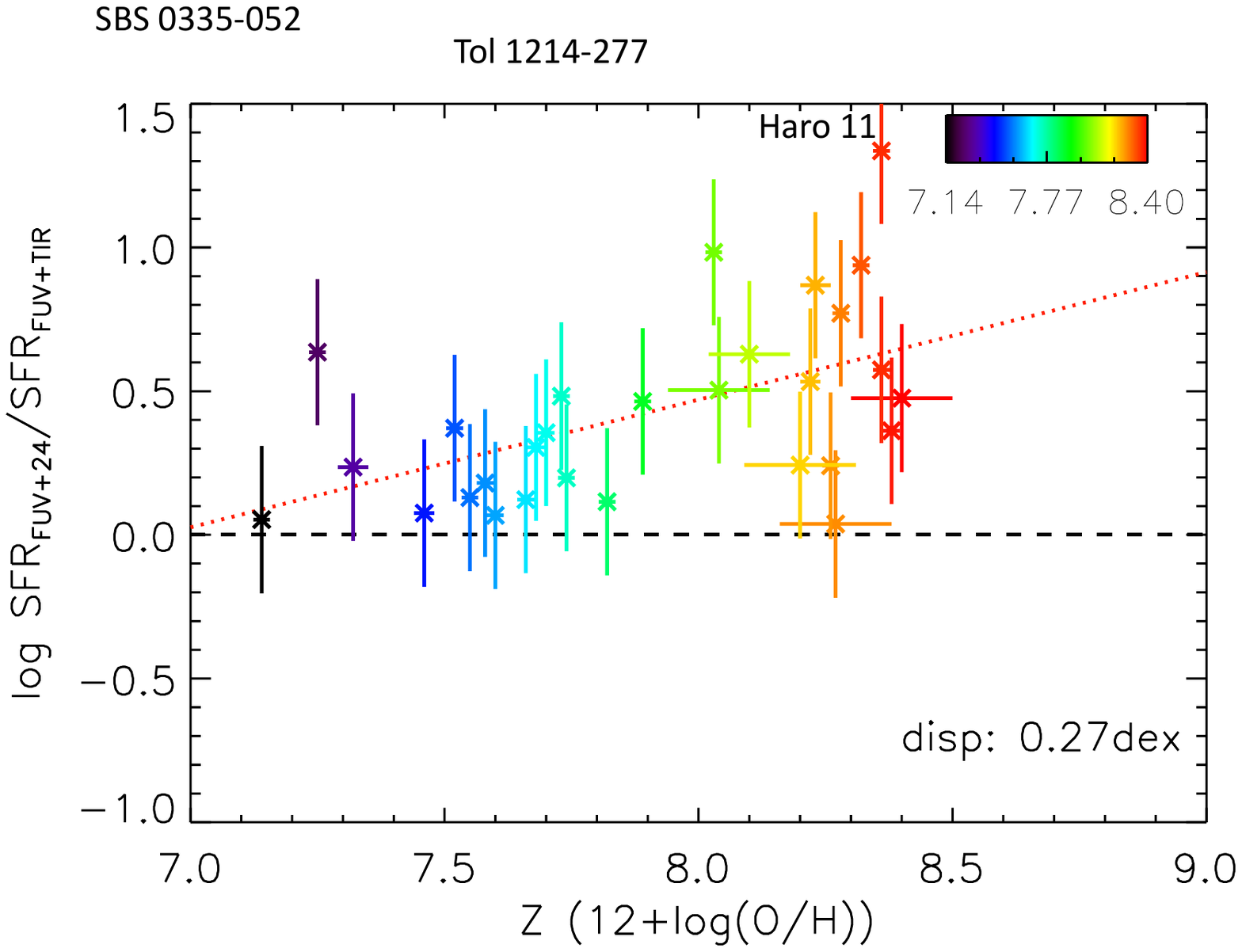}    \\
    \includegraphics[width=8.5cm]{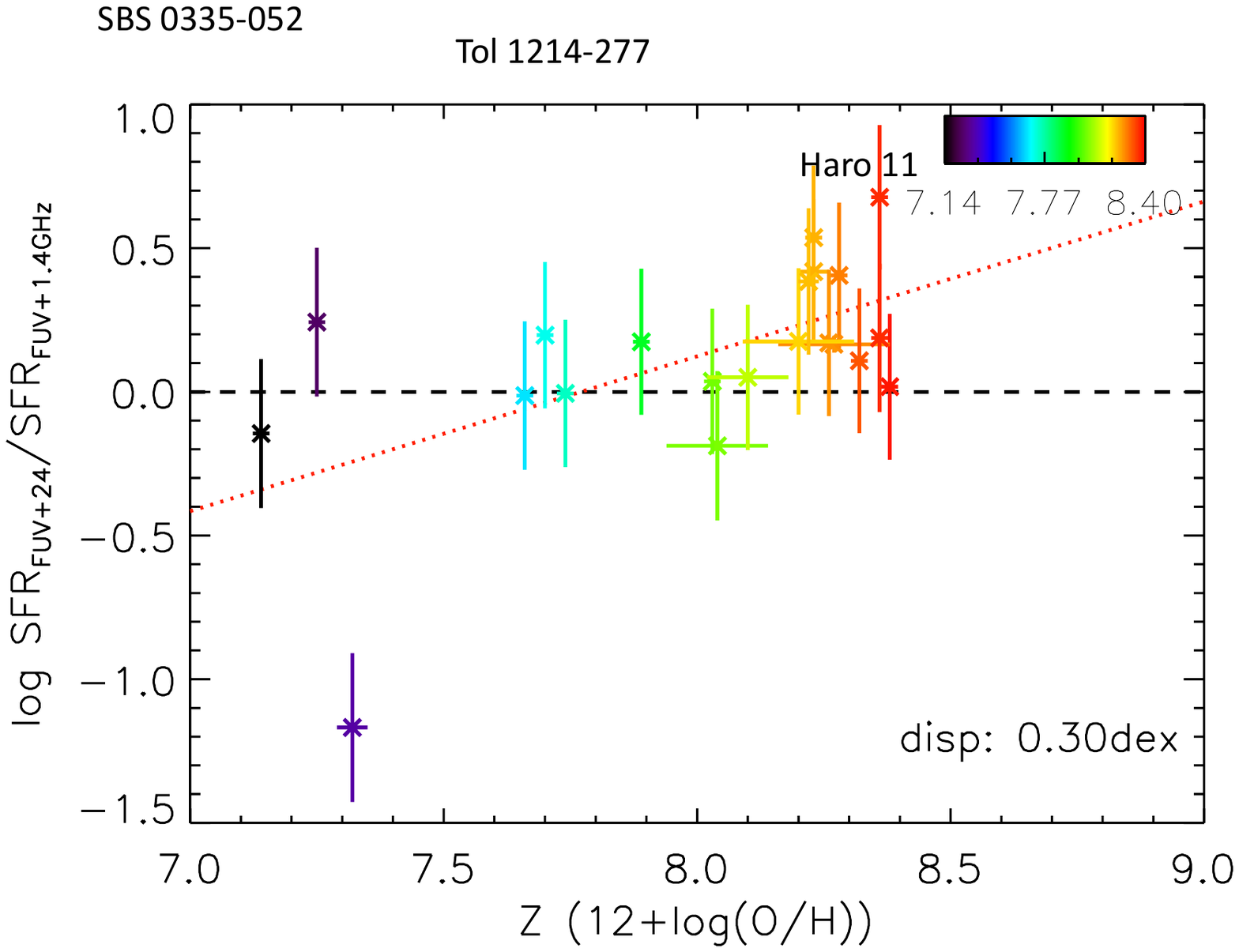}    
   \caption{Comparison between the ratio of the SFR as obtained from the SFR calibrators MIPS\,24\,$\mu$m and PACS\,70\,$\mu$m (top), $FUV$+MIPS\,24\,$\mu$m and $FUV$+TIR (middle) and $FUV$+MIPS\,24\,$\mu$m and $FUV$+1.4\,GHz (bottom) as a function of oxygen abundance. Galaxies are color-coded according to metallicity with increasing oxygen abundances going from black over blue, green and yellow to red colors. The red, dotted and black, dashed line represents the best fit and a perfect one-to-one correlation, respectively. The dispersion around the best fit is indicated in the bottom right corner.}
              \label{plot_mips24vs}
    \end{figure}

Towards longer wavelengths (70, 100, 160\,$\mu$m), the band emission is dominated by larger dust grains and should not depend strongly on the abundance of very small grains. For larger dust grains a significant fraction of the dust heating might, however, be attributed to more evolved stellar populations, making the link between far-infrared continuum emission and star formation more dispersed \citep{2010A&A...518L..65B,2010ApJ...714.1256C,2011AJ....142..111B,2012MNRAS.419.1833B,2012MNRAS.426..892G,2012ApJ...756...40S}. In a similar way, the total-infrared luminosity might be subject to heating from old stars and, therefore, only linked to star formation on much longer timescales.
Some individual studies argue that most of the dust heating is provided by star-forming regions in dwarfs (e.g. \citealt{2010A&A...518L..55G,2012MNRAS.419.1833B}), even at wavelengths long ward of 160 $\mu$m, suggesting that the longer wavelength data (70, 100, 160\,$\mu$m) could potentially be reliable star formation rate tracers. More detailed analyses of the dominant heating sources for dust in dwarfs might however be required to conclude on the applicability of far-infrared continuum bands to trace the SFR. 
In Figure \ref{plot_mips24vs}, we compare the SFR as estimated from the single continuum bands MIPS\,24\,$\mu$m and PACS\,70\,$\mu$m (see top panel).
The best fitting line and dispersion (0.28 dex) in this plot is mainly dominated by three galaxies (SBS\,0335-052, Tol\,1214-277, Haro\,11) which have the peak of their SED at very short wavelengths and, therefore, the 24\,$\mu$m band will overestimate the SFR while the 70\,$\mu$m emission will underestimate the SFR.
The other galaxies seem to follow the one-to-one correlation better with a dispersion of 0.18 dex around the one-to-one correlation (or difference between the two SFR estimates up to 51$\%$). 

In the central panel of Figure \ref{plot_mips24vs}, we make a similar comparison between the composite SFR tracers $FUV$+24\,$\mu$m and $FUV$+TIR.
The best fitting line (with slope $\alpha$ = 0.44) clearly diverges from the one-to-one correlation with the combination of $FUV$ and MIPS\,24\,$\mu$m providing much higher SFR than inferred from $FUV$+TIR. In particular for galaxies with 12+$\log$(O/H) $\geq$ 8.1, the SFR estimates can differ up to one order of magnitude.
We argue that this discrepancy is mainly caused by the fact that the SFR relations for the total infrared luminosity were calibrated based on galaxy samples of normal star-forming galaxies with close to solar abundances. They might, therefore, be less appropriate for metal-poor dwarfs, which are typically characterized by steeply rising mid-infrared (MIR) to far-infrared (FIR) slopes and overall SEDs peaking at wavelengths lower than $\sim$ 60\,$\mu$m (e.g. \citealt{1999ApJ...516..783T,2004ApJS..154..211H,2009A&A...508..645G,2011A&A...532A..56G}).

Radio continuum emission at 1.4\,GHz is dominated by non-thermal synchrotron emission associated with the acceleration of electrons in a galaxy's magnetic field, and, therefore, independent of the grain composition in a galaxy's ISM.  
The emission at 1.4\,GHz is often used as a tracer of star formation, since the optically thin radio synchrotron emission correlates well with the FIR emission (i.e. the FIR-radio correlation, e.g. \citealt{1985A&A...147L...6D,1985ApJ...298L...7H}). However, the 1.4 GHz is not really a tracer of obscured star formation since it probes the energy from supernovae associated with star forming regions and, therefore, rather correlates with the total energy output from star-forming regions, including both obscured and unobscured star formation. 
Figure \ref{plot_mips24vs} (bottom panel) shows the ratio of SFR estimated from $FUV$+24\,$\mu$m and $FUV$+1.4\,GHz as a function of oxygen abundance (or metallicity), with the black dashed line representing the one-to-one correlation.
In general, the SFR obtained from the composite tracer $FUV$+MIPS\,24\,$\mu$m is higher than the SFR estimated from $FUV$+1.4\,GHz.
We argue that this deviation can again be attributed to the bursty star formation history in dwarf galaxies.
Although the emission at 24\,$\mu$m and 1.4\,GHz is sensitive to stars of ages up to at least 100 Myr (see Table \ref{SFRref}), the MIPS\,24\,$\mu$m band is dominated by the emission of stars younger than 10 Myr. In view of the recent trigger of star formation that characterizes most dwarf galaxies in our sample, the 1.4\,GHz emission will typically underestimate the SFR since it was calibrated for models of constant SFR (or supernova rate) over the past 100 Myr. 
One exception is galaxy HS 0822+3542, for which the SFR seems to be underestimated based on $FUV$+MIPS 24\,$\mu$m data. 
The strange behavior of this galaxy in the PACS wavebands \citep{2013A&A...557A..95R} supports the peculiarity of this object. 
Another outlier is Haro\,11, which shows the highest ratio based on the composite tracer $FUV$+24\,$\mu$m. We believe that the warmer dust temperatures with a peak in its SED at very short wavelengths ($\sim$ 40\,$\mu$m, \citealt{2009A&A...508..645G}) in this Luminous InfraRed Galaxy (LIRG) causes an overestimation of the SFR based on MIPS\,24\,$\mu$m. 

Based on the above comparison between different tracers of obscured star formation, we opted to use the reference SFR tracers $FUV$ and MIPS\,24\,$\mu$m for the analysis of the DGS sample. By estimating the SFR from $FUV$ and MIPS\,24\,$\mu$m emission, we should be tracing the emission of young stars with ages up to 100 Myr. However, it is possible to get diffuse emission that is not locally heated by star forming regions in both $FUV$ and MIPS\,24\,$\mu$m bands originating from heating by the diffuse interstellar radiation field. This could potentially cause an overestimation of the SFR in diffuse regions and might bias the interpretation of the observed relations between the SFR and FIR line emission, in particular for the spatially resolved analysis of Section \ref{Res.sec}. With the Dwarf Galaxy Survey often not mapping the nearby galaxies completely in the FIR lines but rather focusing on the brightest star forming regions, we argue that the contribution from diffuse regions and the heating by evolved stellar populations most likely will be limited in many cases.
Although, we should keep in mind the possible contribution of diffuse $FUV$ and 24\,$\mu$m emission to the SFR estimates upon analyzing the SFR relations.  
Similarly, the $FUV$ band might contain residual starlight from evolved stellar populations due to the low level of obscuration in some of the dwarfs. 
However, the presence of only few evolved stars in dwarfs (compared to more massive early-type galaxies with large spheroids) makes it unlikely that residual starlight of old stars will contribute significantly to the $FUV$ emission.

\subsection{Comparison ISO-\textit{Herschel}}    
\label{Compare.sec}

Since the literature sample is composed of \textit{Herschel} and ISO fluxes, we need to verify whether the spectroscopic flux calibration of the \textit{Herschel} PACS and ISO LWS instruments are compatible. Hereto, we assemble the ISO fluxes from the \citet{2008ApJS..178..280B} catalog for galaxies in our literature sample with \textit{Herschel} measurements. We only consider galaxies which are classified as unresolved with respect to the LWS beam in  \citet{2008ApJS..178..280B} to minimize the influence of different beam sizes in our comparison. We gather a sample of 44, 19 and 10 sources with [C{\sc{ii}}], [O{\sc{i}}]$_{63}$ and [O{\sc{iii}}]$_{88}$ fluxes from both \textit{Herschel} and ISO observations. Those galaxies are indicated with a cross behind their name in Tables \ref{table2amin}, \ref{table2a} and  \ref{table2b} of the Appendix. Figure \ref{compare_HerschelvsISO} presents the \textit{Herschel}-to-ISO line ratios for [C{\sc{ii}}] (top), [O{\sc{i}}]$_{63}$ (middle) and [O{\sc{iii}}]$_{88}$ (bottom). Based on the 44 measurements for [C{\sc{ii}}], we find that the \textit{Herschel} fluxes are on average 1.19$\pm$0.43 higher compared to the ISO measurements. The best fitting line (see red solid line in Figure \ref{compare_HerschelvsISO}, top panel) lies close to the one-to-one correlation. Overall, the \textit{Herschel} and ISO [C{\sc{ii}}] fluxes are consistent with each other within the error bars, in particular for the higher luminosity sources. Based on the 19 [O{\sc{i}}]$_{63}$ and 10 [O{\sc{iii}}]$_{88}$ measurements, we infer that the \textit{Herschel} measurements are on average lower by a factor of 0.86$\pm$0.16 and 0.69$\pm$0.18, respectively, compared to the ISO fluxes. For the [O{\sc{i}}]$_{63}$ line, the \textit{Herschel} PACS and ISO LWS measurements still agree within the error bars, but for the [O{\sc{iii}}]$_{88}$ line the discrepancy between \textit{Herschel} and ISO measurements is considered significant. Although the larger ISO LWS beam ($\sim$ 80$\arcsec$) could collect more flux compared to the PACS beam for the [O{\sc{i}}]$_{63}$ and [O{\sc{iii}}]$_{88}$ lines (FWHM $\sim$ 9.5$\arcsec$), we would expect to see a similar behavior for the [C{\sc{ii}}] line (PACS FWHM $\sim$ 11.5$\arcsec$) if the difference in beam size is driving the offset between \textit{Herschel} and ISO fluxes. The absence of any beam size effects for [C{\sc{ii}}] makes us conclude that the discrepancy between \textit{Herschel} and ISO for [O{\sc{iii}}]$_{88}$ can likely be attributed to a difference in calibration. Given the significant difference in [O{\sc{iii}}]$_{88}$ line fluxes, we will only take the [O{\sc{iii}}]$_{88}$ measurements derived from \textit{Herschel} observations into account for the SFR calibrations.

        \begin{figure}
   \centering
    \includegraphics[width=8.45cm]{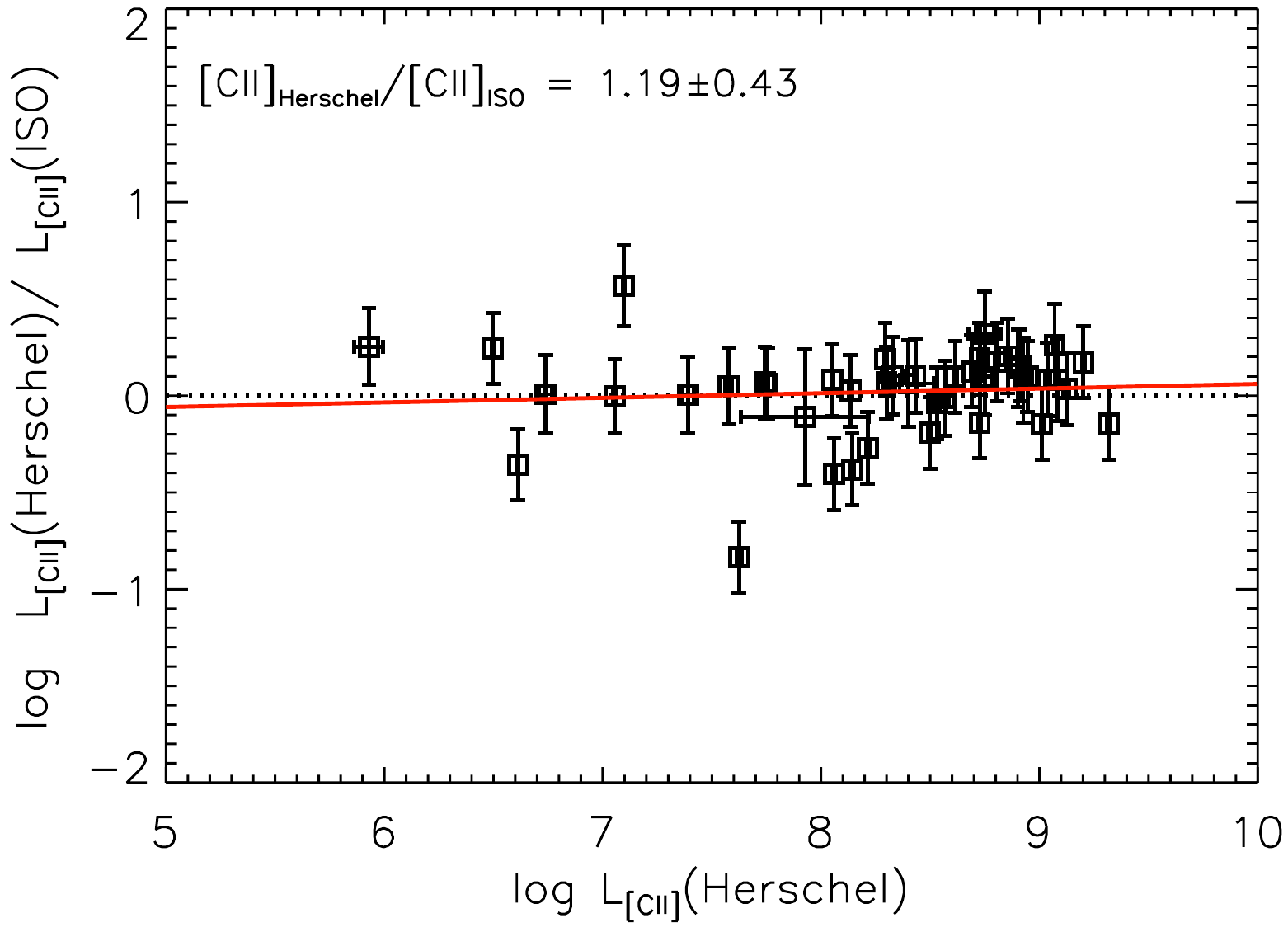}   \\
         \includegraphics[width=8.45cm]{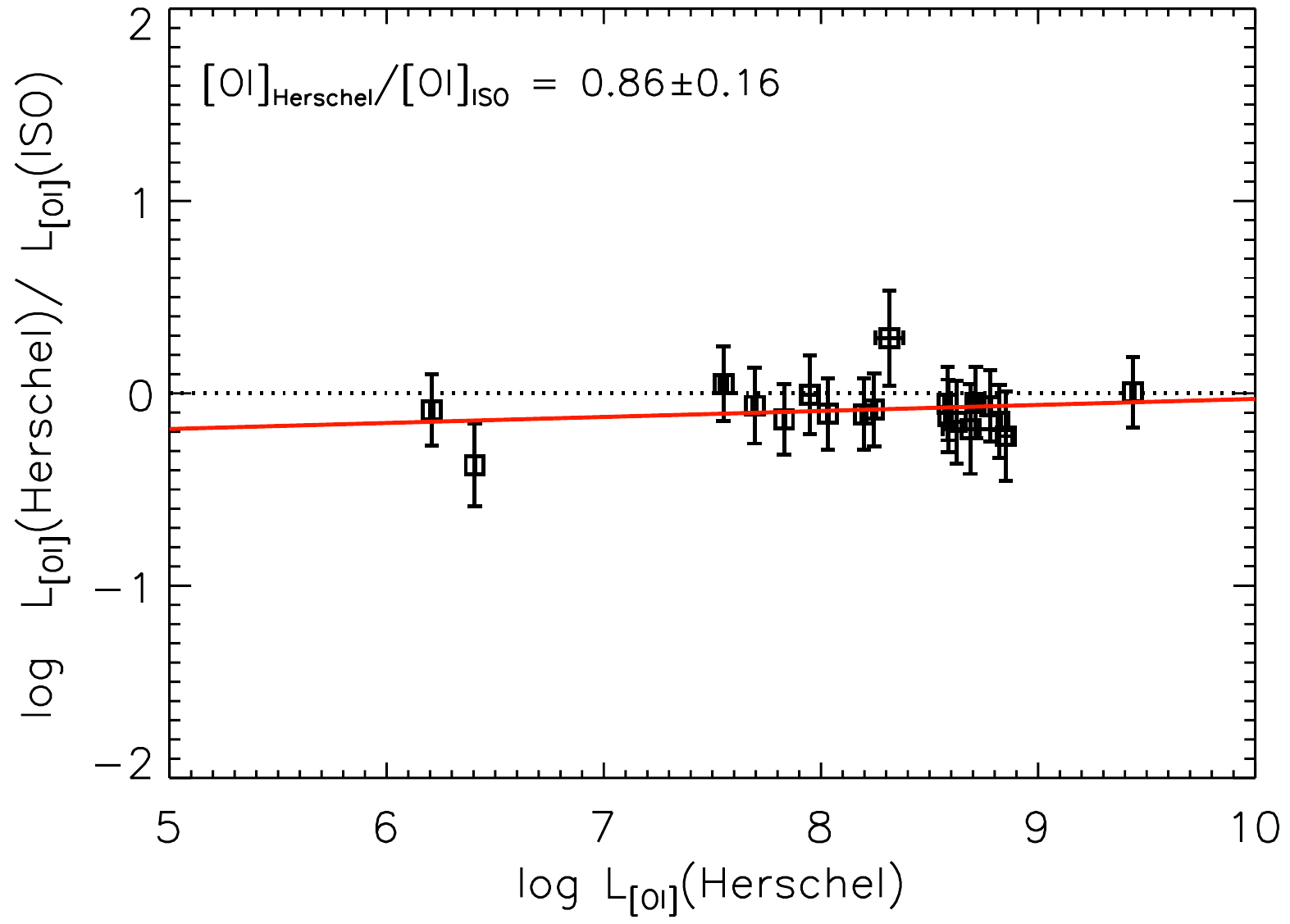}    \\
    \includegraphics[width=8.45cm]{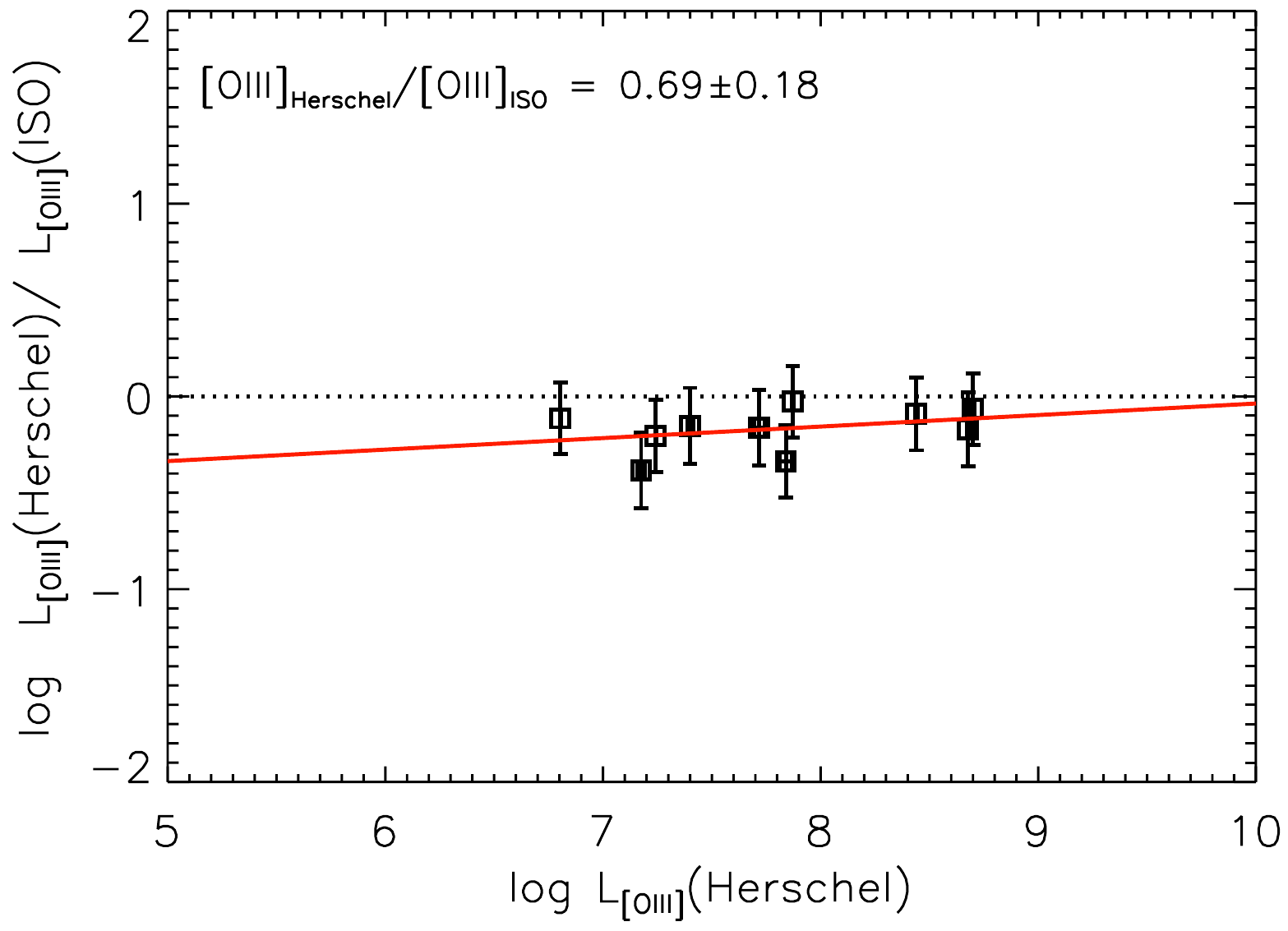}      
   \caption{Comparison between the \textit{Herschel} and ISO line fluxes, i.e. the ratio of the \textit{Herschel} and ISO measurements for [C{\sc{ii}}] (top), [O{\sc{i}}]$_{63}$ (middle) and [O{\sc{iii}}]$_{88}$ (bottom) as a function of \textit{Herschel} line luminosities. The mean and standard deviation of the \textit{Herschel}-to-ISO flux ratio is indicated in the top left corner of each panel. The best fitting line and a perfect one-to-one correlation are indicated as red, solid and black, dotted lines, respectively. } 
                 \label{compare_HerschelvsISO}%
    \end{figure}

\newpage

\section{Tables}
\label{tables}
\bigskip
\begin{table*}[!h]
\caption{Overview of DGS sources used in the SFR calibrations presented in this paper. Different galaxies are ordered in increasing order of oxygen abundance. The first, second and third column specify the name of the source, distance and oxygen abundance, respectively. Galaxies from the complete galaxy sample (i.e. with FIR line detections for [C{\sc{ii}}], [O{\sc{i}}]$_{63}$ and [O{\sc{iii}}]$_{88}$) are indicated with an asterisk behind their name. GALEX $FUV$ and MIPS\,24\,$\mu$m photometry results are presented in columns 4 and 5.
The last column gives the SFR, as estimated from the GALEX $FUV$ and MIPS\,24\,$\mu$m measurements and the calibrations presented in \citet{2011ApJ...741..124H} and \citet{2009ApJ...703.1672K}.}
\label{table2}
\centering
\begin{tabular}{|lccccc|}
\hline
Source & D\tablefootmark{a} & 12+$\log$(O/H)\tablefootmark{a} & $FUV$ & MIPS\,24\,$\mu$m\tablefootmark{b} & SFR\\
 & [Mpc] & & [mJy] & [mJy] & [M$_{\odot}$ yr$^{-1}$]\\
\hline\hline
I\,Zw\,18* & 18.2 & 7.14  & 1.482 $\pm$ 0.220 & 6.1 $\pm$ 0.3* & 0.057 $\pm$ 0.021 \\
SBS\,0335-052* & 56.0 & 7.25  & 0.805 $\pm$ 0.119 & 76.8 $\pm$ 3.2* & 0.890 $\pm$ 0.313 \\
HS\,0822+3542 & 11.0 & 7.32 & 0.166 $\pm$ 0.025 & 3.2 $\pm$ 0.3* & 0.0031 $\pm$ 0.0011 \\
UGC\,4483 & 3.2 & 7.46 & 1.011 $\pm$ 0.150  & 10.1 $\pm$ 0.5* & 0.0014 $\pm$ 0.0005 \\
 &  &  & [1.762 $\pm$ 0.261] &  &  \\
Tol\,1214-277* & 120.5 & 7.52 & 0.188 $\pm$ 0.028 & 6.8 $\pm$ 0.3* &  0.543 $\pm$ 0.194 \\
SBS\,1415+437* & 13.6 & 7.55 & 1.124 $\pm$ 0.167 & 16.8 $\pm$ 1.7 & 0.030 $\pm$ 0.011 \\
SBS\,1211+540 & 19.3 & 7.58 & 0.198 $\pm$ 0.029 & 3.3 $\pm$ 0.3* & 0.011 $\pm$ 0.004  \\
HS\,1442+4250 & 14.4 & 7.60 &  0.590 $\pm$ 0.087  & 6.6 $\pm$ 0.3* & 0.016 $\pm$ 0.006 \\
 &  &  & [1.145 $\pm$ 0.170]  &  &  \\
VII\,Zw\,403* & 4.5 & 7.66 & 2.485 $\pm$ 0.368 & 24.9 $\pm$ 2.4 & 0.007 $\pm$ 0.002 \\
SBS\,1249+493 & 110.8 & 7.68 & 0.088 $\pm$ 0.013 & 4.3 $\pm$ 0.3* & 0.251 $\pm$ 0.089 \\
NGC\,2366* & 3.2 & 7.70 &  8.627 $\pm$ 1.278 & 483.7 $\pm$ 13.3 & 0.022 $\pm$ 0.008 \\
UM\,461* & 13.2 & 7.73 &  0.483 $\pm$ 0.072 & 34.4 $\pm$ 3.2* & 0.024 $\pm$ 0.009  \\
Mrk\,209* & 5.8 & 7.74 &  1.920 $\pm$ 0.284 & 40.1 $\pm$ 2.3 & 0.010 $\pm$ 0.004  \\
UM\,133* & 22.7 & 7.82 &  0.560 $\pm$ 0.083  & 9.4 $\pm$ 0.5* &  0.043$\pm$ 0.016 \\
 &  &  & [0.978 $\pm$ 0.145] &  &  \\
NGC\,4861* & 7.5 & 7.89 &  7.911 $\pm$ 1.172 & 262.7 $\pm$ 10.0 & 0.085 $\pm$ 0.030 \\
NGC\,6822-Hubble V* & 0.5 & 7.96  & 7.619 $\pm$ 1.129 & 970.6 $\pm$ 18.1 & 0.0008 $\pm$ 0.0003 \\
NGC\,1569*$^{\dagger}$ & 3.1 & 8.02 &  179.284 $\pm$ 26.562 & 6709.9 $\pm$ 154.4 &  0.348 $\pm$ 0.124 \\
Mrk\,930* & 77.8 & 8.03 & 1.100 $\pm$ 0.163 & 152.4 $\pm$ 5.6 & 3.095 $\pm$ 1.085 \\
HS\,0052+2536* & 191.0 & 8.04 & 0.262 $\pm$ 0.039  & 20.7 $\pm$ 1.2* & 2.963 $\pm$ 1.045 \\
Mrk\,1089* & 56.6 & 8.10 & 3.509 $\pm$ 0.520 & 391.4 $\pm$ 7.6 & 4.436 $\pm$ 1.558 \\
NGC\,4449* & 4.2 & 8.20 &  73.385 $\pm$ 8.379 & 1967.1 $\pm$ 19.7 & 0.226 $\pm$ 0.080 \\
NGC\,625*$^{\dagger}$ & 3.9 & 8.22&  6.253 $\pm$ 0.926 & 717.1 $\pm$ 16.6 & 0.038 $\pm$ 0.013 \\
II\,Zw\,40*$^{\dagger}$ & 12.1 & 8.23  & 20.374 $\pm$ 3.018 & 1361.3 $\pm$ 54.4 & 0.831 $\pm$ 0.294 \\
Haro\,2* & 21.7 & 8.23 & 2.888 $\pm$ 0.428 & 663.3 $\pm$ 12.8 & 0.954 $\pm$ 0.334 \\
NGC\,4214* &  2.9 & 8.26 &  34.892 $\pm$ 5.169 & 1495.7 $\pm$ 29.1 & 0.063 $\pm$ 0.023 \\
NGC\,1705* & 5.1 & 8.27& 12.768 $\pm$ 1.892 & 43.9 $\pm$ 0.9 & 0.038 $\pm$ 0.014   \\
Haro\,3* & 19.3 &  8.28 & 3.794 $\pm$ 0.562 & 714.7 $\pm$ 25.3 & 0.840 $\pm$ 0.294 \\
UM\,448* & 87.8 & 8.32& 1.890 $\pm$ 0.280 & 522.5 $\pm$ 18.2 & 11.994 $\pm$ 4.193 \\
UM\,311* & 23.5 & 8.36&  1.422 $\pm$ 0.211 & 145.2 $\pm$ 6.5 &  0.291 $\pm$ 0.102 \\
Haro\,11*$^{\dagger}$ & 92.1 & 8.36 &  2.398 $\pm$ 0.355 & 1854.4 $\pm$ 96.3 & 43.010 $\pm$ 15.022  \\
NGC\,1140* & 20.0 & 8.38 &  5.923 $\pm$ 0.877 & 271.0 $\pm$ 9.1 & 0.530 $\pm$ 0.189 \\
HS\,2352+2733 & 116.7 & 8.40 & 0.044 $\pm$ 0.007 & 2.6 $\pm$ 0.3* & 0.155 $\pm$ 0.055 \\
\hline
\end{tabular}
\tablefoot{
\tablefoottext{a}{\footnotesize Distances and oxygen abundances are adopted from \citet{2013PASP..125..600M}.\\}
\tablefoottext{b}{\footnotesize MIPS\,24\,$\mu$m flux densities with an asterisk correspond to total galaxy fluxes and were retrieved from \citet{2012MNRAS.423..197B}.}}
\end{table*}

\begin{table*}
\caption{Overview of literature data for high-redshift galaxies used for the SFR calibrations presented in this paper. The first, second and third column specify the name of the source, redshift and reference to the literature work, respectively. The subsequent columns provide the [C{\sc{ii}}], [O{\sc{i}}]$_{63}$, [O{\sc{iii}}]$_{88}$, total infrared (8-1000\,$\mu$m) luminosities as well as star formation rate, respectively. Upper limits for undetected FIR lines at high-redshift correspond to 3$\sigma$ upper limits.}
\label{table3}
\centering
\begin{tabular}{|lccccccc|}
\hline
Source & z & Ref\tablefootmark{a} & [C{\sc{ii}}] & [O{\sc{i}}]$_{63}$ & [O{\sc{iii}}]$_{88}$  & $\log L_{\text{TIR}}$ & $\log SFR$\\
 & & & [$\log$ L$_{\odot}$] & [$\log$ L$_{\odot}$]  & [$\log$ L$_{\odot}$]  &  [$\log$ L$_{\odot}$]  & [$\log$ M$_{\odot}$ yr$^{-1}$] \\
\hline\hline
IRAS F16413+3954 & 0.59 & 1 & - & 9.52\tablefootmark{b} & - & 13.47 &  3.65 \\
IRAS F04207$-$0127 & 0.91 & 1 & - & $\leq$ 10.61 & - & 13.67 & 3.84 \\
LESS J033217$.$6$-$275230 & 1.10 & 2 & - & $\leq$ 9.87 & - & 11.67 & 1.84 \\
IRAS F10026+4949 & 1.12 & 3 & 10.44 & - & - & 14.02 & 4.20 \\
3C 368 & 1.13 & 3 & 10.00 & - & - & 13.14 & 3.31 \\
PG 1206+459 & 1.16 & 3 & 9.94 & - & - &  13.74 & 3.92 \\
SMM J22471$-$0206 & 1.16 & 3 & 10.28 & - & - & 13.56 & 3.74 \\
3C 065 & 1.18 & 3 & $\leq$ 9.22 & - & - & 13.09 & 3.26 \\
SMM J123634$.$51+621241.0 & 1.22 & 3 & 10.20 & - & - & 12.77 & 2.94 \\
LESS J0333294$.$9$-$273441 & 1.24 & 2 & - & $\leq$ 9.99 & - & 11.76  &  1.94 \\
PG 1241+176 & 1.27 & 3 & 10.60 & - & - & 13.80 & 3.97\\
LESS J033155$.$2$-$275345 & 1.27 & 2 & - & 9.69 & - & 11.91 &  2.08 \\
MIPS J142824$.$0+352619\tablefootmark{c}  & 1.30 & 3, 4, 5 & 9.85 & 9.53 & - & 12.29 & 2.47 \\ 
LESS J033331$.$7$-$275406 & 1.32 & 2 & - & 9.71 & - & 12.33 & 2.50 \\
IRAS F16348+7037 & 1.33 & 1 & - & $\leq$ 10.64 & - & 13.86 & 4.04 \\
IRAS F22231-0512 & 1.40 & 1 & - & $\leq$ 10.82 & - & 14.66 & 4.83 \\
3C 446 & 1.40 & 3 & 11.13 & - & - & 14.66 & 4.83 \\
LESS J033150.84$-$274438 & 1.61 & 2 & - & $\leq$ 10.43 & - & 12.68 & 2.86 \\
LESS J033140.1$-$275631 & 1.62 & 2 & - & 10.27 & - & 12.29 & 2.47 \\
PKS 0215+015 & 1.72 & 3 & 10.70 & - & - & 14.27 & 4.44 \\
H-ATLAS J091043.1-000322\tablefootmark{d} & 1.79 & 6 &  9.33 & 9.40 & - & 12.33 & 2.75 \\
RX J094144.51+385434.8 & 1.82 & 3 & 10.38  & - & - & 13.69 & 3.86 \\
SDSS J100038.01+020822.4 & 1.83 & 3 & 10.06 & - & - & 13.13 & 3.30 \\
SWIRE J104704$.$97+592332.3 & 1.95 & 3 & 10.31 & - & - & 13.13 & 3.30 \\
SWIRE J104738$.$32+591010.0 & 1.96 & 3 & 10.09 & - & - & 12.73 & 2.91 \\
SMM J2135-0102\tablefootmark{e} & 2.33 & 7 & 9.75 & $\leq$ 10.18 & $\leq$ 9.91 & 12.37 & 2.55 \\
H-ATLAS ID130\tablefootmark{f} & 2.63 & 8 & - & $\leq$ 10.31 & $\leq$ 10.22 & 12.97 & 3.14 \\
SMM J02399$-$0136\tablefootmark{g} & 2.81 & 9 & - & - & 10.66 & 13.34 & 3.51 \\
H-ATLAS ID81\tablefootmark{h} & 3.04 & 8 & 10.11  & - & 9.76 & 12.36 & 2.53 \\
APM 08279+5255\tablefootmark{i} & 3.91 & 9 & - & - & 9.07 & 12.59 & 2.77 \\
H-ATLAS ID141\tablefootmark{j} & 4.24 & 10 & 9.49 & - & - & 12.79 & 2.87 \\
BRI\,1335-0417 & 4.41 & 11 & 10.22 & - & - & 13.74 & 3.91 \\
ALESS 65.1 & 4.42 & 12 & 9.52 & - & - & 12.32 & 2.49 \\
BRI\,0952-0115\tablefootmark{k} & 4.43 & 13 & 9.66 & - & - & 12.67 &  2.85 \\
ALESS 61.1 & 4.44 & 12 & 9.19 & - & - & 12.34 & 2.51 \\
BR 1202$-$0725 N & 4.69 & 14, 15 & 10.00 & - & - & 13.15 & 3.33 \\
BR 1202$-$0725 S & 4.69 & 14, 15 & 9.82 & - & - & 13.48 & 3.66 \\
LESS J033229.4 & 4.76 & 16 & 10.01 & - & - & 12.78 & 2.96 \\
CI 1358+62\tablefootmark{l} & 4.93 & 17 & $\leq$ 9.98 & - & - & 12.38 & 2.56 \\
HLSJ091828.6+514223\tablefootmark{m}  & 5.24 & 18 & 10.19 & - & - & 13.27 & 3.44 \\
HFLS3 & 6.34 & 19 & 10.19 & - & - & 13.70 & 3.87 \\
SDSS J1148+52511 & 6.42 & 20 & 9.64 & - & - & 13.41 & 3.58 \\
Himiko & 6.60 & 21 & $\leq$ 7.75 & - & - & $\leq$ 10.80 & 2.00\tablefootmark{n} \\
\hline
\end{tabular}
\tablefoot{
\tablefoottext{a}{\footnotesize References: (1) \citet{2004ApJ...604..565D}; (2) \citet{2012MNRAS.427..520C}; (3) \citet{2010ApJ...724..957S}; (4) \citet{2010ApJ...714L.162H}; (5) \citet{2010A&A...518L..36S}; (6) \citet{2014ApJ...780..142F};
(7) \citet{2010A&A...518L..35I}; (8) \citet{2011MNRAS.415.3473V};  (9) \citet{2010ApJ...714L.147F}; (10) \citet{2011ApJ...740...63C}; (11) \citet{2010A&A...519L...1W}; (12) \citet{2012MNRAS.427.1066S}; (13) \citet{2009A&A...500L...1M}; (14) \citet{2006ApJ...645L..97I}; (15) \citet{2012ApJ...752L..30W}; (16) \citet{2011A&A...530L...8D}; (17) \citet{2005MNRAS.359...43M}; (18) \citet{2013arXiv1310.4090R}; (19) \citet{2013Natur.496..329R}; (20) \citet{2005A&A...440L..51M}; (21) \citet{2013ApJ...778..102O}.}
\tablefoottext{b}{\footnotesize Only a 2.7$\sigma$ detection.} 
\tablefoottext{c}{\footnotesize Luminosities have not been corrected for a possible lensing factor, but the galaxy could possibly be lensed up to a factor of 8 \citep{2010ApJ...724..957S}.} 
\tablefoottext{d}{\footnotesize Luminosities have been corrected assuming a lensing factor of 18, following \citet{2014ApJ...780..142F}.} 
\tablefoottext{e}{\footnotesize We apply the luminosities corrected for lensing reported by \citet{2010A&A...518L..35I}.} 
\tablefoottext{f}{\footnotesize Luminosities have been corrected assuming a lensing factor of 6, following the best fit for the lens model reported in \citet{2011MNRAS.415.3473V}.} 
\tablefoottext{g}{\footnotesize Luminosities have been corrected assuming a lensing factor of 2.38, following \citet{2010MNRAS.404..198I}.} 
\tablefoottext{h}{\footnotesize Luminosities have been corrected assuming a lensing factor of 25, following the improved lens model presented in \citet{2011MNRAS.415.3473V}.} 
\tablefoottext{i}{\footnotesize Luminosities have been corrected assuming a lensing factor of 90, following the most extreme estimate of the lensing factor ($\mu$ $\sim$ 4-90) reported in \citet{2009ApJ...690..463R}.} 
\tablefoottext{j}{\footnotesize Luminosities have been corrected assuming a lensing factor of 20, which is the median of the range in amplification factors reported in \citet{2011ApJ...740...63C}} 
\tablefoottext{k}{\footnotesize Luminosities have been corrected assuming a lensing factor of 4.5, following \citet{2009A&A...500L...1M}.} 
\tablefoottext{l}{\footnotesize Luminosities have been corrected assuming a lensing factor of 11, following the most extreme lensing predictions of \citet{1997ApJ...486L..75F}.} 
\tablefoottext{m}{\footnotesize Luminosities have been corrected assuming a lensing factor of 8.9, following \citet{2013arXiv1310.4090R}.} 
\tablefoottext{n}{\footnotesize The SFR is not calculated from $L_{\text{TIR}}$ since Himiko is an optically bright galaxy with low metal and dust content. Therefore, we adapt the SFR estimate ($\sim$ 100 M$_{\odot}$ yr$^{-1}$) obtained from the stellar population synthesis in \citet{2013ApJ...778..102O}.}} 
\end{table*}

\begin{table*}
\caption{Overview of literature data for dwarf galaxies used for the SFR calibrations presented in this paper. The dwarf galaxies from the Dwarf Galaxy Survey are also part of the literature sample but they are not repeated here as they correspond to the sources in Table \ref{table2}. The first, second and third column specify the name of the source, luminosity distance and reference to the literature work, respectively. The subsequent columns provide the GALEX $FUV$ and MIPS\,24\,$\mu$m or IRAS\,25\,$\mu$m flux densities and star formation rates (estimated from the GALEX $FUV$ and MIPS\,24\,$\mu$m/IRAS\,25\,$\mu$m measurements and the calibrations presented in \citet{2011ApJ...741..124H} and \citet{2009ApJ...703.1672K}), respectively.}
\label{table2amin}
\centering
\begin{tabular}{|lccccc|}
\hline
Source & $D_{\text{L}}$\tablefootmark{a}   & Ref\tablefootmark{b} & $FUV$ & MIPS\,24\,$\mu$m/IRAS\,25\,$\mu$m & $\log SFR$ \\
& [Mpc] & &  [mJy] & [mJy] & [$\log$ M$_{\odot}$ yr$^{-1}$] \\
\hline\hline
NGC\,4713 & 19.9  & 1 & 13.308 $\pm$ 1.972 & 186.5  $\pm$ 50.4 & -0.13 \\
NGC\,4490 &	8.7	& 1 &  10.248 $\pm$	1.518 & 1394.8 $\pm$ 55.9 &	-0.45 \\ 
NGC\,1156 &	7.1 & 1 &	23.307 $\pm$ 3.453 & 447.1 $\pm$ 18.0 & -0.74 \\
NGC\,4299 &	18.7	& 1 & 8.996 $\pm$ 1.333 & 210.4 $\pm$ 8.5 &	 -0.28 \\ 
NGC\,693 & 23.4 & 1 & 0.973 $\pm$ 0.144 & 550.0 $\pm$ 50.0 & -0.08 \\ 
NGC\,814 & 24.1 & 1 & 0.321 $\pm$ 0.048 & 897.2 $\pm$ 62.8 & 0.14 \\
NGC\,4294 & 17 & 1 & 5.249 $\pm$ 0.778 & 192.5 $\pm$ 57.8 & -0.52 \\
VCC\,1326 & 17 & 1 & 0.290 $\pm$ 0.043 & 418.4 $\pm$ 54.4 & -0.49 \\ 
NGC\,4522 & 17 & 1 & 2.766 $\pm$ 0.410 & 211.2 $\pm$ 48.6 & -0.62 \\ 
NGC\,6821 & 25.1 & 1 & 6.920 $\pm$ 1.025 & 307.4 $\pm$ 46.1 & -0.02 \\ 
\hline
\end{tabular}
\tablefoot{
\tablefoottext{a}{Distances for galaxies with redshifts $z<0.01$ are obtained from the Nearby Galaxies Catalog \citep{1988ngc..book.....T} (if available), while for other galaxies distances are derived from the redshifts reported on the NASA Extragalactic Database (NED). For Virgo cluster galaxies, we retrieve distances from the Goldmine database \citep{2003A&A...400..451G}, based on results reported in \citet{1999MNRAS.304..595G}. For galaxies with redshifts $z\geq0.01$, the luminosity distances are calculated for a spatially flat cosmology with $H_{\text{0}}$ = 67.3 km s$^{-1}$ Mpc$^{-1}$, $\Omega_{\lambda}$ = 0.685 and $\Omega_{\text{m}}$ = 0.315 \citep{2013arXiv1303.5076P}.}\\
\tablefoottext{b}{References: (1) \citet{2008ApJS..178..280B}}}
\end{table*}

\begin{center}
\onecolumn
\tablefirsthead{
\hline
Source & $D_{\text{L}}$\tablefootmark{a}   & Ref\tablefootmark{b} & $\log L_{\text{TIR}}$ & $\log SFR$ \\
& [Mpc] & & [$\log$ L$_{\odot}$] & [$\log$ M$_{\odot}$ yr$^{-1}$] \\
\hline}
\tablehead{
 \multicolumn{5}{r}{\textit{Continued from previous page}} \\
\hline
Source & $D_{\text{L}}$\tablefootmark{a}   & Ref\tablefootmark{b} & $\log L_{\text{TIR}}$ & $\log SFR$ \\
& [Mpc] & & [$\log$ L$_{\odot}$] & [$\log$ M$_{\odot}$ yr$^{-1}$] \\
\hline}
\tabletail{
\hline
\multicolumn{5}{l}{Continued on next page}\\}
\tablelasttail{\hline}
\topcaption{Overview of literature data for galaxies with H{\sc{ii}} or starburst source classification used for the SFR calibrations presented in this paper. The first, second and third column specify the name of the source, luminosity distance and reference to the literature work, respectively. The subsequent columns provide the total infrared (8-1000\,$\mu$m) luminosity and star formation rate, respectively.}\label{table2a}
\begin{supertabular}{|lcccc|}
NGC\,5713		&	33.9	& 1 &	10.91	&	1.09 \\
NGC\,7714		&	41.1	& 1 &	10.75	&	0.92  \\
Cartwheel	&	137.5	& 1 &	10.73	&	0.91 \\
NGC\,0986		&	25.9	& 1 &	10.76	&	0.94 \\	 
NGC\,1317		&	18.8	& 1 &	9.69	&	-0.14 \\
NGC\,4041		&	25.3	& 1 &	10.49	&	0.67 \\
NGC\,4189		&	18.7	& 1 &	9.64	&	-0.18 \\
NGC\,278 & 13.2 & 1 & 10.15 & 0.33 \\
NGC\,1022 & 21.7 & 1 & 10.40 & 0.57 \\
NGC\,1155 & 69.8  & 1 & 10.69 & 0.86 \\
NGC\,1222 & 34.5 & 1 & 10.66 & 0.83 \\
IC\,1953 & 24.6 & 1 & 10.13 & 0.31 \\
NGC\,1385 & 19.5 & 1 & 10.37 & 0.54 \\
UGC\,02855 & 22.6 & 1 & 10.88 & 1.05 \\
NGC\,1482 & 21.8 & 1 & 10.68 & 0.85 \\
NGC\,1546 & 14.9 & 1 & 9.82 & -0.002 \\
ESO\,317$-$G023 & 43.3 & 1 & 10.85 & 1.03 \\
NGC\,3583 & 37.9 & 1 & 10.60 & 0.77 \\
NGC\,3683 & 31.6 & 1 & 10.69 & 0.86 \\
NGC\,3885 & 31.0 & 1 & 10.51 & 0.69 \\
NGC\,3949 & 18.9 & 1 & 10.16 & 0.34 \\
NGC\,4027 & 28.5 & 1 & 10.51 & 0.69 \\
NGC\,4041 & 25.3 & 1 & 10.49 & 0.67 \\
VCC\,873 & 17 & 1 & 9.82 & -0.002 \\
NGC\,4519 & 17 & 1 & 9.60 & -0.23 \\
VCC\,1972 & 17 & 1 & 9.80 & -0.03 \\
VCC\,1987 & 17 & 1 & 10.15 & 0.33 \\
NGC\,4691 & 25.1 & 1 & 10.46 & 0.63 \\
IC\,3908 & 19.3 & 1 & 10.00 & 0.18 \\
NGC\,4818 & 24.0 & 1 & 10.54 & 0.72 \\
NGC\,5430 & 44.3 & 1 & 10.84 & 1.01 \\
NGC\,5433 & 65.4 & 1 & 10.94 & 1.11 \\
NGC\,5937 & 42.0 & 1 & 10.77 & 0.95 \\
NGC\,5962 & 35.4 & 1 & 10.63 & 0.80 \\
NGC\,7218 & 24.5 & 1 & 10.00 & 0.18 \\
NGC\,7418 & 19.8 & 1 & 9.97 & 0.14 \\
III\,Zw\,107 & 86.4  & 1 & 10.57 & 0.74 \\
M\,51\tablefootmark{c} & 9.9 & 2 & 9.89 & 0.07  \\ 
NGC\,253				&	3.3	& 3 & 10.61	&	0.79 \\
NGC\,1808			&	12.0	& 3 & 10.73	&	0.91	\\  
NGC\,3256$^{\dagger}$			&	41.7	& 3 & 11.82	&	1.99	 \\
NGC\,4945			&	5.8	& 3 & 10.92	&	1.09	 \\
NGC\,7552$^{\dagger}$			&	21.7	& 3 & 11.10	&	1.28	 \\
Arp\,299$^{\dagger}$				&	47.2	& 3 & 11.94	&	2.11	 \\
IRAS\,F17208$-$0014*$^{\dagger}$	&	197.0	& 3 & 12.68	&	2.85	 \\
IRAS\,F20551$-$4250*$^{\dagger}$	&	197.5	& 3 & 12.24	&	2.41	 \\
IRAS\,F23128$-$5919*$^{\dagger}$	&	205.3	& 3 & 12.21	&	2.38	 \\
IRAS\,F10565$+$2448$^{\dagger}$		&	198.2	& 3 & 12.28	& 2.45	\\
IRAS\,F19297$-$0406*$^{\dagger}$	&	405.4	& 3 & 12.61	&	2.79	 \\
IRAS\,F09022$-$3615*	&	277.6	& 3 & 12.47	&	2.65	 \\
M83 center			&	5.2	& 3 & 10.06	&	0.23 \\
NGC\,4038			&	28.3	& 3 & 10.56	&	0.74 \\
IRAS\,15206+3342*$^{\dagger}$\tablefootmark{d}           &   610.0	& 3 & 12.30 & 2.47 \\
IRAS\,F22491$-$1808*	&	366.2	& 3 & 12.42	&	2.59	 \\
M83 bar				&	5.2	& 3 & 9.33 &	-0.50 \\
M83 arm				&	5.2	  & 3 & 8.72	&	-1.11\\
NGC\,4039			&	28.3	& 3 & 10.77	&	0.95 \\
NGC\,4038/4039 Overlap	&	28.3	& 3  & 10.71	&	0.89 \\
MCG$-$02$-$01$-$051   &   123.2 & 4 &     11.50  &    1.68 \\
ESO244$-$G012   &   103.8  & 4 &  11.47  &   1.65 \\
UGC01385  &     84.9  & 4 &   11.11   &   1.29 \\
CGCG052$-$037  &    111.2  & 4 &  11.43   &   1.61 \\
ESO286$-$G035  &   78.5   & 4 &  11.15   &  1.33 \\
MCG$-$01$-$60$-$022    &  105.2  & 4 &   11.23   &   1.41 \\ 
CGCG381$-$051    &  139.9   & 4 &  11.24    &  1.42 \\
NGC\,6907		& 48.1			& 5 &	10.55	  &	  0.73 \\
ESO286$-$G035	& 79.0			& 5 &	10.96	  &	   1.14 \\
ESO343$-$IG013$\_$S	& 86.8			& 5 &	10.59	  &	  0.77 \\
IRAS\,21101$+$5810	& 182.6			& 5 &	11.67	  &	   1.85 \\
CGCG453$-$062	&	115.3		& 5 &	11.32	  &	   1.50 \\
Arp84$\_$N		&  52.6			& 5 &	10.86	  &	  1.04 \\
VV705		&  186.9		& 5 &	11.84	  &	   2.02 \\
MCG$+$12$-$02$-$001$^{\dagger}$	&  71.2			& 5 &	11.34	  &	   1.52 \\
UGC12150	&	    97.8	& 5 &	11.37	  &	   1.55 \\
NGC\,5331$\_$S	&	   153.5	& 5 &	11.51	  &	   1.69 \\
NGC\,5331$\_$N	&	   153.5	& 5 &	10.73	  &	  0.91 \\
UGC08387$^{\dagger}$	&	   107.0	& 5 &	11.80	  &	   1.98 \\
ESO320$-$G030	&	    48.5	& 5 &	11.34	  &	   1.52 \\
Arp240$\_$W	&	  103.8		& 5 &	10.62	  &	  0.80 \\
Arp240$\_$E	&		   103.8	& 5 &	10.86	  &	   1.04 \\
ESO440$-$IG058$\_$S	 & 106.5		& 5 &	11.24	  &	   1.42 \\
ESO264$-$G057	&	    78.1	& 5 &	10.86	  &	   1.04 \\
CGCG043$-$099	&	   175.1	& 5 &	11.64	  &	   1.82 \\
NGC\,1614$^{\dagger}$		&	    72.6	& 5 &	11.50	  &	   1.68 \\
MCG$-$02$-$33$-$098$\_$W	 &   72.1		& 5 &	10.79	  &	  0.97 \\
MCG$-$02$-$33$-$098$\_$E	&	    72.1	& 5 &	10.30	  &	  0.48 \\
NGC\,5104		&	    84.9	& 5 &	11.21	  &	   1.39\\
ESO267$-$G030$\_$W	&	    84.5	& 5 &	10.96	  &	   1.14 \\
ESO267$-$G030$\_$E	&	    84.5	& 5 &	10.90	  &	   1.08 \\
ESO173$-$G015$^{\dagger}$	&	    44.0	& 5 &	11.42	  &	   1.60 \\
NGC\,6621		&	    94.6	& 5 &	11.24	  &	   1.42 \\
NGC\,6090		&	     138.1	& 5 &	11.44	  &	   1.62 \\
CGCG049$-$057$^{\dagger}$	&	    58.9	& 5 &	11.20	  &	   1.38 \\
IRAS\,13052$-$5711	&	   97.3		& 5 &	11.17	  &	   1.35\\ 
ESO255$-$IG007$\_$E	&	   181.7	& 5 &	10.74	  &	  0.92 \\
ESO255$-$IG007$\_$W	&	   181.7	& 5 &	11.75	  &	   1.93 \\
ESO255$-$IG007$\_$S	&	   181.7	& 5 &	10.75	  &	  0.93 \\
IRAS\,F05187$-$1017	&	   130.6	& 5 &	11.40	  &	   1.58 \\
IRAS\,F03217$+$4022	&	   107.5	& 5 &	11.42	  &	   1.60 \\
ESO069$-$IG006	* &	   219.6	& 5 &	12.05	  &	   2.23 \\
NGC\,2342		&	    80.4	& 5 &	10.89	  &	   1.07 \\
NGC\,2388		&	    62.6	& 5 &	11.23	  &	   1.41 \\
IC4734		&	    70.8	& 5 &	11.33	  &	   1.51 \\
UGC11041	&	    74.0	& 5 &	10.82	  &	  1.00 \\
IRAS\,18090$+$0130$\_W$ &	   133.4	& 5 &	10.98	  &	   1.16\\ 
IRAS\,18090$+$0130$\_E$ &	  133.4		& 5 &	11.47	  &	   1.65 \\
VIIZw031*	 &	    249.8	& 5 &	12.03	&	   2.21 \\
NGC6286$\_$N	&	    83.6	& 5 &	10.51	  &	  0.69 \\
NGC6286$\_$S	&	    83.6	& 5 &	11.01	  &	   1.19 \\
CGCG142$-$034$\_$E	&	    85.4	& 5 &	10.84	  &	  1.02 \\
UGC04881$\_$W	&	   185.0	& 5 &	10.86	  &	   1.04 \\
UGC03410$\_$W	&	    59.4	& 5 &	9.84	  &	0.02\\
UGC03410$\_$E	&	    59.4	& 5 &	10.51	  &	  0.69\\ 
ESO557$-$G002$\_S$	&	    97.3	& 5 &	10.40	  &	  0.58 \\
ESO557$-$G002$\_N$	&	    97.3	& 5 &	11.16	  &	   1.34 \\
ESO432$-$IG006$\_W$	&	    73.5	& 5 &	10.70	  &	  0.92 \\
UGC03608	&	    97.8	& 5 &	11.20	  &	   1.38 \\
IRAS\,F18293$-$3413	&	    82.7	& 5 &	11.85	  &	   2.03 \\
VV414$\_$W		&	  115.3		& 5 &	10.93	  &	   1.11 \\
NGC\,3110		&	    76.7	& 5 &	10.95	  &	   1.13 \\
Arp303$\_$S	&	    91.4	& 5 &	10.51	  &	  0.69 \\
Arp303$\_$N	&	    91.4	& 5 &	10.39	  &	  0.57 \\
ESO593$-$IG008	&	  231.1		& 5 &	11.87	  &	   2.05 \\
ESO467$-$G027	&	    79.4	& 5 &	10.47	  &	  0.65 \\
CGCG247$-$020	&	  119.5		& 5 &	11.35	  &	   1.53 \\
ESO077$-$IG014$\_$W	&	   195.4	& 5 &	11.11	  &	   1.29 \\
IC5179		&	    51.7	& 5 &	10.77	  &	  0.95 \\
MCG$+$04$-$48$-$002	&	    63.0	& 5 &	10.88	  &	   1.06 \\
IRAS\,20351$+$2521	&	   156.7	& 5 &	11.45	  &	   1.63 \\
VV250a$\_$W	&	   144.1	& 5 &	10.87	  &	   1.05 \\
VV250a$\_$E	&	   144.1	& 5 &	11.70	  &	   1.88 \\
UGC08739	&	    76.2	& 5 &	10.52	  &	  0.70 \\
NGC\,4194$^{\dagger}$		&	    43.6	& 5 &	10.94	  &	   1.12 \\
NGC\,7592$\_$E	&	   112.1	& 5 &	11.10	  &	   1.28 \\
NGC\,6670$\_$W	&	   132.0	& 5 &	11.21	  &	   1.39 \\
NGC\,6670$\_$E	&	   132.0	& 5 &	11.27	  &	   1.45 \\
NGC\,3221		&	    62.1	& 5 &	9.93	  &	 0.11 \\
MCG$+$07$-$23$-$019	 &  160.5		& 5 &	11.64	  &	   1.82 \\
Arp256		&	   125.5	& 5 &	11.44	  &	   1.62 \\
VV340a$\_$S	&	   156.7	& 5 &	10.59	  &	  0.77 \\
VV340a$\_$N	&	  156.7		& 5 &	11.37	  &	   1.55 \\
NGC\,0023		&	    68.9	& 5 &	10.93	  &	   1.11 \\
NGC\,6701		&	    59.8	& 5 &	10.65	  &	  0.83 \\
MCG$-$03$-$04$-$014	 & 155.8		& 5 &	11.63	  &	   1.81\\ 
NGC\,7679		&	    78.1	& 5 &	10.88	  &	   1.06 \\
UGC01385	&	    85.4	& 5 &	11.03	  &	   1.21 \\
NGC\,6052$^{\dagger}$		&	    71.7	& 5 &	10.52          &	  0.70 \\
ESO244$-$G012	&	    96.4	& 5 &	11.40	  &	   1.58 \\
ESO507$-$G070	&	   99.2		& 5 &	11.59	  &	   1.77 \\
IC4280		&	    74.0	& 5 &	10.65	  &	  0.83 \\
Arp86$\_$S		&	    77.6	& 5 &	10.50	  &	  0.68 \\
NGC0232$\_$W	&	    101.5	& 5 &	11.47	  &	   1.65 \\
NGC7771$\_$N	&	    64.8	& 5 &	11.29	  &	   1.47 \\
Mrk\,331$^{\dagger}$		&	    84.5	& 5 &	11.56	  &	   1.74 \\
NGC\,5936		&	    60.3	& 5 &	10.79	  &	  0.97 \\
NGC\,5653		&	    53.9	& 5 &	10.59	  &	  0.77 \\
ESO099$-$G004	&	   135.3	& 5 &	11.67	  &	   1.85 \\
ESO221$-$IG010	 &   46.7		& 5 &	10.66	  &	  0.84 \\
NGC\,695$^{\dagger}$		&	   150.7	& 6 &	11.49	  &	   1.67 \\
MCG$+$05$-$06$-$036$\_$S	&	   156.7	& 5 &	11.50	  &	   1.67 \\
UGC02369	&	   147.9	&  5 &	11.66	  &	   1.84 \\
UGC02238$^{\dagger}$	&	    100.1	& 5 &	11.26	  &	   1.44 \\
ESO353$-$G020	&	    72.6	& 5 &	11.08	  &	   1.26 \\
NGC\,828		&	    81.7	& 5 &	11.19	  &	   1.37 \\
IRAS\,F16164$-$0746	&	   125.0	& 5 &	11.68	  &	   1.86 \\
NGC0838$\_$E	&	    58.0	& 5 &	10.92	  &	   1.10 \\
NGC0838$\_$S	&	    58.0	& 5 &	11.04	  &	   1.22 \\
UGC01845	&	    70.8	& 5 &	11.10	  &	   1.28 \\
IC0214		&	   139.9	& 5 &	11.16	  &	   1.34 \\
NGC\,992		&	    62.6	& 5 &	10.77	  &	  0.95 \\
CGCG465$-$012$\_$N	&	    101.9	& 5 &	10.43	  &	  0.61 \\ 
CGCG465$-$012$\_$S	&	    101.9	& 5 &	11.06	  &	   1.24 \\
MCG$-$05$-$12$-$006	&	    85.4	& 5 &	11.06	  &	   1.24 \\
CGCG052$-$037	&	   112.5	& 5 &	11.32	  &	   1.50 \\
IRAS\,F16516$-$0948	&	   103.3		& 5 &	10.96	  &	   1.14 \\
UGC02982	&	    80.8	& 5 &	10.95	  &	   1.13 \\
IC4687		&	    79.0	& 5 &	11.21	  &	   1.39 \\
IRAS\,17578$-$0400$\_$N &	    63.5	& 5 &	11.45	  &	   1.63 \\
ESO138$-$G027	&	    95.0	& 5 &	11.26	  &	   1.44\\
IRAS\,05083$+$2441$\_$S &	    105.6	& 5 &	11.16	  &	   1.34\\ 
IRAS\,04271$+$3849	&	    85.9	& 5 &	11.04  &	   1.22 \\
NGC\,1797		&	    67.1	& 5 &	11.01	  &	   1.19\\
IRAS\,F17138$-$1017	&	    79.0	& 5 &	11.35          &	   1.53 \\
CGCG468$-$002$\_$E	&	    83.1	& 5 &	11.17	  &	   1.35 \\
IRAS\,05442$+$1732	&	    84.9	& 5 &	11.21	  &	   1.39 \\
MCG$+$08$-$11$-$002	&	    87.2	& 5 &	11.47	  &	   1.65 \\
IRAS\,05129$+$5128	 &  126.4		& 5 &	11.33	  &	   1.51 \\
UGC03351	&	    67.6	& 5 &	11.22	  &	   1.40 \\
NGC\,2146$^{\dagger}$		&	    19.2	& 5 &	10.83	  &	  1.01 \\
NGC\,317B	&	    82.7	& 5 &	11.32	  &	   1.50 \\
AM0702$-$601$\_$S	  & 145.1		& 5 &	11.37	  &	   1.55\\
RR032$\_$N		&	    79.0	& 5 &	10.71	  &	  0.89 \\
RR032$\_$S		&	    79.0	& 5 &	10.73	  &	  0.91 \\
IRAS\,F11231$+$1456	 & 158.1		& 5 &	11.55	  &	   1.73 \\
IRAS\,01003$-$2238* & 571.0 & 6 &  12.32 & 2.50 \\
IRAS\,20100$-$4156*$^{\dagger}$ & 633.9 & 6 & 12.67 & 2.85 \\
IRAS\,00397$-$1312* & 1378.7 & 6 & 12.90 & 3.08 \\
IRAS\,06035$-$7102* &	 372.5 & 6  &	12.22 & 2.40 \\
IRAS\,20414$-$1651* &	 412.5 & 6 &	12.22 & 2.40 \\
IRAS\,23253$-$5415* & 633.9 & 6 & 12.36 & 2.54\\
\end{supertabular}
\tablefoot{
\tablefoottext{a}{\footnotesize Distances for galaxies with redshifts $z<0.01$ are obtained from the Nearby Galaxies Catalog \citep{1988ngc..book.....T} (if available), while for other galaxies distances are derived from the redshifts reported on the NASA Extragalactic Database (NED). For Virgo cluster galaxies, we retrieve distances from the Goldmine database \citep{2003A&A...400..451G}, based on results reported in \citet{1999MNRAS.304..595G}. For galaxies with redshifts $z\geq0.01$, the luminosity distances are calculated for a spatially flat cosmology with $H_{\text{0}}$ = 67.3 km s$^{-1}$ Mpc$^{-1}$, $\Omega_{\lambda}$ = 0.685 and $\Omega_{\text{m}}$ = 0.315 \citep{2013arXiv1303.5076P}.}\\
\tablefoottext{b}{\footnotesize References: (1) \citet{2008ApJS..178..280B}; (2) \citet{2013ApJ...776...65P}; (3) Graci{\'a}-Carpio et al. (in prep.); (4) \citet{2012ApJ...755..171S}; (5) \citet{2013ApJ...774...68D}; (6) \citet{2013ApJ...776...38F}.}\\
\tablefoottext{c}{\footnotesize The FIR line measurements correspond to the central 80$\arcsec$ region in M\,51.}\\
\tablefoottext{d}{\footnotesize IRAS\,15206+3342 was observed during the Herschel performance verification phase.}}
\end{center}

\begin{center}
\label{table2b}
\onecolumn
\tablefirsthead{
\hline
Source & $D_{\text{L}}$\tablefootmark{a}   & Ref\tablefootmark{b} & $\log L_{\text{TIR}}$ & $\log SFR$ \\
& [Mpc] & & [$\log$ L$_{\odot}$] & [$\log$ M$_{\odot}$ yr$^{-1}$] \\
\hline}
\tablehead{
 \multicolumn{5}{r}{\textit{Continued from previous page}} \\
\hline
Source & $D_{\text{L}}$\tablefootmark{a}   & Ref\tablefootmark{b} & $\log L_{\text{TIR}}$ & $\log SFR$ \\
& [Mpc] & & [$\log$ L$_{\odot}$] & [$\log$ M$_{\odot}$ yr$^{-1}$] \\
\hline}
\tabletail{
\hline
\multicolumn{5}{l}{Continued on next page}\\}
\tablelasttail{\hline}
\topcaption{Overview of literature data for galaxies with composite or AGN source classification used for the SFR calibrations presented in this paper. The first, second and third column specify the name of the source, luminosity distance and reference to the literature work, respectively. The subsequent columns provide the total infrared (8-1000\,$\mu$m) luminosity and star formation rate, respectively.}
\begin{supertabular}{|lcccc|}
NGC\,0660		&	13.2	& 1 &	10.53	&	0.70 \\      
NGC\,1266		&	32.8	& 1 &	10.55	&	0.73  \\
NGC\,4278		&	10.8	& 1 &	8.59	&	-1.24 \\
NGC\,4293		&	18.9	& 1 &	9.74	&	-0.09 \\
NGC\,4651		&	18.7	& 1 &	9.87	&	0.04 \\
NGC\,4698		&	18.7	& 1 &	9.00	&	-0.83 \\ 
NGC\,6221		&	21.6	& 1 &	10.80	&	0.98 \\	 
NGC\,6810		&	28.2	& 1 &	10.70	&	0.88 \\
NGC\,7217		&	17.8	& 1 &	 9.83	&	0.003 \\
NGC\,449 & 71.9  & 1 & 10.74 & 0.91 \\
Mrk\,573 & 77.5  & 1 & 10.58 & 0.76 \\
NGC\,1052 & 22.5 & 1 & 9.30 & -0.52 \\
NGC\,1326 & 18.8 & 1 & 9.95 & 0.12 \\
IC\,356 & 18.1 & 1 & 9.85 & 0.02 \\
IC\,450 & 85.0 & 1 & 10.66 & 0.83 \\
NGC\,3705 & 18.9 & 1 & 9.75 & -0.07 \\
NGC\,4102 & 18.9 & 1 & 10.71 & 0.88 \\
NGC\,4314 & 17 & 1 & 9.53 & -0.30 \\
VCC\,857 & 17 & 1 & 9.17 & -0.65 \\
NGC\,4414 & 17 & 1 & 10.45 & 0.62 \\
VCC\,1003 & 17 & 1 & 9.31 & -0.52 \\
VCC\,1043 & 17 & 1 & 9.60 & -0.23 \\
VCC\,1110 & 17 & 1 & 9.32 & -0.51 \\
NGC\,4486 & 17 & 1 & 9.00 & -0.82 \\
VCC\,1727 & 17 & 1 & 9.78 & -0.05 \\
NGC\,5643 & 18.8 & 1 & 10.38 & 0.56 \\
I\,Zw\,92 &173.2 & 1 & 11.20 & 1.37 \\
NGC\,5866 & 17.1 & 1 & 9.79 & -0.04 \\
NGC\,5953 & 36.8 & 1 & 10.64 & 0.82 \\
NGC\,6217 & 26.6 & 1 & 10.39 & 0.57 \\
NGC\,6574 & 39.0 & 1 & 10.84 & 1.01 \\
NGC\,6764 & 41.2 & 1 & 10.57 & 0.75 \\
Mrk\,509 & 157.2 & 1 & 11.25 & 1.42 \\
NGC\,6958 & 39.5 & 1 & 9.82 & -0.003 \\
IC\,5063 & 51.0 & 1 & 10.88 & 1.06 \\
IC\,1459 & 22.3 & 1 & 9.12 & -0.71 \\
NGC\,1365		& 18.8 & 2 &	11.09	&	  1.26  \\
NGC\,3783		& 42.9 	& 2 &	10.27	&	0.45\\   
NGC\,4051		& 18.9 	& 2 &	 9.63	&	   -0.19  \\
NGC\,4151$^{\dagger}$		& 22.6 	& 2 &	9.95	&	0.12  \\
NGC\,4593		& 44.0 	& 2 &	10.34	&	 0.50\\
NGC\,5033		& 20.8 	& 2 &	10.44	&	0.61\\
NGC\,5506		& 32.0 	& 2 &	10.44	&	0.61  \\ 
NGC\,7469$^{\dagger}$		& 74.2 	& 2 &	11.70	&	1.88	   \\
IC\,4329A$^{\dagger}$			& 72.0	& 2 &	10.59	 &	   0.76  \\
Cen\,A			& 5.5	 & 2 &	9.98	&	0.16 \\
Circinus			& 4.7 	& 2 &	10.39 	  &      0.57\\   
NGC\,1068$^{\dagger}$		& 16.0 	& 2 &	11.26	&	1.43	  \\ 
NGC\,1275$^{\dagger}$		& 79.0 	& 2 &	11.19	&	  1.37  \\
NGC\,1386		& 18.8 	& 2 &	9.89	&	0.07 \\
NGC\,7314$^{\dagger}$		& 20.4 	& 2 &	9.61	&	-0.21  \\
NGC\,7582$^{\dagger}$		& 19.6	& 2 &	10.82	&	1.00 \\	   
Mrk\,3$^{\dagger}$			& 60.8	& 2 &	10.60	&	 0.77 \\ 
IRAS\,F18325$-$5926	& 90.0	& 2 &	10.89	&	1.06	   \\
UGC5101*			& 180.8	& 2 &	12.17	&	2.35\\	   
Mrk\,231*$^{\dagger}$			& 193.9	& 2 &	12.62	&	  2.79  \\
Mrk\,273*$^{\dagger}$			& 173.0	& 2 &	12.36	&	2.53	   \\
NGC\,6240*$^{\dagger}$		& 112.3	& 2 &	12.01	&	2.19	   \\
IRAS\,F23365$+$3604*$^{\dagger}$	& 301.1	& 2 &	12.37	&	2.55\\	   
IRAS\,F15250$+$3609*	& 257.0	& 2 &	12.19	&	  2.37\\  
IRAS\,F12112$+$0305*	& 342.7	& 2 &	12.48	&	2.65	   \\
IRAS\,F13120$-$5453*	& 141.6     & 2 &	12.47	&	2.64	   \\
IRAS\,F14348$-$1447*	& 390.3 	& 2 &	12.60	&	2.77	   \\
IRAS\,F05189$-$2524*$^{\dagger}$	& 196.6    & 2 &	12.25	&	2.43	   \\
IRAS\,F08572$+$3915*	& 270.7 	& 2 &	12.34	&	2.52\\	   
NGC\,3079		& 22.73 & 2 &	10.95	&	1.13 \\
M82\,center		& 5.8 	& 2 &	11.28	&	 1.45 \\ 
IRAS\,F07251$-$0248*\tablefootmark{c} &  277.6 & 2 & 12.47 & 2.65 \\
IRAS\,F19542$-$1110*\tablefootmark{d} & 292.0 & 2 & 12.21 & 2.39  \\   
NGC\,4418$^{\dagger}$	         & 31.9 & 2 & 11.26	&       1.43\\
Arp\,220*$^{\dagger}$			& 81.4	& 2 &	12.46	&	2.64	   \\
IRAS\,F14378$-$3651*	& 318.9 	& 2 &	12.33	&	  2.51  \\
     Mrk\,334		   &   99.2	& 3 &	11.16	&	   1.34\\
    IRAS00199$-$7426*   &   459.9	& 3 &	12.37	&	   2.55\\ 
     E12$-$G21		   &   150.7	& 3 &	11.10	&	   1.25 \\
IRAS\,F00456$-$2904SW*  &   531.1 	& 3 &   12.26 	&	   2.44\\
     MCG$-$03$-$04$-$014   &   160.0	& 3 &	11.67	&	   1.85\\
   NGC\,454		   &   54.8	& 3 &	10.31	&	  0.49\\
     ESO353$-$G020	   &   71.7	& 3 &	11.13	&	   1.31 \\
     IRAS\,F01364$-$1042 &   222.5	& 3 &	11.85	&	   2.03\\
     NGC\,788		   &   61.2	& 3 &	10.21	&	  0.39 \\
   Mrk\,590		   &   111.6	& 3 &	10.74    &	  0.92 \\
     UGC01845		   &   70.3	& 3 &	11.17	 &	   1.35 \\
    IC1816		   &   76.2	& 3 &	10.56	 &	  0.74\\
   NGC\,973		   &   73.1	& 3 &	10.51	 &	  0.69\\
 IRAS\,F02437$+$2122	   &   105.6	& 3 &	11.22   &	   1.40 \\
     UGC02369		   &   142.3	& 3 &	11.67	 &	   1.85\\ 
     Mrk\,1066		   &   53.9	& 3 &	10.97	 &	   1.15\\
 IRAS\,F03217$+$4022	   &   106.1	& 3 &	11.37	 &	   1.55\\
     Mrk\,609		   &   157.7	& 3 &	11.44	 &	   1.62\\
 IRAS\,F03359$+$1523	   &   161.9	& 3 &	11.62	 &	   1.80\\
IRAS\,F03450$+$0055      &  141.3 	& 3 &	11.09	 &	   1.27 \\
IRAS\,04103$-$2838*	   &   568.4	& 3 &	12.25	&	   2.43 \\
 IRAS\,04114$-$5117*	   &   607.5	& 3 &	12.29	&	   2.47 \\
 ESO420$-$G013	   &   53.5	& 3 &	11.15	&	   1.33\\
  3C120	                   &  150.7	& 3 &	11.23	&	   1.41\\ 
   ESO203$-$IG001	   &   245.0	& 3 &	11.90	&	   2.08\\ 
   MCG$-$05$-$12$-$006	   &   84.9	& 3 &	11.24	&	   1.42\\
   Ark120	                   &   149.3	& 3 & 	11.07	&	   1.25 \\
 IRAS\,F05187$-$1017	   &   128.8	& 3 &	11.33	&	   1.50 \\
2MASXJ05580206$-$3820043 & 154.9	& 3 &	11.23	&	   1.41\\ 
IRAS\,F06076$-$2139	   &   171.3	 & 3 &	11.68	&	   1.86 \\
 IRAS\,06301$-$7934*	   &   775.2	& 3 &	12.41	&	   2.59 \\
 IRAS\,06361$-$6217*	   &   792.6	& 3 & 	12.46	&	   2.63 \\
 NGC\,2273    	           &   31.6	& 3 &	10.35	&	  0.53 \\
    UGC03608		   &   96.9	& 3 &	11.24	&	   1.42\\
     IRAS\,F06592$-$6313  &   104.2	& 3 &	11.24	&	   1.42\\ 
   Mrk\,9	           &   183.1	& 3 &	11.20	&	   1.37 \\
  IRAS\,07598$+$6508*	    &  734.1	& 3 &	12.56	&	   2.74\\ 
 Mrk\,622	                    &  105.2	& 3 &	10.77	&	  0.95 \\
   ESO60$-$IG016	    &  209.6	& 3 &	11.77	&	   1.95\\ 
   Mrk\,18	            &  49.9	& 3 &	10.22	&	  0.40 \\
   MCG$-$01$-$24$-$012	    &  88.6	& 3 &	10.43	&	  0.61\\
    Mrk\,705		    &  132.9	& 3 &	10.75	&	  0.93\\ 
   IRAS\,F10038$-$3338     & 156.3	& 3 &	11.74	&	   1.92 \\
   NGC\,3393		    &  56.2	& 3 &	10.47	&	  0.65 \\
     ESO319$-$G022	    &  74.0	& 3 &	11.07	&	   1.25 \\
     IRAS12018$+$1941*   &   841.8	& 3 & 	12.55	&	   2.73 \\
   UGC07064		   &   113.5	& 3 &	11.09	&	   1.27 \\
   NGC\,4507		   &   53.0	& 3 &	10.74	&	  0.92 \\
    PG1244$+$026	    &  222.5	& 3 & 	11.05	&	   1.23\\ 
     ESO507$-$G070	   &   98.3	& 3 &	11.51	&	   1.69 \\
    NGC\,4941		   &   7.1	& 3 &	8.87	&	 -0.95 \\
    ESO323$-$G077	   &   70.3	& 3 &	11.03	&	   1.21\\ 
   MCG$-$03$-$34$-$064	    &  74.4	& 3 &	11.19	&	   1.37 \\
    IRAS\,F13279$+$3401  &   107.9	& 3 &	10.53	&	  0.71\\ 
   M$-$6$-$30$-$15		   &   34.5	& 3 &	10.04	&	  0.22 \\
    IRAS\,13342$+$3932*  &   900.9	& 3 &	12.49	&	   2.67\\ 
   IRAS\,F13349$+$2438*    & 517.1	& 3 &	12.35	&	   2.53 \\
    NGC\,5347		    &  40.9	& 3 &	10.26	&	  0.44 \\  
    OQ$+$208		   &   360.6	& 3 &	11.72	&	   1.90\\ 
    NGC\,5548		  &   77.6	& 3 &	10.68	&	  0.86 \\
     Mrk\,1490		   &   116.7	& 3 & 	11.37	&	   1.55\\ 
   PG1426$+$015	    &  410.0	& 3 & 	11.55	&	   1.73 \\
    PG1440$+$356	    &  367.6	& 3 &	11.57	&	   1.75\\ 
    NGC\,5728		    &  47.0	& 3 &	10.77	&	  0.95 \\
     NGC\,5793		    & 52.1	& 3 &	10.72	&	  0.90\\
    IRAS\,15001$+$1433*     & 809.5	& 3 &	12.52	&	   2.70\\ 
    IRAS\,15225$+$2350*      & 681.6	& 3 &	12.21	&	   2.39\\
    Mrk\,876*	            &  628.6	& 3 &	12.01	&	   2.19 \\
     IRAS\,F16164$-$0746   & 106.5	& 3 &	11.62	&	   1.80\\
     Mrk\,883		    &  173.2	& 3 &  	11.01	&	   1.19 \\
    IRAS\,16334$+$4630*    &  965.2	& 3 &	12.49	&	   2.67 \\
     ESO069$-$IG006	   &   213.9	& 3 &	11.98	&	   2.16 \\
     IRAS\,F16399$-$0937N & 122.7	& 3 &	11.55	&	   1.73 \\
    IRAS\,17044$+$6720*     & 660.3	& 3 &	12.21	&	   2.39\\
    IRAS\,17068$+$4027*     & 899.3	& 3 &	12.44	&	   2.62 \\
    IRAS\,F17132$+$5313    & 235.4	& 3 &	11.94	&	   2.12\\
     ESO138$-$G027	     & 94.1	& 3 &	11.41	&	   1.59\\
     CGCG141$-$034	     & 85.9	& 3 &	11.18	&	   1.36\\
     IC4734	             & 70.3	& 3 &	11.35	&	   1.53 \\
    IRAS18443$+$7433*     & 658.7	& 3 &	12.37	&	   2.55\\ 
    ESO140$-$G043	     & 63.9	& 3 &	10.65	&	  0.83 \\
     H1846$-$786	     & 348.3	& 3 &	11.46	&	   1.64 \\
    ESO593$-$IG008	     & 224.9	& 3 &	11.94	&	   2.12 \\
   ESO$-$141$-$G055	     & 169.9	& 3 &	11.11	&	   1.29 \\
    ESO339$-$G011	    &  86.8	& 3 &	11.18	&	   1.36 \\
    IRAS\,20037$-$1547*  &   971.4	& 3 &	12.66	&	   2.84 \\
   NGC\,6860	           &   67.1	& 3 &	10.44	&	  0.62\\
   NGC\,7213	           &   24.5	& 3 &	10.10	&	  0.28 \\
   ESO602$-$G025	   &   113.5	& 3 &	11.39	&	   1.57\\ 
    UGC12138		   &   113.5	& 3 &  	10.82	&	   1.00 \\
     UGC12150		   &   96.9	& 3 &	11.38	&	   1.56 \\
    ESO239$-$IG002	    &  197.7	& 3 &	11.85	&	   2.03 \\
     Zw453.062		   &   113.9	& 3 &	11.41	&	   1.59\\
    NGC\,7603		   &   134.3	& 3 &	10.90	&	   1.08 \\
ESO339$-$G011	& 87.7			& 4 &	11.00	  &	   1.18 \\
NGC6926		& 89.5			&  4 &	10.34	  &	  0.52 \\
ESO343$-$IG013$\_$N	& 86.8			& 4 &	10.86	  &	   1.04 \\
CGCG448$-$020$\_$E	& 168.5			& 4 &	11.78	  &	   1.96 \\
CGCG448$-$020c	& 168.5			& 4 &	10.66	  &	  0.84 \\
IC1623B		&  91.8			&  4 &	11.47	  &	   1.65 \\
IC2810b		&	   158.1	& 4 &	11.27	  &	   1.45 \\
NGC4922		&	   108.4	& 4 &	11.28	  &	   1.46 \\
CGCG011$-$0767	&	   114.4	& 4 &	11.26	  &	   1.44\\ 
IRAS23436$+$5257	& 158.6			& 4 &	11.54	  &	   1.72\\ 
ESO264$-$G036	&	   96.0		& 4 &	10.97	  &	   1.15 \\
NGC5135		&	    62.1	& 4 &	11.33	  &	   1.51 \\
MCG$-$03$-$34$-$064	&	    75.3	&  4 &	10.85	  &	  1.03\\ 
IRAS08355$-$4944	&	   119.0	&  4 &	11.39	  &	   1.57 \\
ESO203$-$IG001	 &    252.2		&  4 &	11.91	  &	   2.09 \\
IRAS12116$-$5615	&	   125.0	& 4  &	11.63	  &	   1.81\\ 
IRASF06076$-$2139	&	   175.1	& 4  &	11.67	  &	   1.85 \\
NGC2342b	&	    80.4	& 4 &	11.02	  &	   1.20\\
NGC2388b	&	    62.6	& 4 &	9.93	  &	 0.11 \\
NGC2623		&	    84.5	& 4 &	11.71	  &	   1.89\\ 
CGCG142$-$034$\_$W	&	    85.4	& 4 &	10.44	  &	  0.62\\
UGC04881$\_$E	&	   185.0	& 4 &	11.58	  &	   1.76\\
MCG$+$02$-$20$-$003$\_$N	 &  74.0		& 4 &	11.07	  &	   1.25\\
NGC1961 &    59.4 & 4 &     9.66 &    -0.16\\
ESO432$-$IG006$\_$E	&	    73.5	&  4 & 	10.75	  &	  0.93 \\
NGC2369		&	    49.0	& 4 &	11.11	  &	   1.29 \\
IRASF06592$-$6313	&	 105.2		&  4 &	11.34	  &	   1.52 \\ 
IRASF17132$+$5313	&	  242.1		&  4 &	11.75	  &	   1.93 \\
VV414$\_$E		&	  115.3		&  4 &	11.01	  &	   1.19\\
 ESO60$-$IG016	&	  218.7		&  4 &	11.67	  &	   1.86\\
 IRASF10173$+$0828	 & 233.0		&  4 &	11.89	  &	   2.07 \\
NGC5256		&	   128.3	&  4 &	11.30	  &	   1.48\\
MCG$+$08$-$18$-$013	&	  119.5		& 4 &	11.35	  &	   1.53\\ 
ESO077$-$IG014$\_$E	&	   195.4	& 4 &	11.68	  &	   1.86\\
 NGC7592$\_$W	&	   112.1	& 4 &	10.32	  &	  0.50\\
 NGC7674		&	   133.9	& 4 &	11.24	  &	   1.42\\
 IRASF12224$-$0624	  & 121.3		& 4 &	11.41	  &	   1.59\\
 IC0860		&	    50.3	& 4 &	11.08	  &	   1.26\\
 ESO239$-$IG002	 & 202.5		& 4 &	11.90	  &	   2.08\\
 NGC7591		&	    75.3	& 4 &	11.07	  &	   1.25\\
 MCG$-$01$-$60$-$022	 & 106.5		& 4 &	11.12	  &	   1.30\\
 IRASF01364$-$1042	  & 228.7		& 4 &	11.86	  &	   2.04\\
NGC\,34		&	    89.5	& 4 &	11.50	  &	   1.68 \\
ESO319$-$G022	&	    74.4	& 4 &	11.01	  &	   1.19\\
 IC5298		&	   126.4	& 4 &	11.54	  &	   1.72\\
 Arp86$\_$N		&	    77.6	& 4 &	9.94	  &	  0.12\\
 NGC5734$\_$N	&	    62.1      & 4 &	10.78	  &	  0.96 \\
NGC5734$\_$S	&	    62.1	& 4 &	10.43	  &	  0.61\\ 
NGC5990		&	    58.0	& 4 &	10.81	  &	  0.99\\
NGC0232$\_$E	&	    101.5	& 4 &	10.75	  &	  0.93\\
NGC77714$\_$W	&	    64.8	& 4 &	10.21	  &	  0.39\\
ESO420$-$G013	&	    53.9	& 4 &	10.98	  &	   1.16\\
MCG$+$05$-$06$-$036$\_$N	 & 156.7		& 4 &	10.98	  &	   1.16\\ 
CGCG436$-$030	&	   144.6	& 4 &	11.67	  &	   1.85\\
 IIIZw035$^{\dagger}$	&	   128.8	& 4 &	11.66	  &	   1.84 \\
NGC6156$^{\dagger}$		&	    49.0 	& 4 &	10.57	  &	  0.75\\
 NGC0838$\_$W	&	    58.0	& 4 &	10.53	  &	  0.71 \\
NGC0958		&	    87.2	& 4 &	10.00	  &	  0.18\\
NGC0877$\_$S	&	    58.9	& 4 &	10.34	  &	  0.52 \\
NGC0877$\_$N	&	    58.9	& 4 &	9.98	  &	  0.16 \\
UGC02608$\_$N	&	   107.0	& 4 &	11.25	  &	   1.43\\ 
IRASF03359$+$1523	&	   165.2	& 4 &	11.59	  &	   1.77\\ 
NGC1572		&	    93.2	& 4 &	11.39	  &	   1.57 \\
IRASF02437+2122	&	    107.0	& 4 &	11.26	  &	   1.44\\ 
ESO550$-$IG025a	&	  148.8		& 4 &	11.36	  &	   1.54\\ 
ESO550$-$IG025b	&	   148.8	& 4 &	11.09	  &	   1.27 \\
IC4687b		&	    79.0	& 4 &	10.56	  &	  0.74 \\
IC4687c		&	    79.0	& 4 &	10.84	  &	  1.02\\
IC4518ABa	&	    72.1	& 4 &	10.79	  &	  0.97 \\
UGC03094	&	   113.5	& 4 &	11.12	  &	   1.30 \\
IRASF16399-0937	& 124.6			& 4 &	11.58	  &	   1.76\\ 
CGCG141$-$034	&	    90.4	& 4 &	11.19	  &	   1.37 \\
CGCG468$-$002$\_W$	&	    83.1	& 4 &	10.54	  &	  0.72\\ 
IRAS03582$+$6012$\_$E &	    139.0	& 4 &	11.13	  &	   1.31 \\
ESO602$-$G025	&	   115.3	& 4 &	11.31	  &	   1.49 \\
ESO453$-$G005$\_$S	 & 95.0			& 4 &	11.36	  &	   1.54\\ 
AM0702$-$601$\_$N	 & 145.1		& 4 &	11.14	  &	   1.32 \\
IRAS\,08311$-$2459* & 478.2 & 5 &   12.50 & 2.68 \\
IRAS\,11095$-$0238* & 514.1 & 5 &  12.28 & 2.46 \\
IRAS\,23230$-$6926* & 514.1 & 5 &  12.37 & 2.55 \\
3C\,273* & 783.9 & 5  &     12.83 & 3.01 \\
IRAS\,06206$-$6315* &	 437.6 & 5 &	12.23 & 2.41 \\
Mrk\,463		    &	 235.9 & 5 &	11.79 & 1.97\\
IRAS\,19254$-$7245*$^{\dagger}$ &	 293.8 & 5 &	12.09 & 2.27\\
IRAS\,15462$-$0450* & 478.2 & 5 &  12.24 & 2.42 \\
IRAS\,20087$-$0308* & 509.0 & 5 &  12.42 & 2.60 \\
IRAS\,13451$+$1232* & 586.6 & 5 &  12.32 & 2.50 \\
IRAS\,00188$-$0856* & 623.3 & 5 &   12.39 & 2.57 \\
IRAS\,12071$-$0444* & 623.3 & 5 & 12.41 & 2.59 \\
IRAS\,03158$+$4227* & 655.0 & 5 &  12.63 & 2.81 \\
IRAS\,16090$-$0139* & 655.0 & 5 & 12.55 & 2.73 \\
IRAS\,10378$+$1109* & 665.6 & 5 & 12.31 & 2.49 \\
IRAS\,07598$+$6508* & 729.8 & 5 & 12.50 & 2.68 \\
IRAS\,03521$+$0028* & 751.3 & 5 & 12.52 & 2.70 \\
Mrk\,1014* & 811.1 & 5 & 12.62 & 2.80 \\
\end{supertabular}
\tablefoot{
\tablefoottext{a}{\footnotesize Distances for galaxies with redshifts $z<0.01$ are obtained from the Nearby Galaxies Catalog \citep{1988ngc..book.....T} (if available), while for other galaxies distances are derived from the redshifts reported on the NASA Extragalactic Database (NED). For Virgo cluster galaxies, we retrieve distances from the Goldmine database \citep{2003A&A...400..451G}, based on results reported in \citet{1999MNRAS.304..595G}. For galaxies with redshifts $z\geq0.01$, the luminosity distances are calculated for a spatially flat cosmology with $H_{\text{0}}$ = 67.3 km s$^{-1}$ Mpc$^{-1}$, $\Omega_{\lambda}$ = 0.685 and $\Omega_{\text{m}}$ = 0.315 \citep{2013arXiv1303.5076P}.}\\
\tablefoottext{b}{\footnotesize References: (1) \citet{2008ApJS..178..280B}; (2) Graci{\'a}-Carpio et al. (in prep.); (3) \citet{2012ApJ...755..171S}; (4) \citet{2013ApJ...774...68D}; (5) \citet{2013ApJ...776...38F}.}\\
\tablefoottext{c}{\footnotesize IRAS\,F07251$-$0248 did not have an optical source classification in Graci{\'a}-Carpio et al. (in prep.). By relying on the equivalent width of the 6.2\,$\mu$m PAH feature (EW (PAH 6.2\,$\mu$m) = 0.29, \citealt{2013ApJS..206....1S}), we classify the source as an AGN.}\\
\tablefoottext{d}{\footnotesize IRAS\,F19542$-$1110 did not have an optical source classification in Graci{\'a}-Carpio et al. (in prep.). By relying on the equivalent width of the 6.2\,$\mu$m PAH feature (EW (PAH 6.2\,$\mu$m) = 0.09, \citealt{2013ApJS..206....1S}), we classify the source as an AGN.}}
\twocolumn
\end{center}

\end{document}